\documentclass[prd,amsmath,amssymb,superscriptaddress,preprintnumbers,twocolumn,10pt]{revtex4-1}

\pdfoutput=1

\usepackage{graphicx}
\usepackage{dcolumn}
\usepackage{bm}
\usepackage{amssymb}
\usepackage{latexsym}
\usepackage{booktabs}
\usepackage{amsmath}
\usepackage{multirow}
\usepackage{url}

\usepackage{float}
\usepackage[colorlinks=true, linkcolor=red, citecolor=blue]{hyperref}

\newcommand{\ve}{\varepsilon}
\usepackage[normalem]{ulem}
\usepackage{color}
\usepackage{array}
\usepackage{enumerate}

\begin{document}

\title{How can gravitational-wave standard sirens and 21-cm intensity mapping jointly provide a precise late-universe cosmological probe?}

\author{Shang-Jie Jin}
\affiliation{Department of Physics, College of Sciences, Northeastern University, Shenyang 110819, China}
\author{Ling-Feng Wang}
\affiliation{Department of Physics, College of Sciences, Northeastern University, Shenyang 110819, China}
\author{Peng-Ju Wu}
\affiliation{Department of Physics, College of Sciences, Northeastern University, Shenyang 110819, China}
\author{Jing-Fei Zhang}
\affiliation{Department of Physics, College of Sciences, Northeastern University, Shenyang 110819, China}

\author{Xin Zhang\footnote{Corresponding author}}
\email{zhangxin@mail.neu.edu.cn}
\affiliation{Department of Physics, College of Sciences, Northeastern University, Shenyang 110819, China}
\affiliation{MOE Key Laboratory of Data Analytics and Optimization for Smart Industry, Northeastern University, Shenyang 110819, China}

\begin{abstract}
In the next decades, the gravitational-wave (GW) standard siren observations and the neutral hydrogen 21-cm intensity mapping (IM) surveys, as two promising cosmological probes, will play an important role in precisely measuring cosmological parameters.
In this work, we make a forecast for cosmological parameter estimation with the synergy between the GW standard siren observations and the 21-cm IM surveys. We choose the Einstein Telescope (ET) and the Taiji observatory as the representatives of the GW detection projects and choose the Square Kilometre Array (SKA) phase I mid-frequency array as the representative of the 21-cm IM experiments. In the simulation of the 21-cm IM data, we assume perfect foreground removal and calibration. We find that the synergy of the GW standard siren observations and the 21-cm IM survey could break the cosmological parameter degeneracies. The joint ET+Taiji+SKA data give $\sigma(H_0)=0.28\ {\rm km\ s^{-1}\ Mpc^{-1}}$ in the $\Lambda$CDM model, $\sigma(w)=0.028$ in the $w$CDM model, which are better than the results of $Planck$+BAO+SNe, and $\sigma(w_0)=0.077$ and $\sigma(w_a)=0.295$ in the CPL model, which are comparable with the results of $Planck$+BAO+SNe. In the $\Lambda$CDM model, the constraint precision of $H_0$ and $\Omega_{\rm m}$ is less than or rather close to 1\%, indicating that the magnificent prospects for precision cosmology {with these two promising cosmological probes} are worth expecting.

\end{abstract}

\maketitle

\section{Introduction}\label{sec:intro}

The precise measurements of the cosmic microwave background (CMB) anisotropies initiated the era of precision cosmology \cite{Bennett:2003bz,Spergel:2003cb}.
The $\Lambda$CDM model as the standard model of cosmology can fit the CMB data with breathtaking precision. Nevertheless, extra cosmological parameters in the extended cosmological models cannot be tightly constrained by solely using the CMB data due to the strong cosmological parameter degeneracies. Thus, the measurements of the late universe are needed as the supplements of the CMB data to break the cosmological parameter degeneracies. However, there are inconsistencies between the early and late universe. For example, the tension between the values of the Hubble constant inferred from the CMB observation \cite{Aghanim:2018eyx} and the Cepheid-supernova distance ladder measurement \cite{Riess:2020fzl} has now reached $4.2\sigma$ \cite{Riess:2020fzl}. The Hubble tension has been intensively discussed in the literature \cite{cai:2020,Guo:2018ans,Guo:2019dui,Yang:2018euj,Vagnozzi:2019ezj,DiValentino:2019jae,DiValentino:2019ffd,Liu:2019awo,Zhang:2019cww,Ding:2019mmw,Feng:2019jqa,Lin:2020jcb,Li:2020tds,Hryczuk:2020jhi,Gao:2021xnk,Wang:2021kxc,Cai:2021wgv,Vagnozzi:2021tjv,Vagnozzi:2021gjh}. It is now commonly believed that the Hubble tension is a severe crisis for cosmology \cite{Verde:2019ivm,Riess:2020sih}.
To solve the current cosmological tensions, one crucial way is to develop new powerful late-universe cosmological probes, besides conceiving novel cosmological models.
Since the current measurements of the late universe are mainly based on optical observations, it is important to develop other new-type cosmological probes.
Obviously, the gravitational-wave (GW) standard siren observations and the neutral hydrogen (H{\tt I}) 21-cm radio observations are two promising cosmological probes.

The standard siren method could be applied in measuring cosmological parameters by establishing the relation between luminosity distance and redshift, which was first proposed by Schutz \cite{Schutz:1986gp}. The absolute luminosity distance to the GW source could be directly obtained from the analysis of the GW waveform.
If the source's redshift can also be obtained by identifying its electromagnetic (EM) counterpart, then this GW-EM event could be treated as a standard siren for exploring the expansion history of the universe \cite{Holz:2005df}. The first actual application of standard siren is using GW170817 \cite{TheLIGOScientific:2017qsa} and its EM counterpart (GRB 170817A) \cite{Monitor:2017mdv,GBM:2017lvd} to measure the Hubble constant, which gives a result with around 15\% precision \cite{Abbott:2017xzu}. A further forecast analysis shows that the measurement {precision} of the Hubble constant could achieve about 2\% using 50 similar standard siren events \cite{Chen:2017rfc}. It can be anticipated that GWs could help resolve the Hubble tension with the accumulation of observed standard siren events. Recently, the GW standard sirens have been widely discussed \cite{Cai:2017buj,DiValentino:2017clw,Yang:2019vni,Zhao:2018gwk,DiValentino:2018jbh,Gray:2019ksv,Chen:2020dyt,Chen:2020zoq,Chen:2020gek,Chen:2020dyt,Zhang:2019ylr,Wang:2018lun,Zhang:2018byx,Li:2019ajo,Zhang:2019loq,Zhang:2019ple,Wang:2019tto,Zhao:2019gyk,Jin:2020hmc,Wang:2021srv,Qi:2021iic,Yang:2021qge,Yu:2021nvx}.

The development of standard siren method in the next decades depends on the future GW detectors. The third-generation ground-based GW detectors, i.e., the Cosmic Explorer \cite{Evans:2016mbw} and the Einstein Telescope (ET) \cite{Punturo:2010zz}, aimed at high frequency-band (a few hundred hertz) GW detections, will begin observing in the 2030s. At the same time, the space-based GW detectors, i.e., LISA \cite{Audley:2017drz}, TianQin \cite{Luo:2020bls,mei2021tianqin}, and Taiji \cite{Hu:2017mde,Guo:2018npi,Taiji-1}, will open the window of detecting milli-hertz frequency-band GWs. Due to the different detection-frequency bands of the ground-based and space-based detectors, synergistically utilizing these GW detectors allows the standard sirens to be realized with different GW sources, e.g., binary neutron star (BNS), binary black hole (BBH), and massive black hole binary (MBHB). The combination of the standard sirens from different GW sources will definitely improve the capability of constraining cosmological parameters. In this work, we shall simulate the GW standard sirens based on the observations from ET (aimed at detecting BNSs) and Taiji (aimed at detecting MBHBs), and use the combination of them as the GW standard siren data.

Moreover, the H{\tt I} 21-cm radio observation is another promising cosmological probe. In the post-reionization epoch of the universe, H{\tt I} is thought to reside in dense gas clouds embedded in galaxies, so it is essentially a tracer of the galaxy distribution. Actually, it is difficult to detect enough H{\tt I}-emitting galaxies to make an accurate cosmological analysis. However, we can simply measure the total H{\tt I} intensity over comparatively large angular scales to study the large-scale structure of the universe, of which the method is called 21-cm intensity mapping (IM). Using the 21-cm IM technique, one could measure the scale of baryon acoustic oscillations (BAO) that is a cosmological standard ruler, thus accurately measuring the late-time expansion history of the universe. Many 21-cm IM experiments have been proposed to measure the H{\tt I} power spectrum and other features of the large-scale structure, e.g., the baryon acoustic oscillations from integrated neutral gas observations (BINGO) \cite{Battye:2012tg,Dickinson:2014wda}, the five-hundred-meter aperture spherical radio telescope {(FAST)} \cite{Nan:2011um,Li:2012ub,Yu:2017hqz}, the square kilometre array (SKA) \cite{Braun:2015zta,Bull:2015esa,Bacon:2018dui,Braun:2019gdo}, and the Tianlai cylinder array \cite{2011SSPMA..41.1358C,Chen:2012xu,Xu:2014bya}. A series of forecasts indicate that 21-cm IM could play an important role in the cosmological parameter estimation \cite{Bacon:2018dui,Zhang:2019dyq,Zhang:2019ipd,Zhang:2021yof} (see {also} Ref.~\cite{Xu:2020uws} for a brief review).

Actually, as two promising cosmological probes, standard sirens and 21-cm IM have different advantages. Standard sirens allow the direct measurements of $d_{\rm L}(z)$ that is inversely proportional to $H_0$, so a large number of standard sirens could constrain $H_0$ well.
The 21-cm IM survey, compared to the optical survey, has some advantages in such as larger survey volumes, deeper redshifts, higher survey efficiency, and so forth.
In addition, the BAO measurements by 21-cm IM can provide the information of $H(z)$ that is related to $w(z)$ by only one integral, therefore, compared with the distance--redshift relation that is related to $w(z)$ by two integrals, radial BAO may provide better constraints on $w(z)$.
This implies that the combination of standard sirens and 21-cm IM may constrain both $H_0$ and $w(z)$ well. Hence, we wish to investigate the capability of estimating cosmological parameters using the combination of these two promising cosmological probes.

Based on the motivations described above, in this work we focus on the synergy of the GW standard siren observations and the 21-cm IM surveys in cosmological parameter estimation. For the {simulation} of standard sirens, we choose {ET and Taiji} as the representatives of the GW detection projects. For the {simulation} of 21-cm IM observations, we choose SKA as the representative of the 21-cm IM experiments (note that we consider perfect foreground removal and calibration in our simulation).
Since the SKA phase I mid-frequency (SKA1-MID) array focuses on exploring the evolution of the late universe \cite{Bull:2015esa}, we only consider SKA1-MID in this work. For the cosmological models, we take the $\Lambda$CDM, $w$CDM, and CPL models as typical examples. The flat $\Lambda$CDM model is taken as the fiducial model to generate mock data, with the fiducial values of cosmological parameters being set to the constraint results from $Planck$ 2018 TT,TE,EE+lowE \cite{Aghanim:2018eyx}.

This work is organized as follows. In Sec.~\ref{sec:GW}, we introduce the methods of simulating GW standard sirens. In Sec.~\ref{sec:IM}, we briefly describe the methods of simulating 21-cm IM data based on SKA. In Sec.~\ref{sec:re}, we give the constraint results and make some relevant discussions. The conclusion is given in Sec.~\ref{sec:con}. Unless otherwise stated, we adopt the system of units in which $G=c=1$ throughout this paper.

\section{Gravitational wave standard siren observation}\label{sec:GW}

\subsection{Simulation of GW standard sirens from ET}\label{sec:GWET}

The frequency band detected by the ground-based GW detectors corresponds to the mergers of binary stellar-mass black holes, BNSs, or neutron star-black hole binaries. In this work, we assume that all the GW standard siren events detected by ET are produced by the BNS merger events. For the redshift distribution of BNSs, we adopt the form in Ref.~\cite{Zhao:2010sz}.
In this paper, we adopt the restricted post-Newtonian (PN) approximation and calculate the waveform to the 3.5 PN order \cite{Sathyaprakash:2009xs}. The Fourier transform $\tilde{h}(f)$ of the time-domain waveform is given by
\begin{align}
\tilde{h}(f)=\mathcal{A}f^{-7/6}\exp \{i\big(2 \pi f t_{\rm c}-\pi / 4+2 \Psi(f / 2)-\varphi_{(2,0)}\big)\},
\end{align}
where the Fourier amplitude $\mathcal{A}$ is given by
\begin{align}
\mathcal{A}=&~~\frac{1}{d_{\rm L}}\sqrt{F_+^2\big(1+\cos^2(\iota)\big)^2+4F_\times^2\cos^2(\iota)}\nonumber\\
            &~~\times \sqrt{5\pi/96}\pi^{-7/6}\mathcal{M}_{\rm c}^{5/6},
\end{align}
$\Psi(f)$ and $\varphi_{(2,0)}$ are given by \cite{Sathyaprakash:2009xs,Blanchet:2004bb}
\begin{align}
&\Psi(f)=-\psi_{\rm c}+\frac{3}{256 \eta} \sum_{i=0}^{7} \psi_{i}(2\pi Mf)^{(i-5) / 3}, \label{Psi}\\
&\varphi_{(2,0)}=\tan ^{-1}\left(-\frac{2 \cos (\iota) F_{\times}}{\big(1+\cos ^{2}(\iota)\big) F_{+}}\right),
\end{align}
where $d_{\rm L}$ is the luminosity distance to the GW source, $F_{+, \times}$ are antenna pattern functions, $\iota$ is the inclination angle between the binary's orbital angular momentum and the line of sight, $\mathcal{M}_{\rm c}=(1+z)\eta^{3/5}M$ is the observed chirp mass, $M=m_1+m_2$ is the total mass of binary system with component masses $m_1$ and $m_2$, $\eta=m_1 m_2/(m_1+m_2)^2$ is the symmetric mass ratio, $\psi_{\rm c}$ is the coalescence phase, {and the} coefficients $\psi_{i}$ are given by \cite{Sathyaprakash:2009xs}
\begin{align}
\psi_{0} &=1, \quad \psi_{1}=0, \quad \psi_{2}=\frac{3715}{756}+\frac{55}{9} \eta, \quad \psi_{3}=-16 \pi, \nonumber\\
\psi_{4} &=\frac{15293365}{508032}+\frac{27145}{504} \eta+\frac{3085}{72} \eta^{2}, \nonumber\\
\quad \psi_{5}&=\pi\left(\frac{38645}{756}-\frac{65}{9} \eta\right)\left[1+\ln \left(6^{3 / 2} \pi M f\right)\right], \nonumber\\
\psi_{6} &=\frac{11583231236531}{4694215680}-\frac{640}{3} \pi^{2}-\frac{6848}{21} \gamma \nonumber\\
&+\left(-\frac{15737765635}{3048192}+\frac{2255}{12} \pi^{2}\right) \eta \nonumber\\
&+\frac{76055}{1728} \eta^{2}-\frac{127825}{1296} \eta^{3}-\frac{6848}{63} \ln (64 \pi M f), \nonumber\\
\psi_{7} &=\pi\left(\frac{77096675}{254016}+\frac{378515}{1512} \eta-\frac{74045}{756} \eta^{2}\right),\label{coefficients}
\end{align}
where $\gamma=0.577$ is the Euler's constant.

The antenna pattern functions of ET are \cite{Zhao:2010sz}
\begin{align}
F_+^{(1)}(\theta, \phi, \psi)=&~~\frac{{\sqrt 3 }}{2}\bigg[\frac{1}{2}\big(1 + {\cos ^2}(\theta )\big)\cos (2\phi )\cos (2\psi ) \nonumber\\
&- \cos (\theta )\sin (2\phi )\sin (2\psi )\bigg],\nonumber\\
F_\times^{(1)}(\theta, \phi, \psi)=&~~\frac{{\sqrt 3 }}{2}\bigg[\frac{1}{2}\big(1 + {\cos ^2}(\theta )\big)\cos (2\phi )\sin (2\psi )\nonumber\\
&+ \cos (\theta )\sin (2\phi )\cos (2\psi )\bigg],
\end{align}
where ($\theta$, $\phi$) are angles describing the location of the source in the sky, and $\psi$ is the polarization angle. {Notice that here $\psi$ is the polarization angle, different from those $\psi_{i}$ in Eqs.~(\ref{Psi}) and (\ref{coefficients}), which are the PN coefficients.} Since ET has three interferometers with $60^\circ$ inclined angles between each other, the other two pattern functions are $F_{+,\times}^{(2)}(\theta, \phi, \psi)=F_{+,\times}^{(1)}(\theta, \phi+2\pi/3, \psi)$ and $F_{+,\times}^{(3)}(\theta, \phi, \psi)=F_{+,\times}^{(1)}(\theta, \phi+4\pi/3, \psi)$.

\begin{figure}[htbp]
\includegraphics[width=0.95\linewidth,angle=0]{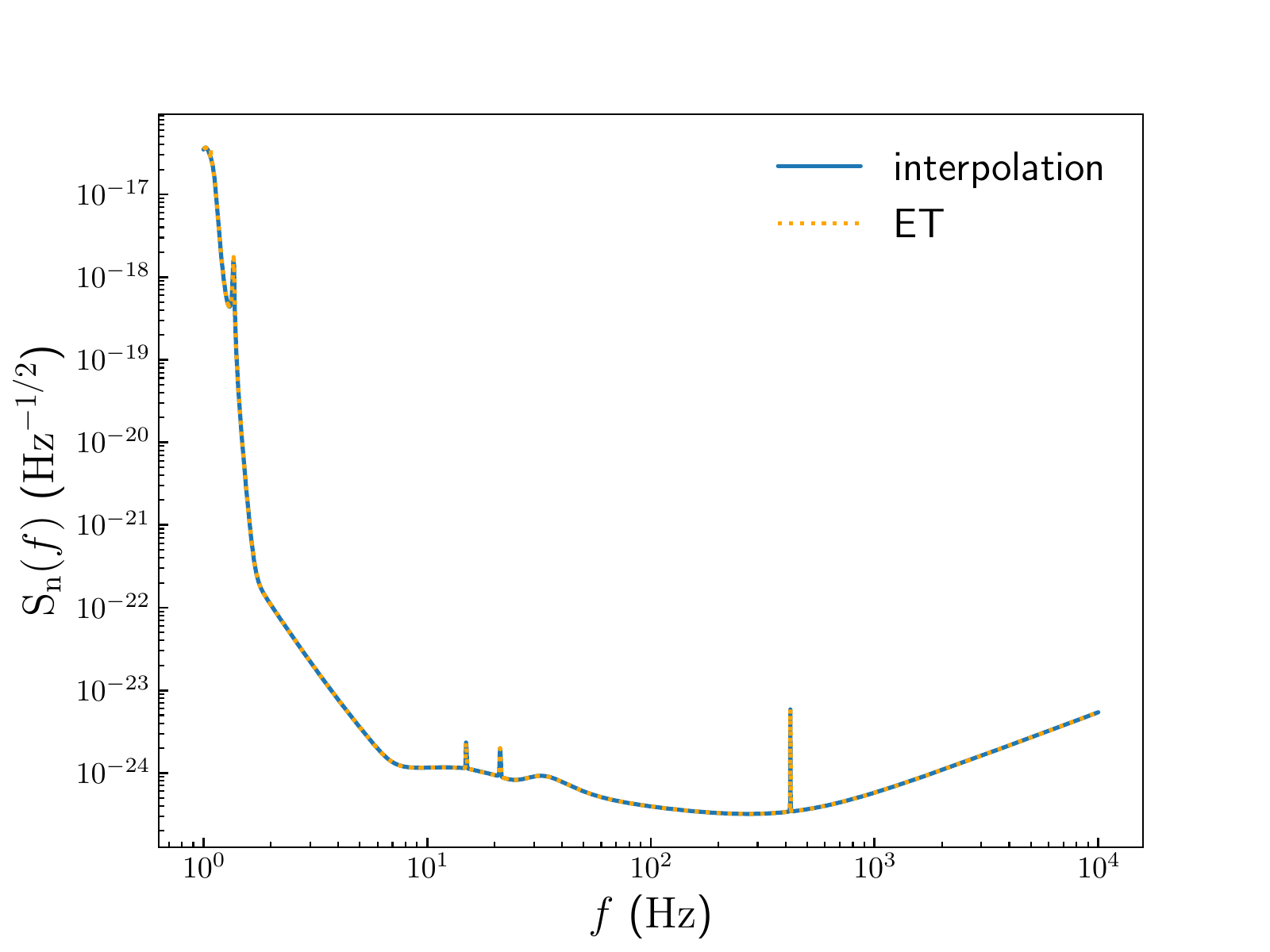}
\caption{\label{curve} The fitting function of $S_{\rm n}(f)$ of ET obtained using the interpolation method.}
\end{figure}

Then we need to select the GW events with signal-to-noise ratios (SNRs) greater than 8 in our simulation. The combined SNR for the detection network of $N$ independent interferometers is given by
\begin{equation}
\rho=\sqrt{\sum\limits_{i=1}^{N}(\rho_{i})^2},\label{snr1}
\end{equation}
where $\rho_{i}=\sqrt{\left\langle \tilde{h}_{i},\tilde{h}_{i}\right\rangle}$. The inner product is defined as
\begin{equation}
\left\langle{a,b}\right\rangle=4\int_{f_{\rm lower}}^{f_{\rm upper}}\frac{a(f) b^\ast(f)+ a^\ast(f) b(f)}{2}\frac{df}{S_{\rm n}(f)}.\label{snr2}
\end{equation}
Here, $f_{\rm lower}=1$ {Hz} is the lower cutoff frequency, $f_{\rm upper}=2/{(6^{3/2}2\pi M_{\rm obs})}$ is the frequency at the last stable orbit with $M_{\rm obs}=(m_1+m_2)(1+z)$ \cite{Zhao:2010sz}, $S_{\rm n}(f)$ is the one-side noise power spectral density (PSD), and we obtain the fitting function of $S_{\rm n}(f)$ using the interpolation method to fit the {sensitivity data of ET} \cite{ETcurve-web}.
The fitting function obtained by the interpolation method and the {sensitivity data of ET} are plotted in Fig.~\ref{curve}.

A few $\times 10^5$ BNS mergers per year could be observed by ET, but only about 0.1\% of them may have $\gamma$-ray bursts toward us \cite{Yu:2021nvx}, which means that a few $\times 10^2$ GW events' redshifts could be obtained per year. Chen \emph{et al.} recently made a forecast {showing} that 910 GW standard siren events could be detected based on the 10-year observation of CE and Swift++ \cite{Chen:2020zoq}.
Therefore, in our forecast in the present work, we simulate 1000 GW standard siren events generated by BNS mergers corresponding to the 10-year operation time of ET.

For ET, we consider three measurement errors of $d_{\rm L}$, consisting of the instrumental error $\sigma_{d_{\rm L}}^{\rm inst}$, the weak-lensing error $\sigma_{d_{\rm L}}^{\rm lens}$, and the peculiar velocity error $\sigma_{d_{\rm L}}^{\rm pv}$. The total error of $d_{\rm L}$ is
\begin{align}
(\sigma_{d_{\rm L}})^2&=(\sigma_{d_{\rm L}}^{\rm inst})^2+(\sigma_{d_{\rm L}}^{\rm lens})^2+(\sigma_{d_{\rm L}}^{\rm pv})^2.\label{total}
\end{align}

First, we need to use the Fisher information matrix to calculate $\sigma_{d_{\rm L}}^{\rm inst}$. For a network including $N$ independent detectors, the Fisher information matrix can be written as
\begin{equation}
\boldsymbol{F}_{ij}=\left\langle\frac{\partial \boldsymbol{h}(f)}{\partial \theta_i},  \frac{\partial \boldsymbol{h}(f)}{\partial \theta_j}\right\rangle,
\end{equation}
with $\boldsymbol{h}$ given by
\begin{equation}
\boldsymbol{h}(f)=\left[\tilde{h}_1 (f),\tilde{h}_2 (f),\cdots,\tilde{h}_N (f)\right],
\end{equation}
where $\theta_i$ denotes nine parameters ($d_L$, $M_c$, $\eta$, $\theta$, $\phi$, $\iota$, $t_c$, $\psi_{c}$, $\psi$) for a GW event. Then we have
\begin{equation}
\Delta \theta_i =\sqrt{(F^{-1})_{ii}},
\end{equation}
where $F_{ij}$ is the total Fisher information matrix for the network of $N$ detectors. Note that here $\sigma_{d_{\rm L}}^{\rm inst}=\Delta \theta_1$.

In our previous works \cite{Wang:2018lun,Zhang:2018byx,Zhang:2019ple,Wang:2019tto,Zhang:2019loq,Li:2019ajo,Zhao:2019gyk,Jin:2020hmc}, we calculated SNRs of GW events to obtain $\sigma_{d_{\rm L}}^{\rm inst}$ using the relation $\sigma_{d_{\rm L}}^{\rm inst}=2d_{\rm L}/{\rho}$. Actually, the randomness of the GW source's parameters may lead to the randomness of $\sigma_{d_{\rm L}}^{\rm inst}$. In this work, in order to {remove this randomness in the} result, we randomly choose the source parameters to perform the Fisher matrix analysis for 100 times, and calculate the average of the 100 matrices.

In addition to $\sigma_{d_{\rm L}}^{\rm inst}$, the measurement of luminosity distance is also affected by the weak lensing and we adopt the form in Ref.~\cite{Hirata:2010ba}
\begin{equation}
\sigma_{d_{\rm L}}^{\rm lens}(z)=d_{\rm L}(z)\times 0.066\bigg[\frac{1-(1+z)^{-0.25}}{0.25}\bigg]^{1.8}.\label{lens}
\end{equation}
{In this work, we consider a delensing factor. We use dedicated matter surveys along the line of sight of the GW event in order to estimate the lensing magnification distribution, which can remove part of the uncertainty due to weak lensing. This reduces the weak lensing uncertainty. Following Ref.~\cite{Speri:2020hwc}, we realistically assume that 30\% of delensing could be achieved at redshift 2 and we thus adopt the following delensing factor,
\begin{equation}
F_{\rm {delens}}(z)=1-\frac{0.3}{\pi / 2} \arctan \left(z / z_{*}\right),
\end{equation}
with $z_{*}=0.073$. The final lensing uncertainty on $d_{\rm L}$ is
\begin{equation}
\sigma_{d_{\rm L}}^{\rm lensing}(z)=F_{\rm {delens}}(z)\sigma_{d_{\rm L}}^{\rm lens}(z).\label{lens}
\end{equation}
We consider the delensing uncertainty of $d_{\rm L}$, i.e., we use $\sigma_{d_{\rm L}}^{\rm lensing}$ to replace $\sigma_{d_{\rm L}}^{\rm lens}$ in Eq.~(\ref{total}).}

The error caused by the peculiar velocity of the GW source is given by \cite{Kocsis:2005vv}
\begin{equation}
\sigma_{d_{\rm L}}^{\rm pv}(z)=d_{\rm L}(z)\times \bigg[ 1+ \frac{c(1+z)^2}{H(z)d_{\rm L}(z)}\bigg]\frac{\sqrt{\langle v^2\rangle}}{c},\label{pv}
\end{equation}
where $H(z)$ is the Hubble parameter. $\sqrt{\langle v^2\rangle}$ is the peculiar velocity of the GW source and we roughly set $\sqrt{\langle v^2\rangle}=500\ {\rm km\ s^{-1}}$.

For each simulated GW source, the sky location ($\theta$, $\phi$), the masses of NSs ($m_1$, $m_2$), the binary inclination $\iota$, the coalescence phase $\psi_{\rm c}$, and the polarization angle $\psi$ are evenly sampled in the ranges of [0, $\pi$], [0, $2\pi$], $[1, 2]\ M_{\odot}$, $[1, 2]\ M_{\odot}$, [0, $\pi/9$], [0, $2\pi$], and [0, $2\pi$], respectively, where $M_{\odot}$ is the solar mass. The merger time is chose to $t_{\rm c}=0$ for simplicity. In this work, we assume that the EM counterparts could be detected through the detections of short $\gamma$-ray bursts (SGRBs) to determine sources' redshifts. The maximal inclination angle that could be detected is about $\iota=20^\circ$ \cite{li2015extracting}, so we set the inclination angle to be in the range of [0, $\pi/9$].
\subsection{Simulation of GW standard sirens from Taiji}\label{sec:GWTJ}

The frequency band detected by the space-based GW detectors corresponds to MBHB mergers. The unknown birth mechanisms of MBHB lead to the uncertainties in predicting the event rate of MBHB.
Based on a semianalytical galaxy formation model, three population models of MBHBs, i.e., the pop III, Q3d, and Q3nod models are proposed, based on the various combinations of the mechanisms of seeding and delay \cite{Klein:2015hvg}. In Ref.~\cite{Zhao:2019gyk}, it is found that the Q3nod model gives the best constraints on cosmological parameters since the Q3nod model yields the most data points among these three models. In this paper, we simulate standard siren events only based on the Q3nod model.

The response functions of Taiji are given by
\begin{align}
 F_{+}(t;\theta, \phi, \psi) =& \frac{1}{2}\Big({\rm cos}(2\psi)D_{+}(t;\theta, \phi)-{\rm sin}(2\psi)D_{\times}(t;\theta, \phi)\Big),  \nonumber\\
 F_{\times}(t;\theta, \phi, \psi) =& \frac{1}{2}\Big({\rm sin}(2\psi)D_{+}(t;\theta, \phi)+{\rm cos}(2\psi)D_{\times}(t;\theta, \phi)\Big)  .
\end{align}

Based on the low-frequency approximation, the forms of $D_{+,\times}$ are given by \cite{Ruan:2020smc}
\begin{align}
D_{+}(t;\theta, \phi) =& \frac{\sqrt{3}}{64}\bigg[-36{\rm sin}^2\theta\,{\rm sin}\big(2\alpha(t)-2\beta\big)+\big(3+{\rm cos(2\theta)}\big)\nonumber \\
&\times\bigg({\rm cos}(2\phi)\Big(9\sin(2\beta)-{\rm sin}\big(4\alpha(t)-2\beta\big)\Big)\nonumber \\
&+{\rm sin}(2\phi)\Big({\rm cos}\big(4\alpha(t)-2\beta\big)-9\cos(2\beta)\Big)\bigg)\nonumber \\
&-4\sqrt{3}{\rm sin}(2\theta)\Big({\rm sin}\big(3\alpha(t)-2\beta-\phi\big)-3{\rm sin}\big(\alpha(t)\nonumber \\
&-2\beta+\phi\big)\Big)\bigg]  \,,\label{Dplus}  \\
D_{\times}(t;\theta, \phi) =& \frac{1}{16}\bigg[\sqrt{3}{\rm cos}\theta\Big(9{\rm cos}(2\phi-2\beta)-{\rm cos}\big(4\alpha(t)-2\beta\nonumber \\
&-2\phi\big)\Big)-6{\rm sin}\theta\Big({\rm cos}\big(3\alpha(t)-2\beta-\phi\big)\nonumber \\
&+3{\rm cos}\big(\alpha(t)-2\beta+\phi\big)\Big)\bigg]  \,,\label{Dcros}
\end{align}
where $\alpha=2\pi f_m t+\kappa$ is the orbital phase of the guiding center, and $\beta=0$ is the relative phase of three spacecraft. Here $\kappa=0$ is the initial ecliptic longitude of the guiding center and $f_m=1/{\rm yr}$. Following Ref.~\cite{Cutler:1997ta}, we equivalently consider Taiji as a combination of two independent interferometers with an azimuthal difference of $\pi/4$. Another equivalent antenna pattern function is $F_{+,\times}^{(2)}(t;\theta, \phi, \psi)=F_{+,\times}^{(1)}(t;\theta, \phi-\pi/4, \psi)$.

In order to study the signal in the Fourier space, we replace the observation time $t$ by \cite{Krolak:1995md,Buonanno:2009zt}
\begin{equation}
t(f) = t_{\rm c} - \frac{5}{256} M_{\rm c} ^{-5/3}(\pi f)^{-8/3},
\end{equation}
where $t_{\rm c}$ is the coalescence time of MBHB. In our analysis, we set $t_{\rm c}=0$.

We calculate SNR of each GW event using Eqs.~(\ref{snr1}) and (\ref{snr2}), and we choose the SNR threshold of 8 for Taiji. In Eq.~(\ref{snr2}), Taiji's PSD is taken from Ref.~\cite{Guo:2018npi}. $f_{\rm lower}=10^{-4}$ {Hz} is the lower frequency cutoff, and $f_{\rm upper}=c^3/{6\sqrt{6}\pi GM_{\rm obs}}$ is the innermost stable circular orbit \cite{Feng:2019wgq} with $M_{\rm obs}=(m_1+m_2)(1+z)$.
Following Ref.~\cite{Zhao:2019gyk}, we assume that Taiji's detection rate of MBHB based on the Q3nod model is identical to that of LISA, i.e., 41 standard siren events based on the 5-year operation time \cite{Tamanini:2016uin} are considered in this work. We adopt the redshift distribution given in Ref.~\cite{Tamanini:2016uin} and simulate 41 standard siren events. For each GW source, the sky position ($\theta$, $\phi$), the masses of MBHs ($m_1$, $m_2$), the inclination angle $\iota$, the coalescence phase $\psi_{\rm c}$, and the polarization angle $\psi$ are evenly sampled in the ranges of [0, $\pi$], [0, $2\pi$], $[10^4, 10^7]\ M_{\odot}$, $[10^4, 10^7]\ M_{\odot}$, [0, $\pi$], [0, $2\pi$], and [0, $2\pi$], respectively.

MBHBs may produce EM signals since they are expected to merge in a gas-rich environment that may power EM emissions through jets, disk winds, or accretions. These EM signals can be applied in identifying the redshifts of the GW sources. If the redshift is measured spectroscopically, the redshift error could be ignored. While if the redshift is measured photometrically, the redshift error should be taken into account. Since the spectroscopic redshift in the range of $z>2$ is almost unavailable \cite{Dahlen:2013fea,Speri:2020hwc}, we assume that the redshifts of GW events with $z>2$ are measured photometrically, while those with $z<2$ are measured spectroscopically. Hence, for the GW events with $z>2$, we take into account the redshift error $\sigma_{d_{\rm L}}^{\rm reds}$ \cite{Speri:2020hwc} in Eq.~(\ref{total}). We estimate the error on the redshift measurement as $(\Delta z)_{n}\simeq 0.03(1+z_{n})$ \cite{Ilbert:2013bf} and propagate it to the error on $d_{\rm L}$,
\begin{equation}
\sigma_{d_{\rm L}}^{\rm reds}=\frac{\partial d_{\rm L}}{\partial z} (\Delta z)_n.
\end{equation}
Here $n$ represent the $n$th GW event.

\begin{figure}[htbp]
\includegraphics[width=0.9\linewidth,angle=0]{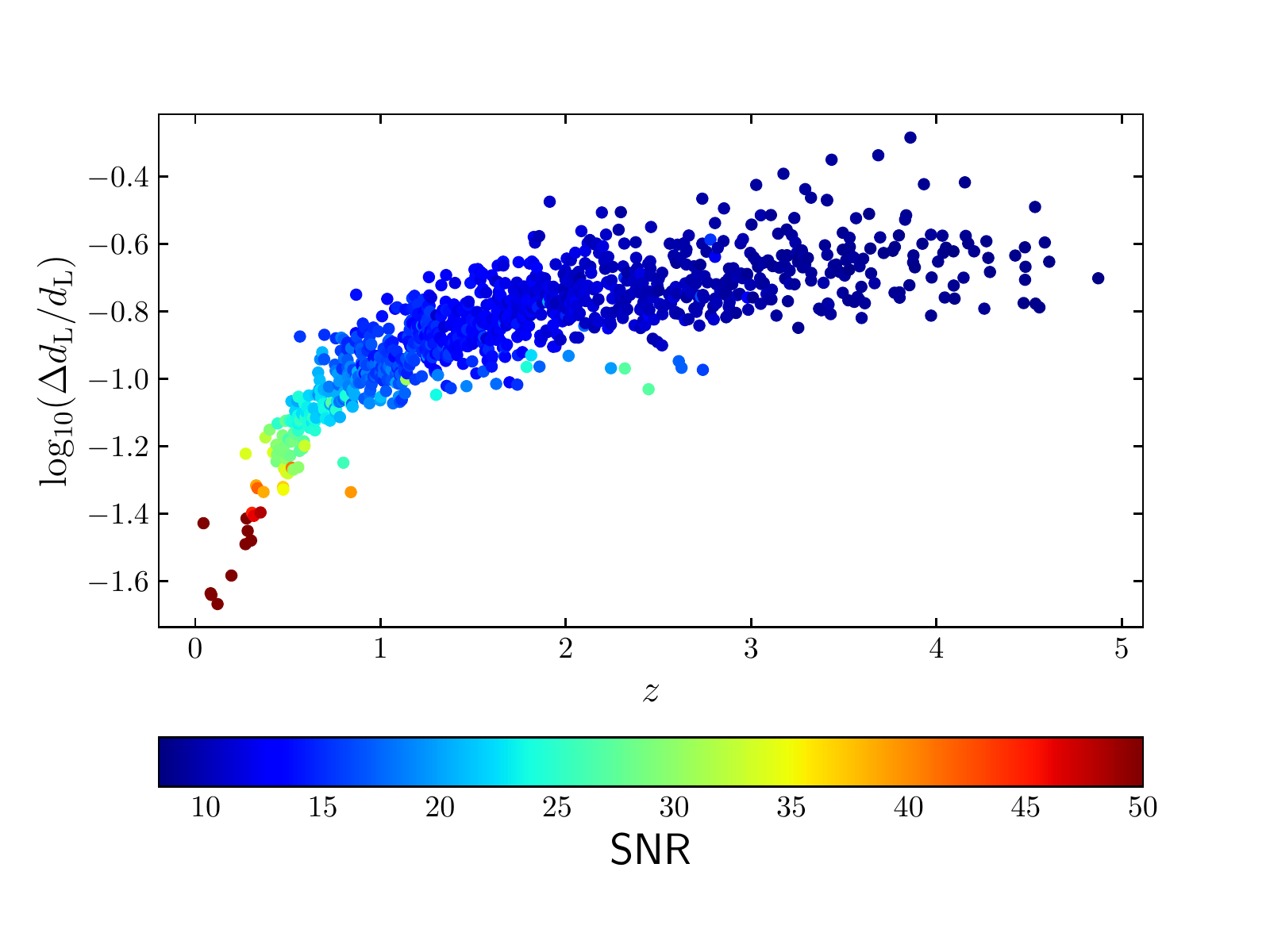}
\includegraphics[width=0.9\linewidth,angle=0]{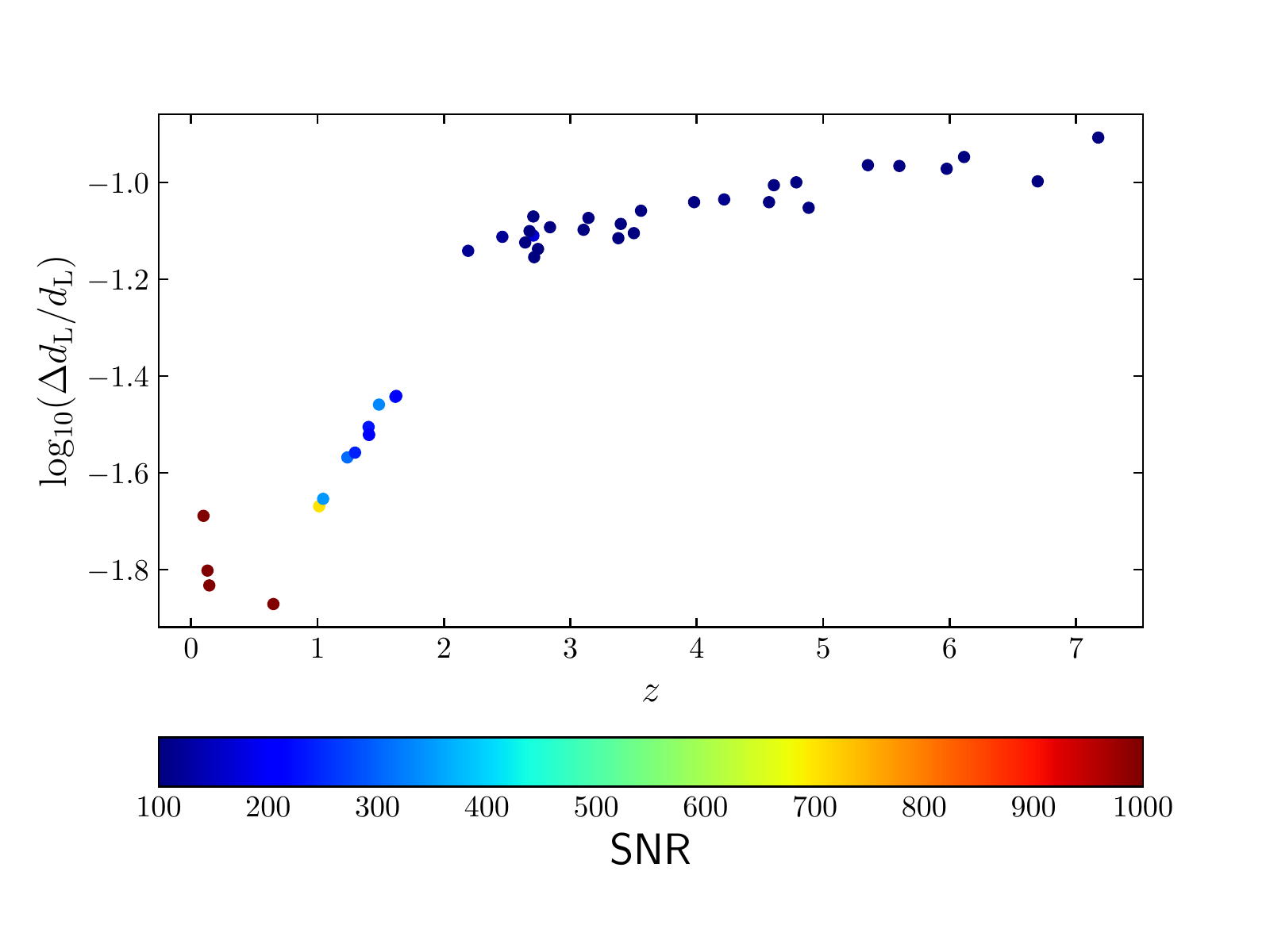}
 \caption{\label{snr} Distribution of $\Delta d_{\rm L}/d_{\rm L}$ as a function of redshift. The color indicates SNRs of the simulated GW standard sirens. {Upper panel}: the 1000 standard sirens within the 10-year observation of ET. {Lower panel}: the 41 standard sirens within the 5-year observation of Taiji, based on the Q3nod model of MBHB population.}
\end{figure}

\begin{figure*}[htbp]
\includegraphics[scale=0.5]{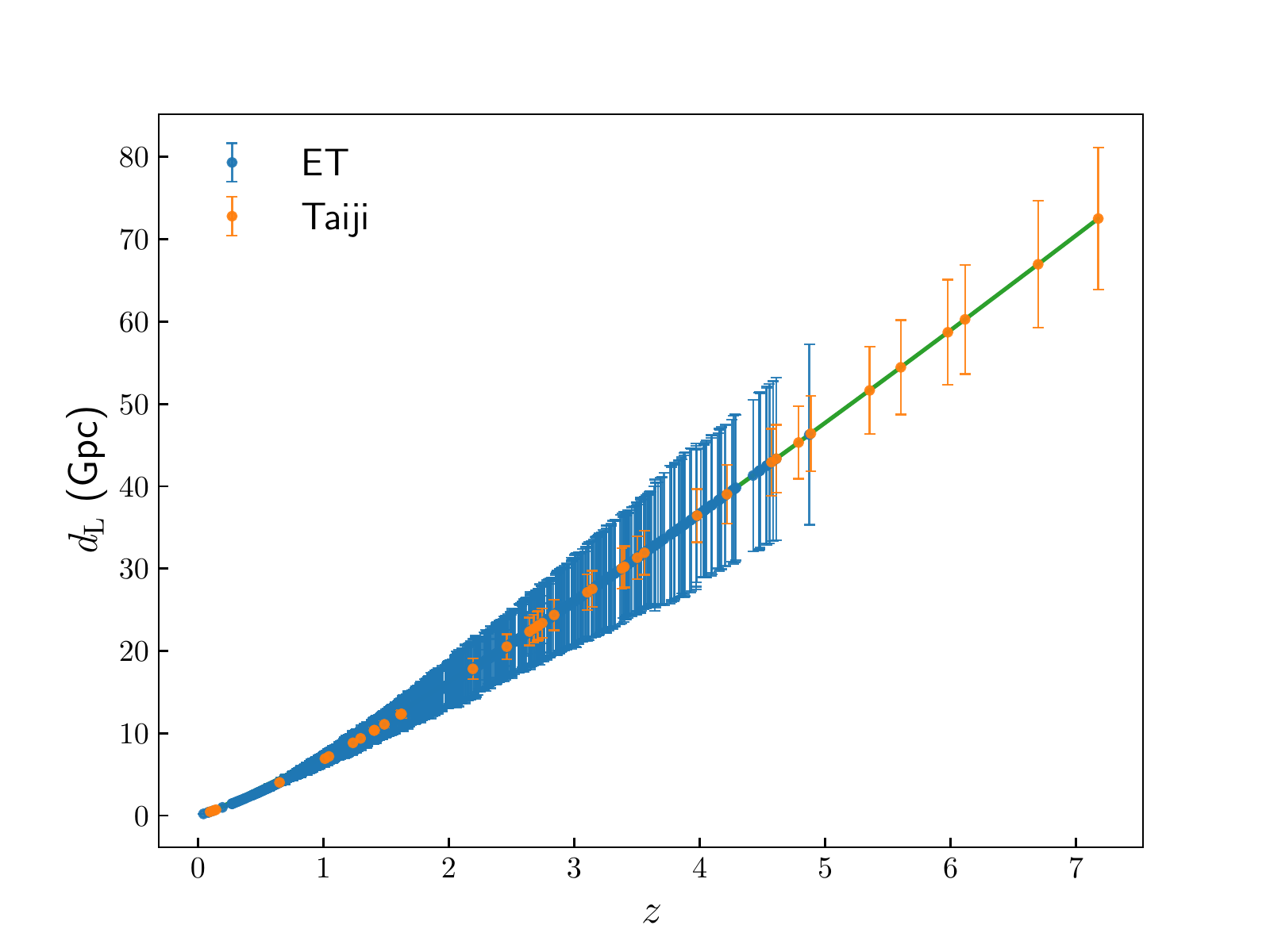}
\includegraphics[scale=0.5]{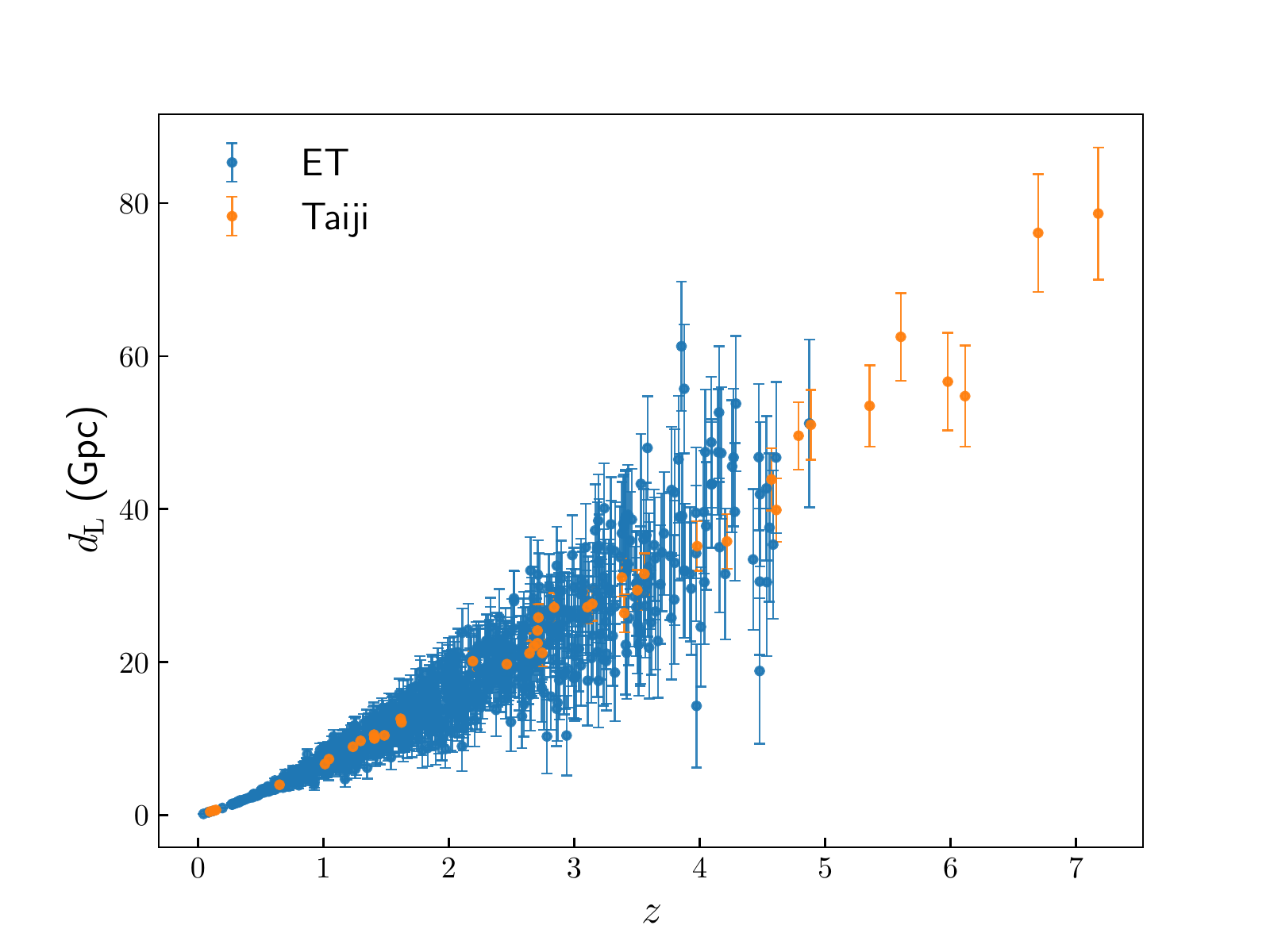}
\caption{\label{zdl} {GW standard sirens simulated for ET and Taiji. The blue data points represent the 1000 standard sirens within the 10-year observation of ET, and the orange data points represent the 41 standard sirens within the 5-year observation of Taiji, based on the Q3nod model of MBHB population. Left panel: the standard siren data points without Gaussian randomness, where the central value of the luminosity distance is calculated by the fiducial cosmological model, and the solid green line represents the $d_{\rm L}(z)$ curve predicted by the fiducial model. Right panel: the standard siren data points with Gaussian randomization, reflecting the fluctuations in measured values resulting from actual observations.}}
\end{figure*}

In Fig.~\ref{snr}, we show the $\Delta d_{\rm L}/d_{\rm L}$ scatter plot of the simulated standard sirens detected by ET (upper panel) and Taiji (lower panel). We can observe the following facts: (i) the number of standard sirens detected by ET is much more than that detected by Taiji, because the event rate of the BNS merger is larger than that of the MBHB merger; (ii) compared with ET, Taiji could detect the GW events at higher redshifts ($z\sim 7$); and (iii) due to the fact that the mass of MBHB is several orders of magnitude larger than the mass of BNS, SNRs of the GW events observed by Taiji are all higher than those observed by ET at the same redshift.

In Fig.~\ref{zdl}, we show the GW standard sirens simulated from ET and Taiji.
The central values and errors of $d_{\rm L}$ are shown in both the left and right panels. The difference between these two panels is that the central values in the left panel are directly obtained by theoretical calculations of the fiducial model, while the central values in the right panel are randomly chosen in the ranges of [$d_{\rm L}-\sigma_{d_{\rm L}}$, $d_{\rm L}+\sigma_{d_{\rm L}}$] with Gaussian distribution. In principle, the right panel is more representative of actual observational data, but the central values of $d_{\rm L}$ have no effect on determining the absolute errors of cosmological parameters. Therefore, we only use the data points in the left panel to constrain the cosmological models, because this is more helpful in investigating how the parameter degeneracies are broken.
We see that the measurement errors of $d_{\rm L}$ from Taiji are smaller than those from ET at similar redshifts. This is due to the fact that the instrumental error of $d_{\rm L}$ is inversely proportional to SNR \cite{li2015extracting}, and SNRs of MBHB merger events are larger than those of BNS merger events. Although we additionally take into account the redshift error for Taiji, there still exist differences of several orders of magnitude in $\Delta d_{\rm L}$ between Taiji and ET.

\begin{figure}[htbp]
\includegraphics[width=0.9\linewidth,angle=0]{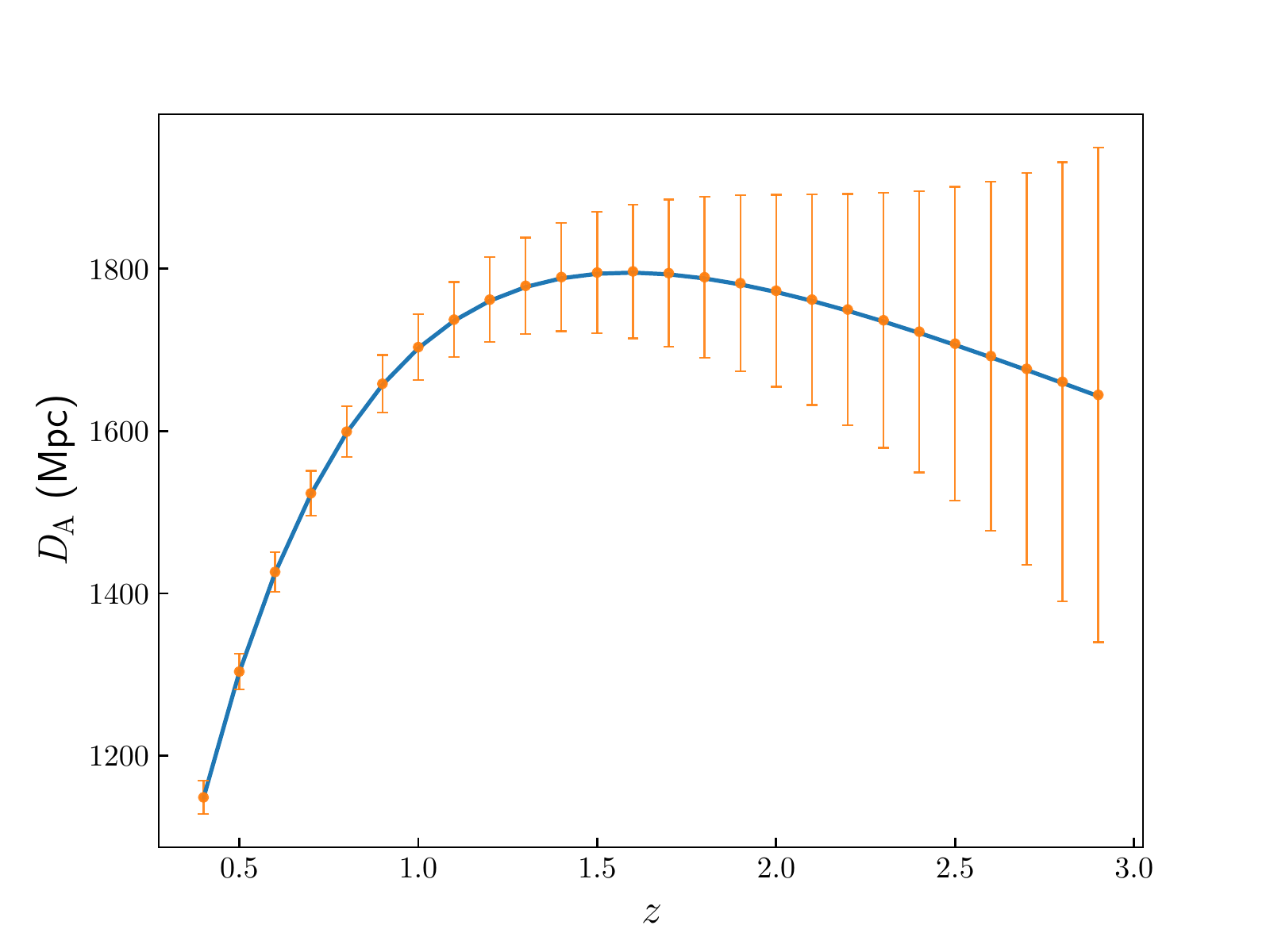}
\includegraphics[width=0.9\linewidth,angle=0]{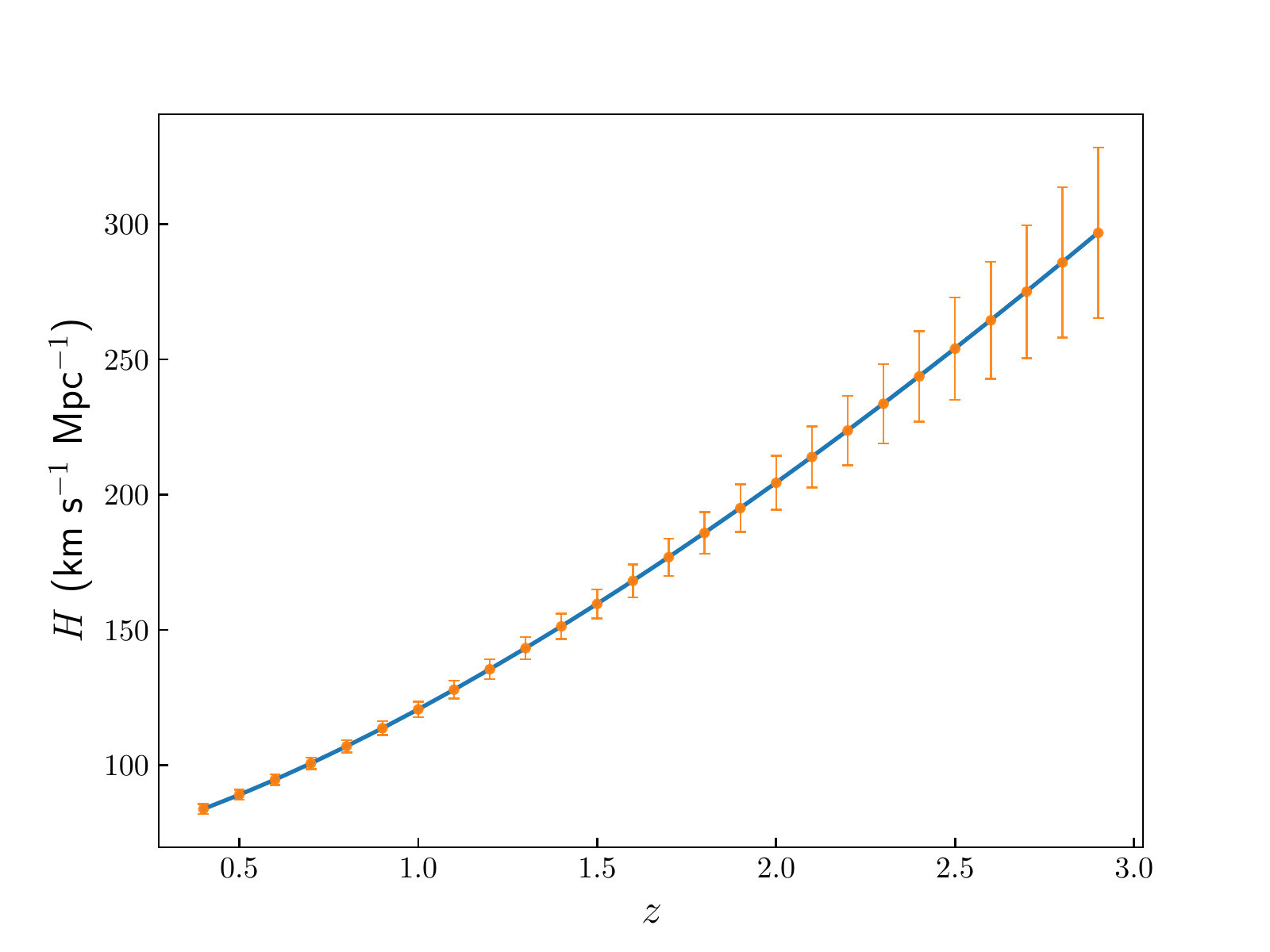}
\includegraphics[width=0.9\linewidth,angle=0]{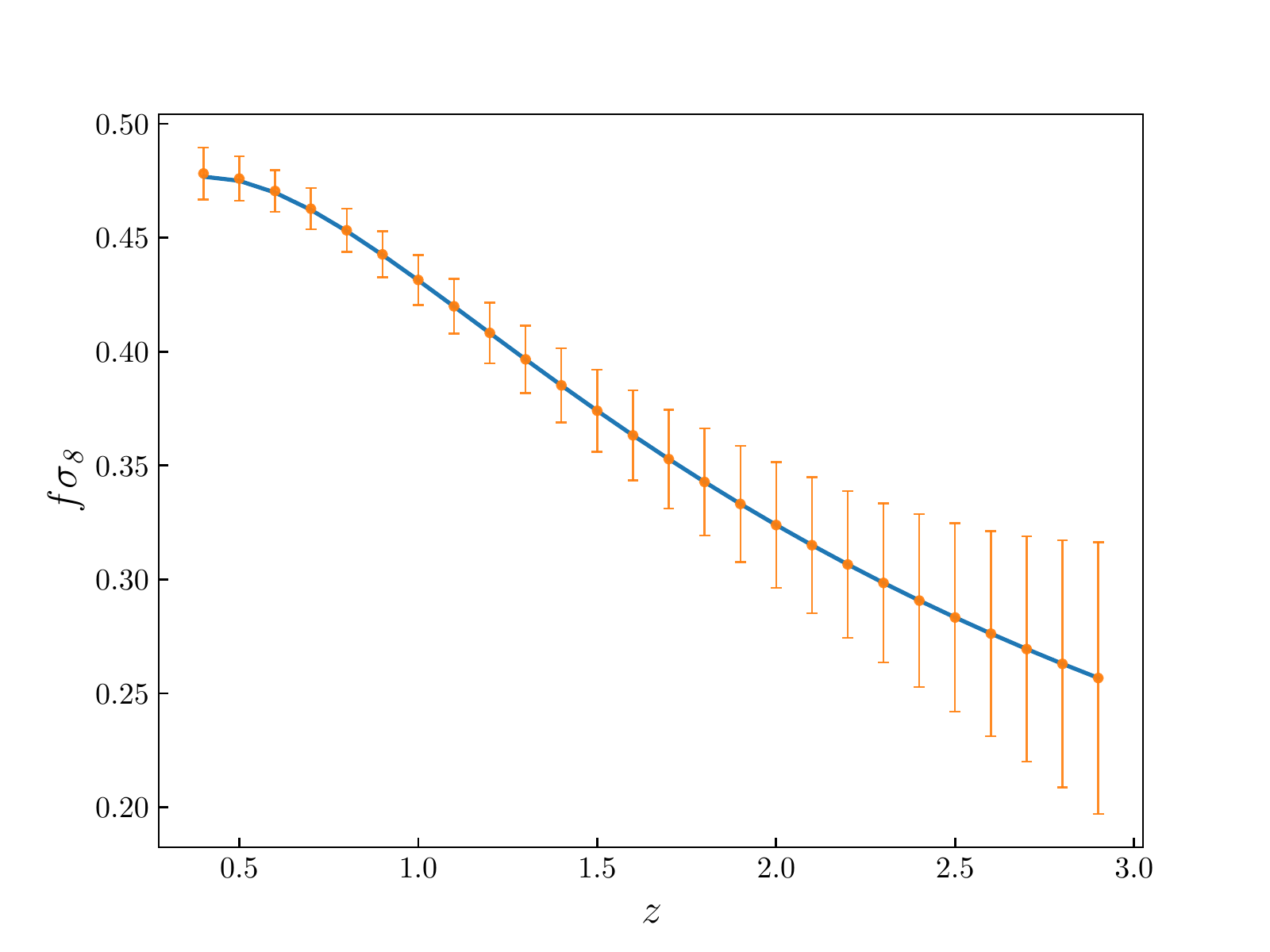}
\caption{\label{dahzfs8} Forecasted data of $D_{\rm A}(z)$, $H(z)$, and [$f\sigma_8](z)$ as functions of redshift based on the 10000-h total integration time of SKA1-MID. The fiducial values are shown as the solid blue lines.}
\end{figure}
\section{21-cm intensity mapping survey}\label{sec:IM}

The 21-cm IM surveys will also be developed into a powerful cosmological probe. In this paper, we consider the SKA1-MID array with 133 15-m SKA dishes and 64 13.5-m MeerKAT dishes as the representatives of the 21-cm IM experiments. For simplicity, we assume both MeerKAT and SKA are 15-meter dishes, and thus directly consider 197 15-m dishes. Note that we only consider the {\it Wide Band 1 Survey} of the SKA1-MID array (with the redshift range of $0.35<z<3$) with perfect foreground removal and calibration.
In the following, we shall briefly introduce the signal power spectrum and the noise power spectrum.

The mean 21-cm brightness temperature is given by \cite{Battye:2012tg}
\begin{equation}
\overline{T}_{\rm b}(z) = 180 \Omega_{\rm H{\tt I}}(z) h \frac{(1+z)^2}{H(z)/H_0} \, {\rm mK},
\end{equation}
where $h$ is the dimensionless Hubble constant. $\Omega_{\rm H{\tt I}}(z)$ is derived from a simulated {\rm H{\tt I}} halo mass function, written as
\begin{equation}
\Omega_{\rm H{\tt I}}(z) \equiv (1+z)^{-3} \rho_{\rm H{\tt I}}(z) / \rho_{\rm c,0},
\end{equation}
where $\rho_{\rm c,0}$ is the critical density today. $\rho_{\rm H{\tt I}}(z)$ is the proper {\rm H{\tt I}} density, calculated by
\begin{equation}
\rho_{\rm H{\tt I}}(z) = \int_{M_{\rm min}}^{M_{\rm max}} dM \frac{dn}{dM} M_{\rm H{\tt I}}(M, z),
\end{equation}
where $M$ is the mass of the dark matter halo, $dn/dM$ is the proper halo mass function, and $M_{\rm H{\tt I}}(M, z)$ is the {\rm H{\tt I}} mass in a halo of mass $M$ at redshift $z$. For detailed calculations, see Ref.~\cite{Bull:2014rha}.

Considering the effect of redshift space distortions (RSDs) \cite{Kaiser:1987qv} caused by the peculiar velocities of the H{\tt I} clouds and the galaxies in which they reside, the signal power spectrum can be written as \cite{Seo:2003pu,Bull:2014rha}
\begin{align}
P^{\rm S}(k_{\rm f}, \mu_{\rm f}, z) =\overline{T}_{\rm b}^2(z) &  \frac{D_{\rm A}^2(z)_{\rm f} H(z)}{D_{\rm A}^2(z) H(z)_{\rm f}}b^2_{\rm H{\tt I}}(z)[1+\beta_{\rm H{\tt I}}(z) \mu^2]^2 \nonumber \\
&\times \exp{(-k^2 \mu^2 \sigma_{\rm NL}^{2})} P(k,z),
\end{align}
where the subscript ``f'' denotes the quantities calculated in the fiducial cosmology and $D_{\rm A}(z)$ is the angular diameter distance. $\mu$ is defined as $\mu=\hat{k}\cdot\hat{z}$. $b_{\rm H{\tt I}}(z)$ is the {\rm H}{\tt I} bias calculated by
\begin{align}
b_{\rm H{\tt I}}(z) = \rho_{\rm H{\tt I}}^{-1}(z) \int_{M_{\rm min}}^{M_{\rm max}} dM \frac{dn}{dM} M_{\rm H{\tt I}}(M, z) b(M, z) ,
\end{align}
where $b(M, z)$ is the halo bias (for the detailed calculation, see Ref.~\cite{Xu:2014bya}). $\beta_{\rm H{\tt I}}\equiv f/b_{\rm H{\tt I}}$ is the RSD parameter, where $f\equiv d {\rm ln}D / d {\rm ln}a$ is the linear growth rate [with $a=1/(1+z)$ being the scale factor]. The exponential term accounts for the ``Fingers of God'' effect and $\sigma_{\rm NL}=7\ \rm{Mpc}$ is the nonlinear dispersion scale \cite{Li:2007rpa}. $P(k,z)=D^2(z)P(k,z=0)$, with $D(z)$ being the growth factor and $P(k,z=0)$ being the matter power spectrum at $z=0$ that can be generated by \texttt{CAMB} \cite{Lewis:1999bs}.

Next, we consider the thermal noise and the effective beams.
The frequency resolution of IM survey performs very well due to the narrow channel bandwidths of SKA's receivers,
so we ignore the instrumental response function in the radial direction and only consider the response due to the finite angular resolution
\begin{equation}
W^2(k)={\rm exp}\left[-k_\perp^2r^2(z)\left(\frac{\theta_{\rm B}}{\sqrt{8{\rm ln}2}}\right)^2\right],
\end{equation}
where $k_\perp$ is the transverse wave vector, $r(z)$ is the comoving radial distance at redshift $z$, and $\theta_{\rm B}$ is the full width at the half-maximum of the beam of an individual dish.

The survey volume of a redshift bin between $z_1$ and $z_2$ can be written as
\begin{equation}
V_{\rm sur}= \Omega_{\rm tot} \int_{z_1}^{z_2} dz \frac{r^2(z)}{H(z)},
\end{equation}
where $\Omega_{\rm tot}=S_{\rm area}$ is the solid angle of the survey area. The pixel volume $V_{\rm pix}$ is also calculated with the similar formula with $\Omega_{\rm tot}$ substituted by $\Omega_{\rm pix} \simeq 1.13\theta^2_{\rm B}$.

For the SKA1-MID array, the pixel noise is given by \cite{Bull:2014rha}
\begin{equation}
\sigma_{\rm pix}=\frac{T_{\rm sys}}{\sqrt{\Delta \nu \, t_{\rm tot}(\theta_{\rm B}^2/S_{\rm area})}}\frac{\lambda^2}{A_{\rm e} \theta_{\rm B}^2}\frac{1}{\sqrt{N_{\rm dish} N_{\rm beam}}},
\end{equation}
where $T_{\rm sys}$ is the system temperature; $N_{\rm dish}=197$ is the number of dishes; $N_{\rm beam}=1$ is the number of beam; $t_{\rm tot}=10000\ \rm h $ is the total integration time; $A_{\rm e}\equiv \eta \pi (D_{\rm dish}/2)^2$ is the effective collecting area of each element; $\theta_{\rm B}\approx \lambda/D_{\rm dish}$; $D_{\rm dish}=15$ m is the diameter of the dish; $\eta$ is an efficiency factor (we adopt $0.7$ in this work) and $S_{\rm area}=20000$ $\rm deg^2$ is the survey area.

The system temperature of the SKA1-MID array can be divided into four parts \cite{Bacon:2018dui},
\begin{equation}
T_{\rm sys} = T_{\rm rec} + T_{\rm spl} + T_{\rm CMB} + T_{\rm gal},
\end{equation}
where $T_{\rm spl}\approx3\ \rm K$ is the contribution from spill-over, $T_{\rm CMB}\approx2.73~\rm K$ is the CMB temperature, $T_{\rm gal} \approx 25\ \rm{K}\times (408~\rm{MHz}/ \nu )^{2.75}$ is the contribution from the Milky Way for a given frequency $\nu$, and $T_{\rm rec}$ is the receiver temperature which is assumed to be \cite{Bacon:2018dui}
\begin{equation}
T_{\rm rec} = 15~\rm{K} + 30~\rm{K} \left(\frac{\nu}{\rm GHz}-0.75 \right)^2.
\end{equation}

Finally, the noise power spectrum is given by
\begin{equation}
P^{\rm N}(k) = \sigma^2_{\rm pix}V_{\rm pix}W^{-2}(k),
\end{equation}
and the Fisher matrix for a set of parameters $\{p\}$ is given by \cite{Tegmark:1997rp}
\begin{equation}
F_{ij}=\frac{1}{8\pi^2}\int^{1}_{-1} d\mu \int^{k_{\rm max}}_{k_{\rm min}} k^2dk \; \frac{\partial {\rm ln}P^{\rm S}}{\partial p_i}\frac{ \partial {\rm ln}P^{\rm S}}{\partial p_j} V_{\rm eff},
\end{equation}
where the ``effective volume" is defined as \cite{Bull:2014rha,Pourtsidou:2016dzn}
\begin{equation}
V_{\rm eff} = V_{\rm sur} \left(\frac{P^{\rm S}}{P^{\rm S}+P^{\rm N}}\right)^2.
\end{equation}

\begin{table*}[!htb]
\caption{The absolute errors (1$\sigma$) and the relative errors of the cosmological parameters in the $\Lambda$CDM, $w$CDM, and CPL models using the ET, ET+Taiji, SKA, and ET+Taiji+SKA data. Here $H_0$ is in units of km s$^{-1}$ Mpc$^{-1}$.}
\label{tab:full}
\setlength{\tabcolsep}{3mm}
\renewcommand{\arraystretch}{1.5}
\begin{center}{\centerline{
\begin{tabular}{ccm{2cm}<{\centering}m{2cm}<{\centering}m{2cm}<{\centering}m{2cm}<{\centering}m{2cm}<{\centering}}
\hline
          Model       & Error&ET & ET+Taiji&SKA&ET+Taiji+SKA  \\ \hline
\multirow{4}{*}{$\Lambda$CDM} & $\sigma(\Omega_{\rm m})$&$0.014$  & $0.012$& $0.006$ & $0.005$  \\
 &$\sigma(H_0)$&$0.55$ &$0.44$&$0.51$&$0.28$ \\
 &$\ve(\Omega_{\rm m})$&$0.044$&$0.038$&$0.020$&$0.015$ \\
 &$\ve(H_0)$&$0.008$&$0.007$&$0.008$&$0.004$ \\ \hline
 \multirow{6}{*}{$w$CDM}& $\sigma(\Omega_{\rm m})$&$0.018$  & $0.016$& $0.007$ & $0.005$  \\
 &$\sigma(H_0)$&$0.92$ &$0.63$&$0.67$&$0.40$ \\
  &$\sigma(w)$&$0.120$ &$0.084$&$0.033$&$0.028$ \\
 &$\ve(\Omega_{\rm m})$&$0.056$&$0.050$&$0.021$&$0.016$ \\
 &$\ve(H_0)$&$0.014$&$0.009$&$0.010$&$0.006$ \\
 &$\ve(w)$&$0.115$&$0.083$&$0.033$&$0.028$ \\ \hline
 \multirow{7}{*}{CPL}& $\sigma(\Omega_{\rm m})$&$0.158$  & $0.157$& $0.015$ & $0.009$  \\
 &$\sigma(H_0)$&$1.42$&$1.11$&$1.00$&$0.63$ \\
  &$\sigma(w_0)$&$0.248$ &$0.216$&$0.105$&$0.077$ \\
  &$\sigma(w_a)$&$1.800$ &$1.565$&$0.410$&$0.295$ \\
 &$\ve(\Omega_{\rm {\rm m}})$&$0.499$&$0.496$&$0.050$&$0.030$ \\
 &$\ve(H_0)$&$0.021$&$0.016$&$0.015$&$0.009$ \\
 &$\ve(w_0)$&$0.248$&$0.215$&$0.119$&$0.075$ \\  \hline
\end{tabular}}}
\end{center}
\end{table*}

In this work, we assume that $b_{\rm H{\tt I}}$ only depends on the redshift $z$. This assumption is appropriate only for large scales, so we impose a nonlinear cutoff at $k_{\rm max} \simeq 0.14 (1+z)^{2/3} \ {\rm Mpc}^{-1}$ \cite{Smith:2002dz}. In addition, the largest scale probed by the survey corresponds to a wave vector $k_{\rm min} \simeq 2\pi/V_{\rm sur}^{1/3}$ \cite{Smith:2002dz}. We choose the parameter set $\{p\}$ as $\{D_{\rm A}(z),H(z),[f\sigma_8](z),[b_{\rm H{\tt I}}\sigma_8](z), \sigma_{\rm NL}\}$, and use only the forecasted observable parameters $\{D_{\rm A}(z),H(z),[f\sigma_8](z)\}$ to constrain cosmological models.

The method of making forecast for cosmological parameter estimation using 21-cm IM surveys has been described in detail in Refs.~\cite{Bull:2014rha,Witzemann:2017lhi}, and we follow the methods described in Refs.~\cite{Bull:2014rha,Witzemann:2017lhi} to perform the forecast for SKA1-MID. First, we measure the full anisotropic power spectrum to obtain the constraints on the angular diameter distance $D_{\rm A}(z)$, the Hubble parameter $H(z)$, and the RSD observable $[f\sigma_8](z)$, which are considered to be independent in each redshift bin. Then, we invert the Fisher matrix to obtain covariance matrices for \{$D_{\rm A}(z_j)$, $H(z_j)$, and $[f\sigma_8](z_j)$; $j=1...N$\} in a series of $N$ redshift bins \{$z_j$\}. Finally, we use these covariance matrices and the fiducial cosmology to generate the mock data of SKA1-MID. The forecasted data of $D_{\rm A}(z)$, $H(z)$, and [$f\sigma_8](z)$ are shown in Fig.~\ref{dahzfs8}.

\section{Cosmological parameter estimation}\label{sec:re}

\begin{figure}[htbp]
\includegraphics[width=7.2cm,height=5.4cm]{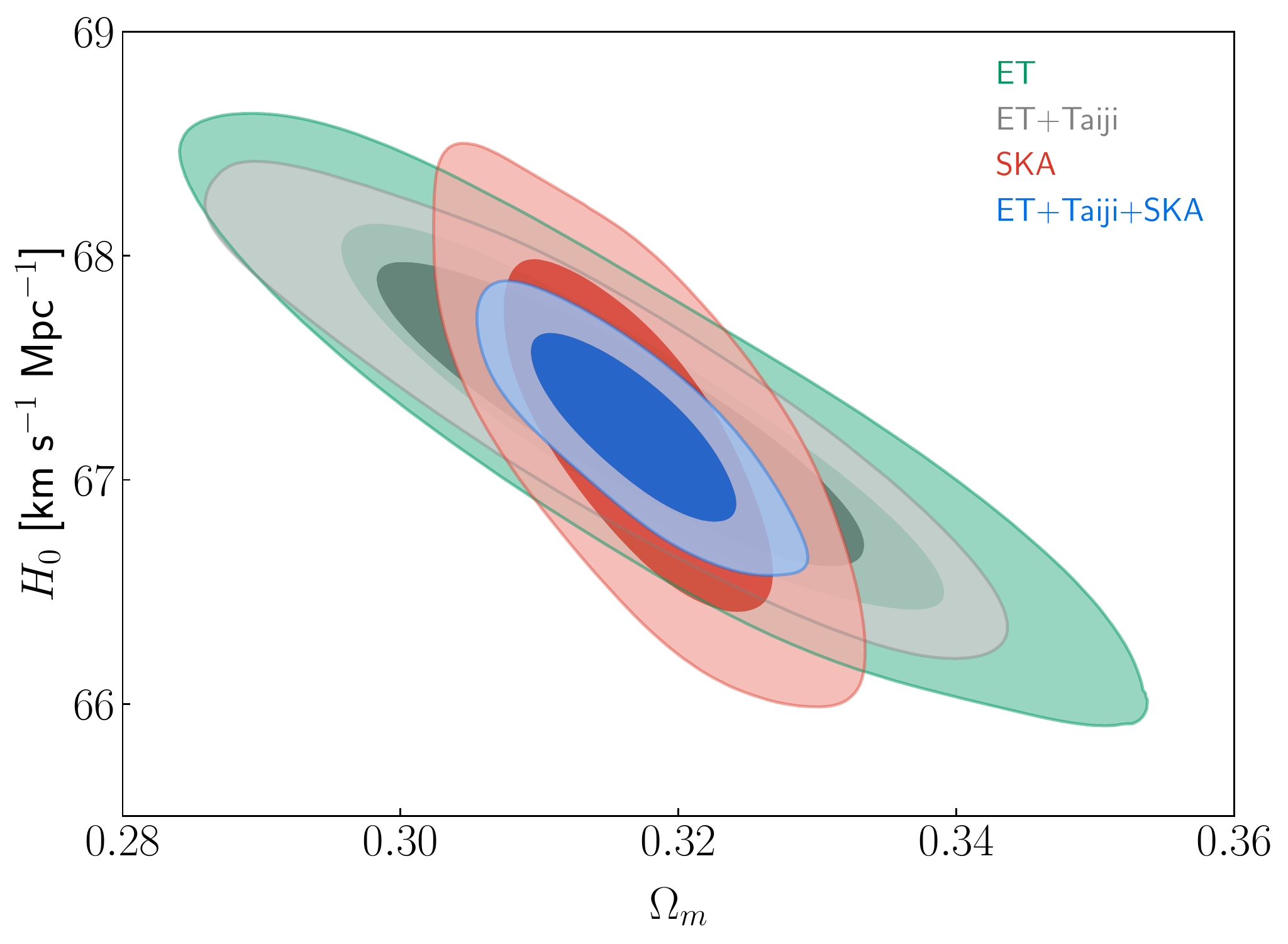}
 \caption{\label{lcdm} Two-dimensional marginalized contours (68.3\% and 95.4\% confidence level) in the $\Omega_{\rm m}$--$H_0$ plane by using the ET, ET+Taiji, SKA, and ET+Taiji+SKA data.}
\end{figure}

\begin{figure}[htbp]
\includegraphics[width=7.2cm,height=5.4cm]{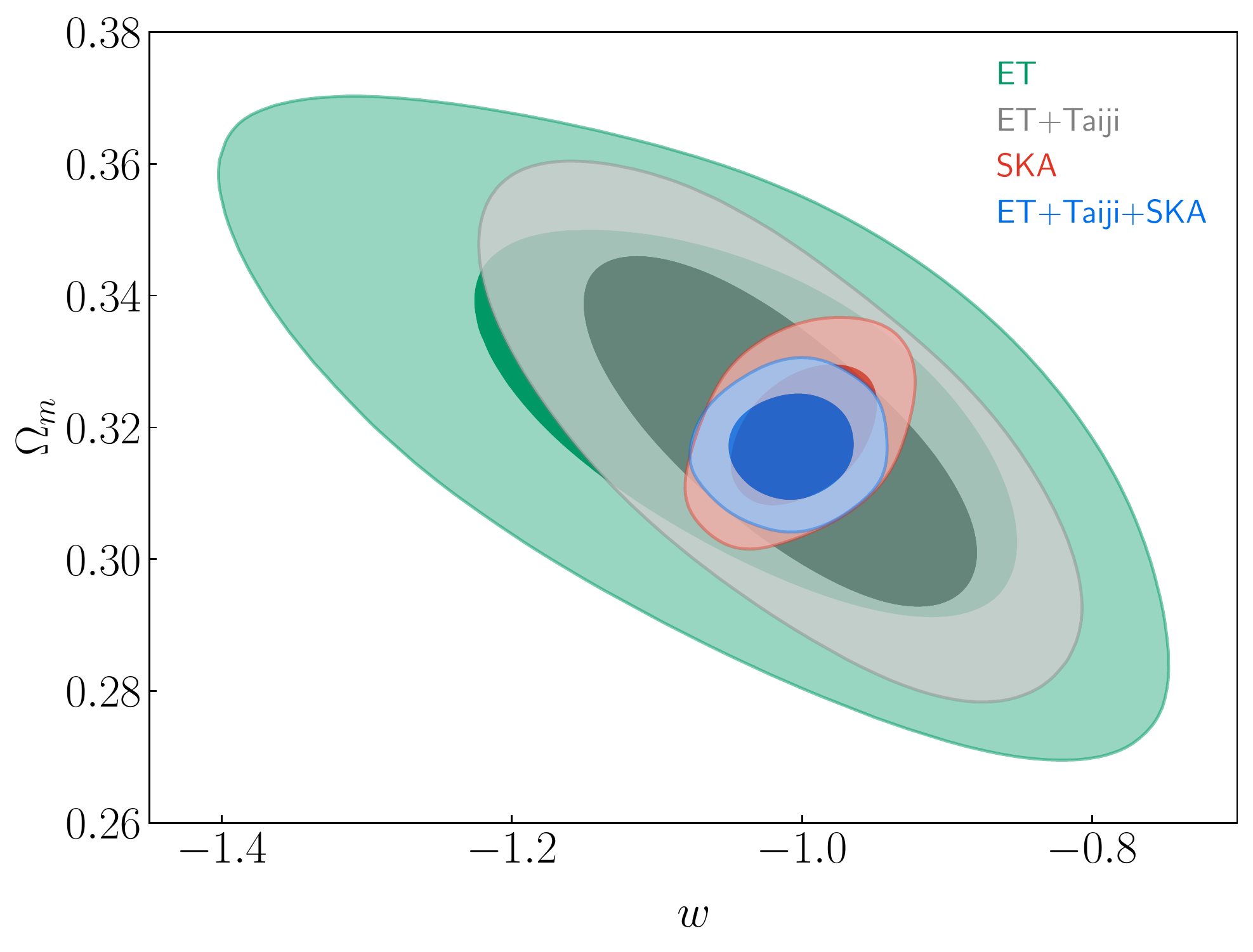}
\includegraphics[width=7.2cm,height=5.4cm]{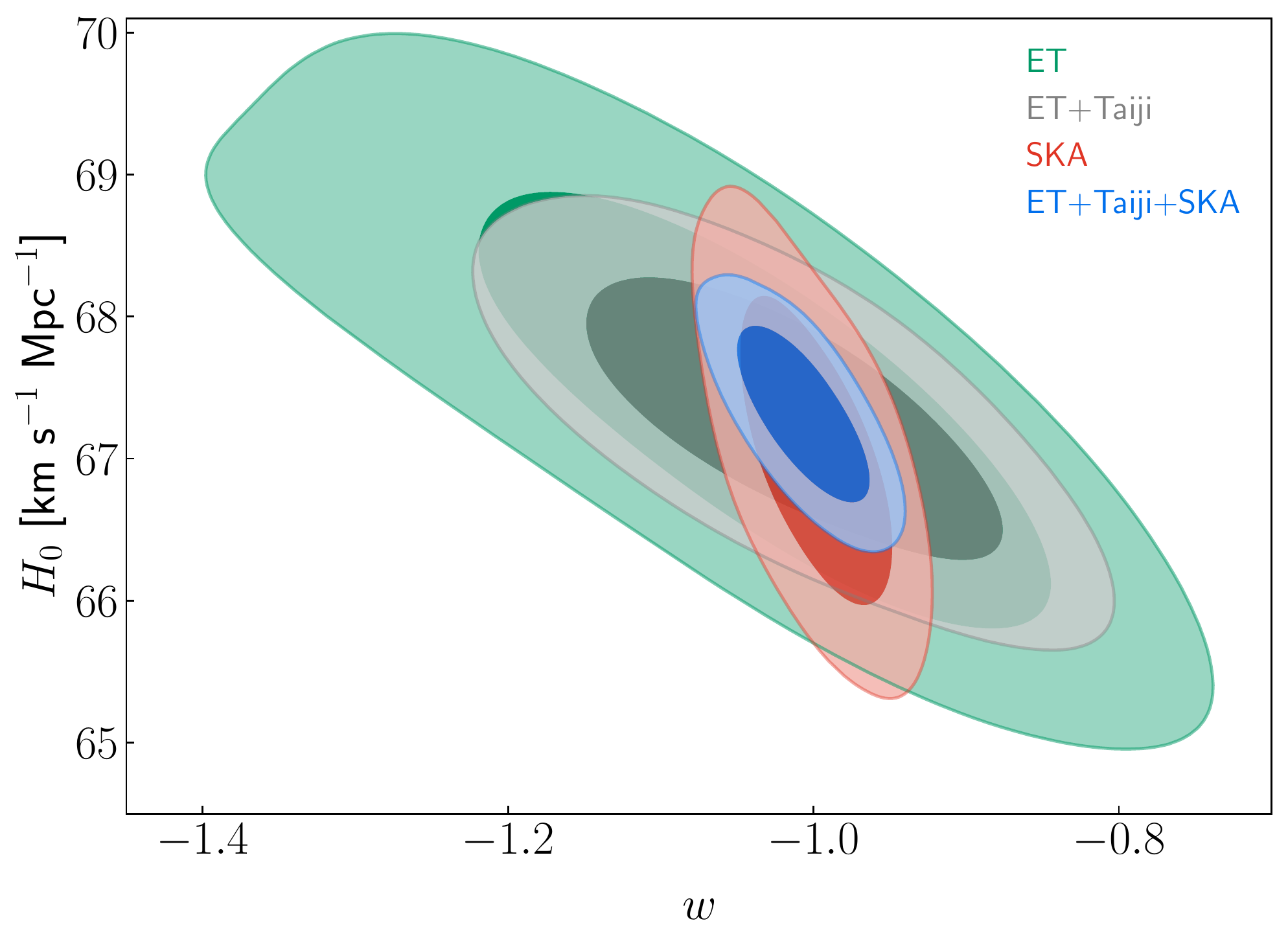}
\centering
 \caption{\label{wcdm} The two-dimensional marginalized contours (68.3\% and 95.4\% confidence level) in the $w$--$\Omega_{\rm m}$ and $w$--$H_0$ planes by using the ET, ET+Taiji, SKA, and ET+Taiji+SKA data.}
\end{figure}

\begin{figure}[htbp]
\includegraphics[width=7.2cm,height=5.4cm]{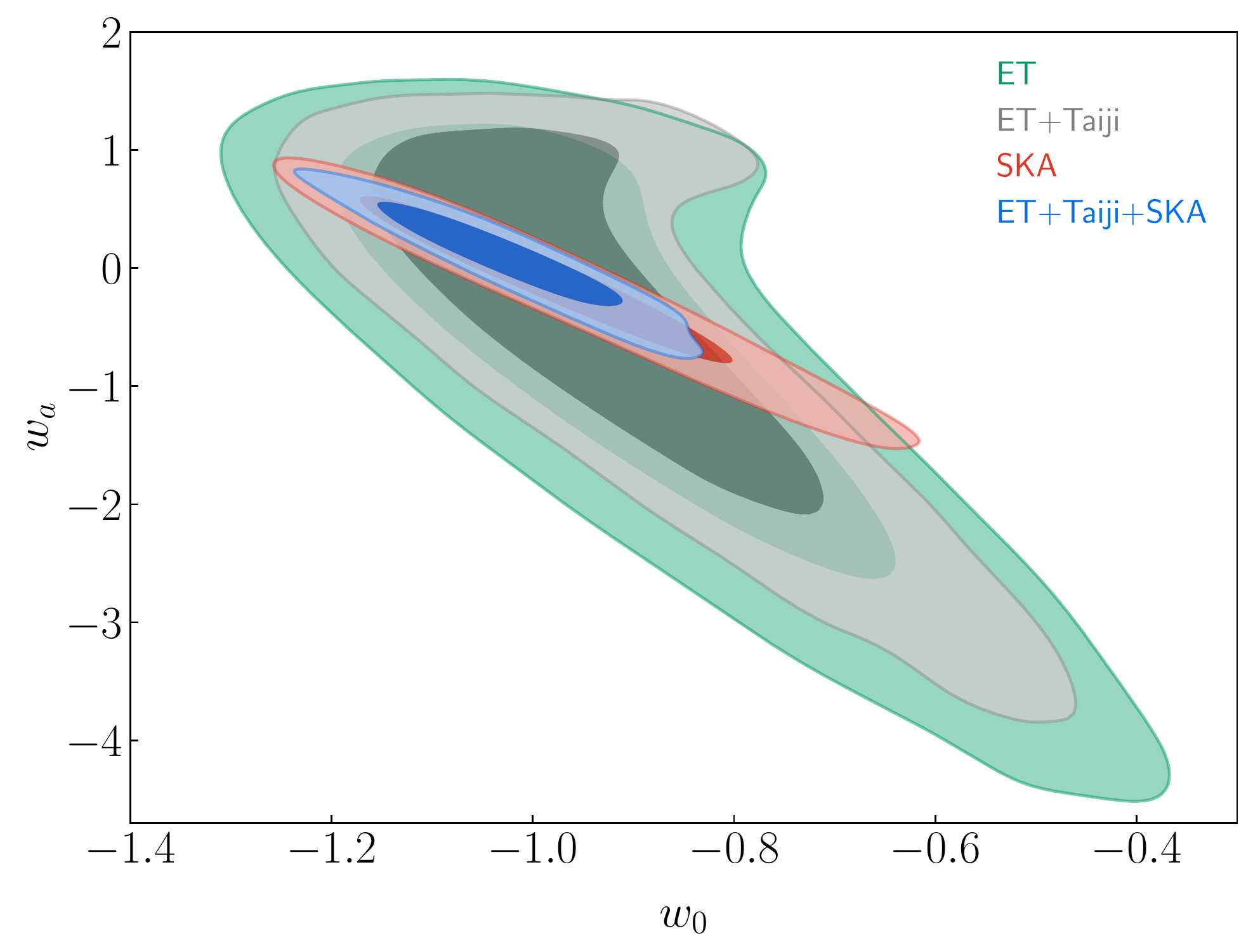}
\centering
 \caption{\label{cpl} The two-dimensional marginalized contours (68.3\% and 95.4\% confidence level) in the $w_0$--$w_a$ plane by using the ET, ET+Taiji, SKA, and ET+Taiji+SKA data.}
\end{figure}

In this section, we shall report the constraint results of cosmological parameters. Here we consider three typical dark energy models, i.e., the $\Lambda$CDM model [$w(z)=-1$], the $w$CDM model [$w(z)=$ constant], and the CPL model [$w(z)=w_0+w_a z/(1+z)$]. We use the simulated standard siren data and the 21-cm IM data to perform the Markov-chain Monte Carlo analysis \cite{Lewis:2002ah} to constrain these three cosmological models. The constraint results are shown in Figs.~\ref{lcdm}--\ref{cpl} and summarized in Table \ref{tab:full}. Note here that we use SKA to denote the 21-cm IM surveys of the SKA1-MID array, and use ET+Taiji to denote the combination of the standard siren observations from ET and Taiji. We use $\sigma(\xi)$ and $\varepsilon(\xi)$ to represent the absolute and relative errors of the parameter $\xi$, respectively, with $\varepsilon(\xi)$ defined as $\varepsilon(\xi)=\sigma(\xi)/\xi$.

From Figs.~\ref{lcdm}--\ref{cpl}, we see that the addition of the Taiji data to the ET data could improve the constraints on the cosmological parameters to some extent. Concretely, for example, in the $\Lambda$CDM model, the constraint on $H_0$ could be improved by 20.0\% when adding the Taiji data to the ET data. For the dark-energy EoS parameters, ET+Taiji could improve the constraints on $w$ by 30.0\% in the $w$CDM model, and on $w_0$ and $w_a$ by 12.9\% and 13.1\% in the CPL model, compared with the ET data.

In Fig.~\ref{lcdm}, we show the constraints on the $\Lambda$CDM model in the $\Omega_{\rm m}$--$H_0$ plane from the ET, ET+Taiji, SKA, and ET+Taiji+SKA data.
The contours of SKA and ET+Taiji show different degeneracy orientations and thus the combination of them could break the parameter degeneracies. We also see that ET+Taiji could provide a tight constraint on $H_0$, {$\sigma(H_0)=0.44\ {\rm km\ s^{-1}\ Mpc^{-1}}$, comparable with the result of $\sigma(H_0)=0.42\ {\rm km\ s^{-1}\ Mpc^{-1}}$ by $Planck$ 2018 TT,TE,EE+lowE+lensing+BAO data. SKA gives $\sigma(\Omega_{\rm m})=0.006$ that is comparable with the result of $\sigma(\Omega_{\rm m})\approx 0.006$ by the $Planck$ 2018 TT,TE,EE+lowE+lensing+BAO data \cite{Aghanim:2018eyx}.} The combination of ET+Taiji and SKA gives tighter constraints on both $H_0$ and $\Omega_{\rm m}$. The joint constraint gives the results of $\sigma(H_0)=0.28\ {\rm km\ s^{-1}\ Mpc^{-1}}$ and $\sigma(\Omega_{\rm m})=0.005$, which are better than {the results of $\sigma(H_0)=0.40\ {\rm km\ s^{-1}\ Mpc^{-1}}$ and $\sigma(\Omega_{\rm m})=0.0054$ by the $Planck$ 2018 TT,TE,EE+lowE+lensing+SNe+BAO data} \cite{Yang:2021qge}. What's more, with the joint data, the constraint {precision} of $H_0$ is 0.4\%, and the constraint {precision} of $\Omega_{\rm m}$ is 1.5\% (rather close to 1\%), indicating that standard sirens and 21-cm IM could jointly provide a precise late-universe cosmological probe.

In Fig.~\ref{wcdm}, we show the constraint results for the $w$CDM model in the $w$--$\Omega_{\rm m}$ and $w$--$H_0$ planes. We clearly see that the parameter degeneracy orientations of SKA and ET+Taiji are almost orthogonal in the $w$--$\Omega_{\rm m}$ plane and thus the combination of them could significantly break the parameter degeneracies. Also, we see that SKA could tightly constrain $\Omega_{\rm m}$ and $w$, while ET+Taiji could tightly constrain $H_0$, and thus the combination of them could tightly constrain all of these three parameters. Concretely, with the ET+Taiji data, the constraint {precision} of $\Omega_{\rm m}$, $H_0$, and $w$ {is} 5.0\%, 0.9\%, and 8.3\%, respectively. With the SKA data, the constraint {precision} of $\Omega_{\rm m}$, $H_0$, and $w$ {is} 2.1\%, 1.0\%, and 3.3\%, respectively. The joint constraint gives $\sigma(w)=0.028$, which is better than {the result of  $\sigma(w)=0.032$ by the $Planck$ 2018 TT,TE,EE+lowE+lensing+SNe+BAO data} \cite{Aghanim:2018eyx}. With the joint data, the constraint {precision} of $\Omega_{\rm m}$, $H_0$, and $w$ {is} 1.6\%, 0.6\%, and 2.8\%, respectively.

In Fig.~\ref{cpl}, we show the case for the CPL model in the $w_0$--$w_a$ plane. We find that ET+Taiji and SKA show different parameter degeneracy orientations and thus the combination of them could break the parameter degeneracies.
Concretely, the joint constraints give the results {$\sigma(w_0)=0.077$ and $\sigma(w_a)=0.295$, which are comparable with the results of $\sigma(w_0)=0.077$ and $\sigma(w_a)=0.290$ by the $Planck$ 2018 TT,TE,EE+lowE+lensing+SNe+BAO data} \cite{Aghanim:2018eyx}.

In the next decade, some other promising cosmological probes, e.g., fast radio bursts (FRBs), time-delay cosmography, galaxy clustering (GC), and weak lensing (WL), will also be greatly developed. Some forecasts for cosmological parameter estimation using these cosmological probes have been made. For example, the constraint precision of $w$ in the $w$CDM could reach 4.3\% using the combination of 10000 localized FRBs and CMB \cite{Zhao:2020ole}; the constraint precision of $H_0$ in the $\Lambda$CDM model could achieve 1.3\% using the measurements of $D_{\Delta t}$ and $D_{\rm d}$ of 20 lensed supernovae \cite{Suyu:2020opl}. Using all cosmological probes considered in the $Euclid$ analysis [GCs (spectroscopic galaxy clustering)+WL+GCp (photometric galaxy clustering)+$\rm GCp\times WL$ (cross-correlations between GCp and WL)], the constraint precision of $w_0$ in the CPL model could achieve 3.8\%, and the constraint precision of $H_0$ in the $\Lambda$CDM model could achieve 0.54\% \cite{Euclid:2021qvm}. In our future works, we will study how the combinations of these promising cosmological probes could break the cosmological parameter degeneracies.

\section{Conclusion}\label{sec:con}

As two promising cosmological probes, standard sirens and 21-cm IM could play a crucial role in the cosmological parameter estimation. Hence, we wish to investigate the capability of estimating cosmological parameters using the combination of these two promising cosmological probes.
In this work, we simulate the standard siren data based on the 10-year operation time of ET and the 5-year operation time of Taiji, and simulate the 21-cm IM data based on the 10000-h total integration time of SKA assuming perfect foreground removal and calibration. By comparing the results of ET+Taiji+SKA with those of ET+Taiji and SKA, we find that standard sirens and 21-cm IM could jointly provide a precise late-universe cosmological probe. In the $\Lambda$CDM model, using the joint data, the constraint {precision} of $H_0$ is 0.4\% (less than 1\%), and the constraint {precision} of $\Omega_{\rm m}$ is 1.5\% (around 1\%), indicating that the precision cosmology using these two promising cosmological probes is worth expecting.

In addition, we find that these two cosmological probes could effectively break the parameter degeneracies. Taking the $w$CDM model as an example, the parameter degeneracy orientations of ET+Taiji and SKA are almost orthogonal in the $w$--$H_0$ plane and thus the combination of them could significantly break the parameter degeneracies. This implies that standard sirens and 21-cm IM could complement each other. Actually, the standard siren could directly measure $d_{\rm L}(z)$, thus providing a powerful constraint on $H_0$, while the measurement of BAO by 21-cm IM with a large survey volume could provide the information of $H(z)$, which could constrain $w(z)$ well. Hence, the combination of standard siren and 21-cm IM could tightly constrain both the Hubble constant and EoS of dark energy. The joint data of ET+Taiji+SKA could give the constraint $\sigma(w)=0.028$, which is better than the result of $Planck$ 2018 TT,TE,EE+lowE+lensing+SNe+BAO, providing a powerful late-universe cosmological probe.

The improvements of cosmological constraints due to the synergy between standard sirens and 21-cm IM could also be seen in the CPL model. The joint data of ET+Taiji+SKA could give tight constraints on $H_0$, $w_0$, and $w_a$ at the same time, with $\sigma(H_0)=0.63\ {\rm km\ s^{-1}\ Mpc^{-1}}$, $\sigma(w_0)=0.077$, and $\sigma(w_a)=0.295$, which are comparable with the results of $Planck$ 2018 TT,TE,EE+lowE+lensing+SNe+BAO.
Therefore, we can conclude that standard sirens and 21-cm IM could jointly provide a precise late-universe cosmological probe.

In the next decades, the fourth-generation dark-energy programs such as LSST \cite{Abell:2009aa}, Euclid \cite{Laureijs:2011gra}, and WFRST \cite{Spergel:2013tha} will be implemented, and the cosmological probes based on the optical observations will be greatly developed.
In addition, the lower frequency (nano-Hz) GWs produced by the inspiralling of supermassive black hole binaries could be detected by the global network of pulsar timing array \cite{Rajagopal:1994zj}. The multiband GW observations combined with the optical, near-infrared, and radio observations will usher in a new era of cosmology.


\begin{acknowledgments}
We thank Ze-Wei Zhao, Jing-Zhao Qi, Hai-Li Li, Ming Zhang, {Ji-Guo Zhang, and Yu Cui} for helpful discussions. This work was supported by the National Natural Science Foundation of China (Grants No. 11975072, No. 11835009, No. 11875102, and No. 11690021), the Liaoning Revitalization Talents Program (Grant No. XLYC1905011), the Fundamental Research Funds for the Central Universities (Grant No. N2005030), and the National 111 Project of China (Grant No. B16009).

\end{acknowledgments}

\bibliography{gwim}

\begin{thebibliography}{112}%
\makeatletter
\providecommand \@ifxundefined [1]{%
 \@ifx{#1\undefined}
}%
\providecommand \@ifnum [1]{%
 \ifnum #1\expandafter \@firstoftwo
 \else \expandafter \@secondoftwo
 \fi
}%
\providecommand \@ifx [1]{%
 \ifx #1\expandafter \@firstoftwo
 \else \expandafter \@secondoftwo
 \fi
}%
\providecommand \natexlab [1]{#1}%
\providecommand \enquote  [1]{``#1''}%
\providecommand \bibnamefont  [1]{#1}%
\providecommand \bibfnamefont [1]{#1}%
\providecommand \citenamefont [1]{#1}%
\providecommand \href@noop [0]{\@secondoftwo}%
\providecommand \href [0]{\begingroup \@sanitize@url \@href}%
\providecommand \@href[1]{\@@startlink{#1}\@@href}%
\providecommand \@@href[1]{\endgroup#1\@@endlink}%
\providecommand \@sanitize@url [0]{\catcode `\\12\catcode `\$12\catcode
  `\&12\catcode `\#12\catcode `\^12\catcode `\_12\catcode `\%12\relax}%
\providecommand \@@startlink[1]{}%
\providecommand \@@endlink[0]{}%
\providecommand \url  [0]{\begingroup\@sanitize@url \@url }%
\providecommand \@url [1]{\endgroup\@href {#1}{\urlprefix }}%
\providecommand \urlprefix  [0]{URL }%
\providecommand \Eprint [0]{\href }%
\providecommand \doibase [0]{http://dx.doi.org/}%
\providecommand \selectlanguage [0]{\@gobble}%
\providecommand \bibinfo  [0]{\@secondoftwo}%
\providecommand \bibfield  [0]{\@secondoftwo}%
\providecommand \translation [1]{[#1]}%
\providecommand \BibitemOpen [0]{}%
\providecommand \bibitemStop [0]{}%
\providecommand \bibitemNoStop [0]{.\EOS\space}%
\providecommand \EOS [0]{\spacefactor3000\relax}%
\providecommand \BibitemShut  [1]{\csname bibitem#1\endcsname}%
\let\auto@bib@innerbib\@empty
\bibitem [{\citenamefont {Bennett}\ \emph {et~al.}(2003)\citenamefont {Bennett}
  \emph {et~al.}}]{Bennett:2003bz}%
  \BibitemOpen
  \bibfield  {author} {\bibinfo {author} {\bibfnamefont {C.~L.}\ \bibnamefont
  {Bennett}} \emph {et~al.} (\bibinfo {collaboration} {WMAP}),\ }\href
  {\doibase 10.1086/377253} {\bibfield  {journal} {\bibinfo  {journal}
  {Astrophys. J. Suppl.}\ }\textbf {\bibinfo {volume} {148}},\ \bibinfo {pages}
  {1} (\bibinfo {year} {2003})},\ \Eprint
  {http://arxiv.org/abs/astro-ph/0302207} {arXiv:astro-ph/0302207} \BibitemShut
  {NoStop}%
\bibitem [{\citenamefont {Spergel}\ \emph {et~al.}(2003)\citenamefont {Spergel}
  \emph {et~al.}}]{Spergel:2003cb}%
  \BibitemOpen
  \bibfield  {author} {\bibinfo {author} {\bibfnamefont {D.~N.}\ \bibnamefont
  {Spergel}} \emph {et~al.} (\bibinfo {collaboration} {WMAP}),\ }\href
  {\doibase 10.1086/377226} {\bibfield  {journal} {\bibinfo  {journal}
  {Astrophys. J. Suppl.}\ }\textbf {\bibinfo {volume} {148}},\ \bibinfo {pages}
  {175} (\bibinfo {year} {2003})},\ \Eprint
  {http://arxiv.org/abs/astro-ph/0302209} {arXiv:astro-ph/0302209} \BibitemShut
  {NoStop}%
\bibitem [{\citenamefont {Aghanim}\ \emph {et~al.}(2020)\citenamefont {Aghanim}
  \emph {et~al.}}]{Aghanim:2018eyx}%
  \BibitemOpen
  \bibfield  {author} {\bibinfo {author} {\bibfnamefont {N.}~\bibnamefont
  {Aghanim}} \emph {et~al.} (\bibinfo {collaboration} {Planck}),\ }\href
  {\doibase 10.1051/0004-6361/201833910} {\bibfield  {journal} {\bibinfo
  {journal} {Astron. Astrophys.}\ }\textbf {\bibinfo {volume} {641}},\ \bibinfo
  {pages} {A6} (\bibinfo {year} {2020})},\ \Eprint
  {http://arxiv.org/abs/1807.06209} {arXiv:1807.06209 [astro-ph.CO]}
  \BibitemShut {NoStop}%
\bibitem [{\citenamefont {Riess}\ \emph {et~al.}(2021)\citenamefont {Riess},
  \citenamefont {Casertano}, \citenamefont {Yuan}, \citenamefont {Bowers},
  \citenamefont {Macri}, \citenamefont {Zinn},\ and\ \citenamefont
  {Scolnic}}]{Riess:2020fzl}%
  \BibitemOpen
  \bibfield  {author} {\bibinfo {author} {\bibfnamefont {A.~G.}\ \bibnamefont
  {Riess}}, \bibinfo {author} {\bibfnamefont {S.}~\bibnamefont {Casertano}},
  \bibinfo {author} {\bibfnamefont {W.}~\bibnamefont {Yuan}}, \bibinfo {author}
  {\bibfnamefont {J.~B.}\ \bibnamefont {Bowers}}, \bibinfo {author}
  {\bibfnamefont {L.}~\bibnamefont {Macri}}, \bibinfo {author} {\bibfnamefont
  {J.~C.}\ \bibnamefont {Zinn}}, \ and\ \bibinfo {author} {\bibfnamefont
  {D.}~\bibnamefont {Scolnic}},\ }\href {\doibase 10.3847/2041-8213/abdbaf}
  {\bibfield  {journal} {\bibinfo  {journal} {Astrophys. J. Lett.}\ }\textbf
  {\bibinfo {volume} {908}},\ \bibinfo {pages} {L6} (\bibinfo {year} {2021})},\
  \Eprint {http://arxiv.org/abs/2012.08534} {arXiv:2012.08534 [astro-ph.CO]}
  \BibitemShut {NoStop}%
\bibitem [{\citenamefont {Cai}(2020)}]{cai:2020}%
  \BibitemOpen
  \bibfield  {author} {\bibinfo {author} {\bibfnamefont {R.-G.}\ \bibnamefont
  {Cai}},\ }\href {\doibase 10.1007/s11433-020-1540-4} {\bibfield  {journal}
  {\bibinfo  {journal} {Sci. China Phys. Mech. Astron.}\ }\textbf {\bibinfo
  {volume} {63}},\ \bibinfo {pages} {290401} (\bibinfo {year}
  {2020})}\BibitemShut {NoStop}%
\bibitem [{\citenamefont {Guo}\ \emph {et~al.}(2019)\citenamefont {Guo},
  \citenamefont {Zhang},\ and\ \citenamefont {Zhang}}]{Guo:2018ans}%
  \BibitemOpen
  \bibfield  {author} {\bibinfo {author} {\bibfnamefont {R.-Y.}\ \bibnamefont
  {Guo}}, \bibinfo {author} {\bibfnamefont {J.-F.}\ \bibnamefont {Zhang}}, \
  and\ \bibinfo {author} {\bibfnamefont {X.}~\bibnamefont {Zhang}},\ }\href
  {\doibase 10.1088/1475-7516/2019/02/054} {\bibfield  {journal} {\bibinfo
  {journal} {JCAP}\ }\textbf {\bibinfo {volume} {02}},\ \bibinfo {pages} {054}
  (\bibinfo {year} {2019})},\ \Eprint {http://arxiv.org/abs/1809.02340}
  {arXiv:1809.02340 [astro-ph.CO]} \BibitemShut {NoStop}%
\bibitem [{\citenamefont {Guo}\ \emph {et~al.}(2020)\citenamefont {Guo},
  \citenamefont {Zhang},\ and\ \citenamefont {Zhang}}]{Guo:2019dui}%
  \BibitemOpen
  \bibfield  {author} {\bibinfo {author} {\bibfnamefont {R.-Y.}\ \bibnamefont
  {Guo}}, \bibinfo {author} {\bibfnamefont {J.-F.}\ \bibnamefont {Zhang}}, \
  and\ \bibinfo {author} {\bibfnamefont {X.}~\bibnamefont {Zhang}},\ }\href
  {\doibase 10.1007/s11433-019-1514-0} {\bibfield  {journal} {\bibinfo
  {journal} {Sci. China Phys. Mech. Astron.}\ }\textbf {\bibinfo {volume}
  {63}},\ \bibinfo {pages} {290406} (\bibinfo {year} {2020})},\ \Eprint
  {http://arxiv.org/abs/1910.13944} {arXiv:1910.13944 [astro-ph.CO]}
  \BibitemShut {NoStop}%
\bibitem [{\citenamefont {Yang}\ \emph {et~al.}(2018)\citenamefont {Yang},
  \citenamefont {Pan}, \citenamefont {Di~Valentino}, \citenamefont {Nunes},
  \citenamefont {Vagnozzi},\ and\ \citenamefont {Mota}}]{Yang:2018euj}%
  \BibitemOpen
  \bibfield  {author} {\bibinfo {author} {\bibfnamefont {W.}~\bibnamefont
  {Yang}}, \bibinfo {author} {\bibfnamefont {S.}~\bibnamefont {Pan}}, \bibinfo
  {author} {\bibfnamefont {E.}~\bibnamefont {Di~Valentino}}, \bibinfo {author}
  {\bibfnamefont {R.~C.}\ \bibnamefont {Nunes}}, \bibinfo {author}
  {\bibfnamefont {S.}~\bibnamefont {Vagnozzi}}, \ and\ \bibinfo {author}
  {\bibfnamefont {D.~F.}\ \bibnamefont {Mota}},\ }\href {\doibase
  10.1088/1475-7516/2018/09/019} {\bibfield  {journal} {\bibinfo  {journal}
  {JCAP}\ }\textbf {\bibinfo {volume} {09}},\ \bibinfo {pages} {019} (\bibinfo
  {year} {2018})},\ \Eprint {http://arxiv.org/abs/1805.08252} {arXiv:1805.08252
  [astro-ph.CO]} \BibitemShut {NoStop}%
\bibitem [{\citenamefont {Vagnozzi}(2020)}]{Vagnozzi:2019ezj}%
  \BibitemOpen
  \bibfield  {author} {\bibinfo {author} {\bibfnamefont {S.}~\bibnamefont
  {Vagnozzi}},\ }\href {\doibase 10.1103/PhysRevD.102.023518} {\bibfield
  {journal} {\bibinfo  {journal} {Phys. Rev. D}\ }\textbf {\bibinfo {volume}
  {102}},\ \bibinfo {pages} {023518} (\bibinfo {year} {2020})},\ \Eprint
  {http://arxiv.org/abs/1907.07569} {arXiv:1907.07569 [astro-ph.CO]}
  \BibitemShut {NoStop}%
\bibitem [{\citenamefont {Di~Valentino}\ \emph
  {et~al.}(2020{\natexlab{a}})\citenamefont {Di~Valentino}, \citenamefont
  {Melchiorri}, \citenamefont {Mena},\ and\ \citenamefont
  {Vagnozzi}}]{DiValentino:2019jae}%
  \BibitemOpen
  \bibfield  {author} {\bibinfo {author} {\bibfnamefont {E.}~\bibnamefont
  {Di~Valentino}}, \bibinfo {author} {\bibfnamefont {A.}~\bibnamefont
  {Melchiorri}}, \bibinfo {author} {\bibfnamefont {O.}~\bibnamefont {Mena}}, \
  and\ \bibinfo {author} {\bibfnamefont {S.}~\bibnamefont {Vagnozzi}},\ }\href
  {\doibase 10.1103/PhysRevD.101.063502} {\bibfield  {journal} {\bibinfo
  {journal} {Phys. Rev. D}\ }\textbf {\bibinfo {volume} {101}},\ \bibinfo
  {pages} {063502} (\bibinfo {year} {2020}{\natexlab{a}})},\ \Eprint
  {http://arxiv.org/abs/1910.09853} {arXiv:1910.09853 [astro-ph.CO]}
  \BibitemShut {NoStop}%
\bibitem [{\citenamefont {Di~Valentino}\ \emph
  {et~al.}(2020{\natexlab{b}})\citenamefont {Di~Valentino}, \citenamefont
  {Melchiorri}, \citenamefont {Mena},\ and\ \citenamefont
  {Vagnozzi}}]{DiValentino:2019ffd}%
  \BibitemOpen
  \bibfield  {author} {\bibinfo {author} {\bibfnamefont {E.}~\bibnamefont
  {Di~Valentino}}, \bibinfo {author} {\bibfnamefont {A.}~\bibnamefont
  {Melchiorri}}, \bibinfo {author} {\bibfnamefont {O.}~\bibnamefont {Mena}}, \
  and\ \bibinfo {author} {\bibfnamefont {S.}~\bibnamefont {Vagnozzi}},\ }\href
  {\doibase 10.1016/j.dark.2020.100666} {\bibfield  {journal} {\bibinfo
  {journal} {Phys. Dark Univ.}\ }\textbf {\bibinfo {volume} {30}},\ \bibinfo
  {pages} {100666} (\bibinfo {year} {2020}{\natexlab{b}})},\ \Eprint
  {http://arxiv.org/abs/1908.04281} {arXiv:1908.04281 [astro-ph.CO]}
  \BibitemShut {NoStop}%
\bibitem [{\citenamefont {Liu}\ \emph {et~al.}(2020)\citenamefont {Liu},
  \citenamefont {Huang}, \citenamefont {Luo}, \citenamefont {Miao},
  \citenamefont {Singh},\ and\ \citenamefont {Huang}}]{Liu:2019awo}%
  \BibitemOpen
  \bibfield  {author} {\bibinfo {author} {\bibfnamefont {M.}~\bibnamefont
  {Liu}}, \bibinfo {author} {\bibfnamefont {Z.}~\bibnamefont {Huang}}, \bibinfo
  {author} {\bibfnamefont {X.}~\bibnamefont {Luo}}, \bibinfo {author}
  {\bibfnamefont {H.}~\bibnamefont {Miao}}, \bibinfo {author} {\bibfnamefont
  {N.~K.}\ \bibnamefont {Singh}}, \ and\ \bibinfo {author} {\bibfnamefont
  {L.}~\bibnamefont {Huang}},\ }\href {\doibase 10.1007/s11433-019-1509-5}
  {\bibfield  {journal} {\bibinfo  {journal} {Sci. China Phys. Mech. Astron.}\
  }\textbf {\bibinfo {volume} {63}},\ \bibinfo {pages} {290405} (\bibinfo
  {year} {2020})},\ \Eprint {http://arxiv.org/abs/1912.00190} {arXiv:1912.00190
  [astro-ph.CO]} \BibitemShut {NoStop}%
\bibitem [{\citenamefont {Zhang}\ and\ \citenamefont
  {Huang}(2020)}]{Zhang:2019cww}%
  \BibitemOpen
  \bibfield  {author} {\bibinfo {author} {\bibfnamefont {X.}~\bibnamefont
  {Zhang}}\ and\ \bibinfo {author} {\bibfnamefont {Q.-G.}\ \bibnamefont
  {Huang}},\ }\href {\doibase 10.1007/s11433-019-1504-8} {\bibfield  {journal}
  {\bibinfo  {journal} {Sci. China Phys. Mech. Astron.}\ }\textbf {\bibinfo
  {volume} {63}},\ \bibinfo {pages} {290402} (\bibinfo {year} {2020})},\
  \Eprint {http://arxiv.org/abs/1911.09439} {arXiv:1911.09439 [astro-ph.CO]}
  \BibitemShut {NoStop}%
\bibitem [{\citenamefont {Ding}\ \emph {et~al.}(2020)\citenamefont {Ding},
  \citenamefont {Nakama},\ and\ \citenamefont {Wang}}]{Ding:2019mmw}%
  \BibitemOpen
  \bibfield  {author} {\bibinfo {author} {\bibfnamefont {Q.}~\bibnamefont
  {Ding}}, \bibinfo {author} {\bibfnamefont {T.}~\bibnamefont {Nakama}}, \ and\
  \bibinfo {author} {\bibfnamefont {Y.}~\bibnamefont {Wang}},\ }\href {\doibase
  10.1007/s11433-020-1531-0} {\bibfield  {journal} {\bibinfo  {journal} {Sci.
  China Phys. Mech. Astron.}\ }\textbf {\bibinfo {volume} {63}},\ \bibinfo
  {pages} {290403} (\bibinfo {year} {2020})},\ \Eprint
  {http://arxiv.org/abs/1912.12600} {arXiv:1912.12600 [astro-ph.CO]}
  \BibitemShut {NoStop}%
\bibitem [{\citenamefont {Feng}\ \emph {et~al.}(2020)\citenamefont {Feng},
  \citenamefont {He}, \citenamefont {Li}, \citenamefont {Zhang},\ and\
  \citenamefont {Zhang}}]{Feng:2019jqa}%
  \BibitemOpen
  \bibfield  {author} {\bibinfo {author} {\bibfnamefont {L.}~\bibnamefont
  {Feng}}, \bibinfo {author} {\bibfnamefont {D.-Z.}\ \bibnamefont {He}},
  \bibinfo {author} {\bibfnamefont {H.-L.}\ \bibnamefont {Li}}, \bibinfo
  {author} {\bibfnamefont {J.-F.}\ \bibnamefont {Zhang}}, \ and\ \bibinfo
  {author} {\bibfnamefont {X.}~\bibnamefont {Zhang}},\ }\href {\doibase
  10.1007/s11433-019-1511-8} {\bibfield  {journal} {\bibinfo  {journal} {Sci.
  China Phys. Mech. Astron.}\ }\textbf {\bibinfo {volume} {63}},\ \bibinfo
  {pages} {290404} (\bibinfo {year} {2020})},\ \Eprint
  {http://arxiv.org/abs/1910.03872} {arXiv:1910.03872 [astro-ph.CO]}
  \BibitemShut {NoStop}%
\bibitem [{\citenamefont {Lin}\ \emph {et~al.}(2020)\citenamefont {Lin},
  \citenamefont {Hu},\ and\ \citenamefont {Raveri}}]{Lin:2020jcb}%
  \BibitemOpen
  \bibfield  {author} {\bibinfo {author} {\bibfnamefont {M.-X.}\ \bibnamefont
  {Lin}}, \bibinfo {author} {\bibfnamefont {W.}~\bibnamefont {Hu}}, \ and\
  \bibinfo {author} {\bibfnamefont {M.}~\bibnamefont {Raveri}},\ }\href
  {\doibase 10.1103/PhysRevD.102.123523} {\bibfield  {journal} {\bibinfo
  {journal} {Phys. Rev. D}\ }\textbf {\bibinfo {volume} {102}},\ \bibinfo
  {pages} {123523} (\bibinfo {year} {2020})},\ \Eprint
  {http://arxiv.org/abs/2009.08974} {arXiv:2009.08974 [astro-ph.CO]}
  \BibitemShut {NoStop}%
\bibitem [{\citenamefont {Li}\ and\ \citenamefont {Zhang}(2020)}]{Li:2020tds}%
  \BibitemOpen
  \bibfield  {author} {\bibinfo {author} {\bibfnamefont {H.}~\bibnamefont
  {Li}}\ and\ \bibinfo {author} {\bibfnamefont {X.}~\bibnamefont {Zhang}},\
  }\href {\doibase 10.1016/j.scib.2020.04.038} {\bibfield  {journal} {\bibinfo
  {journal} {Sci. Bull.}\ }\textbf {\bibinfo {volume} {65}},\ \bibinfo {pages}
  {1419} (\bibinfo {year} {2020})},\ \Eprint {http://arxiv.org/abs/2005.10458}
  {arXiv:2005.10458 [astro-ph.CO]} \BibitemShut {NoStop}%
\bibitem [{\citenamefont {Hryczuk}\ and\ \citenamefont
  {Jod\l{}owski}(2020)}]{Hryczuk:2020jhi}%
  \BibitemOpen
  \bibfield  {author} {\bibinfo {author} {\bibfnamefont {A.}~\bibnamefont
  {Hryczuk}}\ and\ \bibinfo {author} {\bibfnamefont {K.}~\bibnamefont
  {Jod\l{}owski}},\ }\href {\doibase 10.1103/PhysRevD.102.043024} {\bibfield
  {journal} {\bibinfo  {journal} {Phys. Rev. D}\ }\textbf {\bibinfo {volume}
  {102}},\ \bibinfo {pages} {043024} (\bibinfo {year} {2020})},\ \Eprint
  {http://arxiv.org/abs/2006.16139} {arXiv:2006.16139 [hep-ph]} \BibitemShut
  {NoStop}%
\bibitem [{\citenamefont {Gao}\ \emph {et~al.}(2021)\citenamefont {Gao},
  \citenamefont {Zhao}, \citenamefont {Xue},\ and\ \citenamefont
  {Zhang}}]{Gao:2021xnk}%
  \BibitemOpen
  \bibfield  {author} {\bibinfo {author} {\bibfnamefont {L.-Y.}\ \bibnamefont
  {Gao}}, \bibinfo {author} {\bibfnamefont {Z.-W.}\ \bibnamefont {Zhao}},
  \bibinfo {author} {\bibfnamefont {S.-S.}\ \bibnamefont {Xue}}, \ and\
  \bibinfo {author} {\bibfnamefont {X.}~\bibnamefont {Zhang}},\ }\href
  {\doibase 10.1088/1475-7516/2021/07/005} {\bibfield  {journal} {\bibinfo
  {journal} {JCAP}\ }\textbf {\bibinfo {volume} {07}},\ \bibinfo {pages} {005}
  (\bibinfo {year} {2021})},\ \Eprint {http://arxiv.org/abs/2101.10714}
  {arXiv:2101.10714 [astro-ph.CO]} \BibitemShut {NoStop}%
\bibitem [{\citenamefont {Wang}\ \emph
  {et~al.}(2021{\natexlab{a}})\citenamefont {Wang}, \citenamefont {He},
  \citenamefont {Zhang},\ and\ \citenamefont {Zhang}}]{Wang:2021kxc}%
  \BibitemOpen
  \bibfield  {author} {\bibinfo {author} {\bibfnamefont {L.-F.}\ \bibnamefont
  {Wang}}, \bibinfo {author} {\bibfnamefont {D.-Z.}\ \bibnamefont {He}},
  \bibinfo {author} {\bibfnamefont {J.-F.}\ \bibnamefont {Zhang}}, \ and\
  \bibinfo {author} {\bibfnamefont {X.}~\bibnamefont {Zhang}},\ }\href@noop {}
  {\  (\bibinfo {year} {2021}{\natexlab{a}})},\ \Eprint
  {http://arxiv.org/abs/2102.09331} {arXiv:2102.09331 [astro-ph.CO]}
  \BibitemShut {NoStop}%
\bibitem [{\citenamefont {Cai}\ \emph {et~al.}(2021)\citenamefont {Cai},
  \citenamefont {Guo}, \citenamefont {Li}, \citenamefont {Wang},\ and\
  \citenamefont {Yu}}]{Cai:2021wgv}%
  \BibitemOpen
  \bibfield  {author} {\bibinfo {author} {\bibfnamefont {R.-G.}\ \bibnamefont
  {Cai}}, \bibinfo {author} {\bibfnamefont {Z.-K.}\ \bibnamefont {Guo}},
  \bibinfo {author} {\bibfnamefont {L.}~\bibnamefont {Li}}, \bibinfo {author}
  {\bibfnamefont {S.-J.}\ \bibnamefont {Wang}}, \ and\ \bibinfo {author}
  {\bibfnamefont {W.-W.}\ \bibnamefont {Yu}},\ }\href {\doibase
  10.1103/PhysRevD.103.L121302} {\bibfield  {journal} {\bibinfo  {journal}
  {Phys. Rev. D}\ }\textbf {\bibinfo {volume} {103}},\ \bibinfo {pages}
  {121302} (\bibinfo {year} {2021})},\ \Eprint
  {http://arxiv.org/abs/2102.02020} {arXiv:2102.02020 [astro-ph.CO]}
  \BibitemShut {NoStop}%
\bibitem [{\citenamefont {Vagnozzi}\ \emph {et~al.}(2021)\citenamefont
  {Vagnozzi}, \citenamefont {Pacucci},\ and\ \citenamefont
  {Loeb}}]{Vagnozzi:2021tjv}%
  \BibitemOpen
  \bibfield  {author} {\bibinfo {author} {\bibfnamefont {S.}~\bibnamefont
  {Vagnozzi}}, \bibinfo {author} {\bibfnamefont {F.}~\bibnamefont {Pacucci}}, \
  and\ \bibinfo {author} {\bibfnamefont {A.}~\bibnamefont {Loeb}},\ }\href@noop
  {} {\  (\bibinfo {year} {2021})},\ \Eprint {http://arxiv.org/abs/2105.10421}
  {arXiv:2105.10421 [astro-ph.CO]} \BibitemShut {NoStop}%
\bibitem [{\citenamefont {Vagnozzi}(2021)}]{Vagnozzi:2021gjh}%
  \BibitemOpen
  \bibfield  {author} {\bibinfo {author} {\bibfnamefont {S.}~\bibnamefont
  {Vagnozzi}},\ }\href@noop {} {\  (\bibinfo {year} {2021})},\ \Eprint
  {http://arxiv.org/abs/2105.10425} {arXiv:2105.10425 [astro-ph.CO]}
  \BibitemShut {NoStop}%
\bibitem [{\citenamefont {Verde}\ \emph {et~al.}(2019)\citenamefont {Verde},
  \citenamefont {Treu},\ and\ \citenamefont {Riess}}]{Verde:2019ivm}%
  \BibitemOpen
  \bibfield  {author} {\bibinfo {author} {\bibfnamefont {L.}~\bibnamefont
  {Verde}}, \bibinfo {author} {\bibfnamefont {T.}~\bibnamefont {Treu}}, \ and\
  \bibinfo {author} {\bibfnamefont {A.~G.}\ \bibnamefont {Riess}},\ }\href
  {\doibase 10.1038/s41550-019-0902-0} {\bibfield  {journal} {\bibinfo
  {journal} {Nature Astron.}\ }\textbf {\bibinfo {volume} {3}},\ \bibinfo
  {pages} {891} (\bibinfo {year} {2019})},\ \Eprint
  {http://arxiv.org/abs/1907.10625} {arXiv:1907.10625 [astro-ph.CO]}
  \BibitemShut {NoStop}%
\bibitem [{\citenamefont {Riess}(2019)}]{Riess:2020sih}%
  \BibitemOpen
  \bibfield  {author} {\bibinfo {author} {\bibfnamefont {A.~G.}\ \bibnamefont
  {Riess}},\ }\href {\doibase 10.1038/s42254-019-0137-0} {\bibfield  {journal}
  {\bibinfo  {journal} {Nature Rev. Phys.}\ }\textbf {\bibinfo {volume} {2}},\
  \bibinfo {pages} {10} (\bibinfo {year} {2019})},\ \Eprint
  {http://arxiv.org/abs/2001.03624} {arXiv:2001.03624 [astro-ph.CO]}
  \BibitemShut {NoStop}%
\bibitem [{\citenamefont {Schutz}(1986)}]{Schutz:1986gp}%
  \BibitemOpen
  \bibfield  {author} {\bibinfo {author} {\bibfnamefont {B.~F.}\ \bibnamefont
  {Schutz}},\ }\href {\doibase 10.1038/323310a0} {\bibfield  {journal}
  {\bibinfo  {journal} {Nature}\ }\textbf {\bibinfo {volume} {323}},\ \bibinfo
  {pages} {310} (\bibinfo {year} {1986})}\BibitemShut {NoStop}%
\bibitem [{\citenamefont {Holz}\ and\ \citenamefont
  {Hughes}(2005)}]{Holz:2005df}%
  \BibitemOpen
  \bibfield  {author} {\bibinfo {author} {\bibfnamefont {D.~E.}\ \bibnamefont
  {Holz}}\ and\ \bibinfo {author} {\bibfnamefont {S.~A.}\ \bibnamefont
  {Hughes}},\ }\href {\doibase 10.1086/431341} {\bibfield  {journal} {\bibinfo
  {journal} {Astrophys. J.}\ }\textbf {\bibinfo {volume} {629}},\ \bibinfo
  {pages} {15} (\bibinfo {year} {2005})},\ \Eprint
  {http://arxiv.org/abs/astro-ph/0504616} {arXiv:astro-ph/0504616} \BibitemShut
  {NoStop}%
\bibitem [{\citenamefont {Abbott}\ \emph
  {et~al.}(2017{\natexlab{a}})\citenamefont {Abbott} \emph
  {et~al.}}]{TheLIGOScientific:2017qsa}%
  \BibitemOpen
  \bibfield  {author} {\bibinfo {author} {\bibfnamefont {B.}~\bibnamefont
  {Abbott}} \emph {et~al.} (\bibinfo {collaboration} {LIGO Scientific,
  Virgo}),\ }\href {\doibase 10.1103/PhysRevLett.119.161101} {\bibfield
  {journal} {\bibinfo  {journal} {Phys. Rev. Lett.}\ }\textbf {\bibinfo
  {volume} {119}},\ \bibinfo {pages} {161101} (\bibinfo {year}
  {2017}{\natexlab{a}})},\ \Eprint {http://arxiv.org/abs/1710.05832}
  {arXiv:1710.05832 [gr-qc]} \BibitemShut {NoStop}%
\bibitem [{\citenamefont {Abbott}\ \emph
  {et~al.}(2017{\natexlab{b}})\citenamefont {Abbott} \emph
  {et~al.}}]{Monitor:2017mdv}%
  \BibitemOpen
  \bibfield  {author} {\bibinfo {author} {\bibfnamefont {B.~P.}\ \bibnamefont
  {Abbott}} \emph {et~al.} (\bibinfo {collaboration} {LIGO Scientific, Virgo,
  Fermi-GBM, INTEGRAL}),\ }\href {\doibase 10.3847/2041-8213/aa920c} {\bibfield
   {journal} {\bibinfo  {journal} {Astrophys. J. Lett.}\ }\textbf {\bibinfo
  {volume} {848}},\ \bibinfo {pages} {L13} (\bibinfo {year}
  {2017}{\natexlab{b}})},\ \Eprint {http://arxiv.org/abs/1710.05834}
  {arXiv:1710.05834 [astro-ph.HE]} \BibitemShut {NoStop}%
\bibitem [{\citenamefont {Abbott}\ \emph
  {et~al.}(2017{\natexlab{c}})\citenamefont {Abbott} \emph
  {et~al.}}]{GBM:2017lvd}%
  \BibitemOpen
  \bibfield  {author} {\bibinfo {author} {\bibfnamefont {B.}~\bibnamefont
  {Abbott}} \emph {et~al.} (\bibinfo {collaboration} {LIGO Scientific, Virgo,
  Fermi GBM, INTEGRAL, IceCube, AstroSat Cadmium Zinc Telluride Imager Team,
  IPN, Insight-Hxmt, ANTARES, Swift, AGILE Team, 1M2H Team, Dark Energy Camera
  GW-EM, DES, DLT40, GRAWITA, Fermi-LAT, ATCA, ASKAP, Las Cumbres Observatory
  Group, OzGrav, DWF (Deeper Wider Faster Program), AST3, CAASTRO, VINROUGE,
  MASTER, J-GEM, GROWTH, JAGWAR, CaltechNRAO, TTU-NRAO, NuSTAR, Pan-STARRS,
  MAXI Team, TZAC Consortium, KU, Nordic Optical Telescope, ePESSTO, GROND,
  Texas Tech University, SALT Group, TOROS, BOOTES, MWA, CALET, IKI-GW
  Follow-up, H.E.S.S., LOFAR, LWA, HAWC, Pierre Auger, ALMA, Euro VLBI Team, Pi
  of Sky, Chandra Team at McGill University, DFN, ATLAS Telescopes, High Time
  Resolution Universe Survey, RIMAS, RATIR, SKA South Africa/MeerKAT}),\ }\href
  {\doibase 10.3847/2041-8213/aa91c9} {\bibfield  {journal} {\bibinfo
  {journal} {Astrophys. J. Lett.}\ }\textbf {\bibinfo {volume} {848}},\
  \bibinfo {pages} {L12} (\bibinfo {year} {2017}{\natexlab{c}})},\ \Eprint
  {http://arxiv.org/abs/1710.05833} {arXiv:1710.05833 [astro-ph.HE]}
  \BibitemShut {NoStop}%
\bibitem [{\citenamefont {Abbott}\ \emph
  {et~al.}(2017{\natexlab{d}})\citenamefont {Abbott} \emph
  {et~al.}}]{Abbott:2017xzu}%
  \BibitemOpen
  \bibfield  {author} {\bibinfo {author} {\bibfnamefont {B.}~\bibnamefont
  {Abbott}} \emph {et~al.} (\bibinfo {collaboration} {LIGO Scientific, Virgo,
  1M2H, Dark Energy Camera GW-E, DES, DLT40, Las Cumbres Observatory, VINROUGE,
  MASTER}),\ }\href {\doibase 10.1038/nature24471} {\bibfield  {journal}
  {\bibinfo  {journal} {Nature}\ }\textbf {\bibinfo {volume} {551}},\ \bibinfo
  {pages} {85} (\bibinfo {year} {2017}{\natexlab{d}})},\ \Eprint
  {http://arxiv.org/abs/1710.05835} {arXiv:1710.05835 [astro-ph.CO]}
  \BibitemShut {NoStop}%
\bibitem [{\citenamefont {Chen}\ \emph {et~al.}(2018)\citenamefont {Chen},
  \citenamefont {Fishbach},\ and\ \citenamefont {Holz}}]{Chen:2017rfc}%
  \BibitemOpen
  \bibfield  {author} {\bibinfo {author} {\bibfnamefont {H.-Y.}\ \bibnamefont
  {Chen}}, \bibinfo {author} {\bibfnamefont {M.}~\bibnamefont {Fishbach}}, \
  and\ \bibinfo {author} {\bibfnamefont {D.~E.}\ \bibnamefont {Holz}},\ }\href
  {\doibase 10.1038/s41586-018-0606-0} {\bibfield  {journal} {\bibinfo
  {journal} {Nature}\ }\textbf {\bibinfo {volume} {562}},\ \bibinfo {pages}
  {545} (\bibinfo {year} {2018})},\ \Eprint {http://arxiv.org/abs/1712.06531}
  {arXiv:1712.06531 [astro-ph.CO]} \BibitemShut {NoStop}%
\bibitem [{\citenamefont {Cai}\ \emph {et~al.}(2018)\citenamefont {Cai},
  \citenamefont {Liu},\ and\ \citenamefont {Wang}}]{Cai:2017buj}%
  \BibitemOpen
  \bibfield  {author} {\bibinfo {author} {\bibfnamefont {R.-G.}\ \bibnamefont
  {Cai}}, \bibinfo {author} {\bibfnamefont {T.-B.}\ \bibnamefont {Liu}}, \ and\
  \bibinfo {author} {\bibfnamefont {S.-J.}\ \bibnamefont {Wang}},\ }\href
  {\doibase 10.1103/PhysRevD.97.023027} {\bibfield  {journal} {\bibinfo
  {journal} {Phys. Rev. D}\ }\textbf {\bibinfo {volume} {97}},\ \bibinfo
  {pages} {023027} (\bibinfo {year} {2018})},\ \Eprint
  {http://arxiv.org/abs/1710.02425} {arXiv:1710.02425 [hep-ph]} \BibitemShut
  {NoStop}%
\bibitem [{\citenamefont {Di~Valentino}\ and\ \citenamefont
  {Melchiorri}(2018)}]{DiValentino:2017clw}%
  \BibitemOpen
  \bibfield  {author} {\bibinfo {author} {\bibfnamefont {E.}~\bibnamefont
  {Di~Valentino}}\ and\ \bibinfo {author} {\bibfnamefont {A.}~\bibnamefont
  {Melchiorri}},\ }\href {\doibase 10.1103/PhysRevD.97.041301} {\bibfield
  {journal} {\bibinfo  {journal} {Phys. Rev. D}\ }\textbf {\bibinfo {volume}
  {97}},\ \bibinfo {pages} {041301} (\bibinfo {year} {2018})},\ \Eprint
  {http://arxiv.org/abs/1710.06370} {arXiv:1710.06370 [astro-ph.CO]}
  \BibitemShut {NoStop}%
\bibitem [{\citenamefont {Yang}\ \emph {et~al.}(2019)\citenamefont {Yang},
  \citenamefont {Vagnozzi}, \citenamefont {Di~Valentino}, \citenamefont
  {Nunes}, \citenamefont {Pan},\ and\ \citenamefont {Mota}}]{Yang:2019vni}%
  \BibitemOpen
  \bibfield  {author} {\bibinfo {author} {\bibfnamefont {W.}~\bibnamefont
  {Yang}}, \bibinfo {author} {\bibfnamefont {S.}~\bibnamefont {Vagnozzi}},
  \bibinfo {author} {\bibfnamefont {E.}~\bibnamefont {Di~Valentino}}, \bibinfo
  {author} {\bibfnamefont {R.~C.}\ \bibnamefont {Nunes}}, \bibinfo {author}
  {\bibfnamefont {S.}~\bibnamefont {Pan}}, \ and\ \bibinfo {author}
  {\bibfnamefont {D.~F.}\ \bibnamefont {Mota}},\ }\href {\doibase
  10.1088/1475-7516/2019/07/037} {\bibfield  {journal} {\bibinfo  {journal}
  {JCAP}\ }\textbf {\bibinfo {volume} {07}},\ \bibinfo {pages} {037} (\bibinfo
  {year} {2019})},\ \Eprint {http://arxiv.org/abs/1905.08286} {arXiv:1905.08286
  [astro-ph.CO]} \BibitemShut {NoStop}%
\bibitem [{\citenamefont {Zhao}\ \emph {et~al.}(2018)\citenamefont {Zhao},
  \citenamefont {Wright},\ and\ \citenamefont {Li}}]{Zhao:2018gwk}%
  \BibitemOpen
  \bibfield  {author} {\bibinfo {author} {\bibfnamefont {W.}~\bibnamefont
  {Zhao}}, \bibinfo {author} {\bibfnamefont {B.~S.}\ \bibnamefont {Wright}}, \
  and\ \bibinfo {author} {\bibfnamefont {B.}~\bibnamefont {Li}},\ }\href
  {\doibase 10.1088/1475-7516/2018/10/052} {\bibfield  {journal} {\bibinfo
  {journal} {JCAP}\ }\textbf {\bibinfo {volume} {10}},\ \bibinfo {pages} {052}
  (\bibinfo {year} {2018})},\ \Eprint {http://arxiv.org/abs/1804.03066}
  {arXiv:1804.03066 [astro-ph.CO]} \BibitemShut {NoStop}%
\bibitem [{\citenamefont {Di~Valentino}\ \emph {et~al.}(2018)\citenamefont
  {Di~Valentino}, \citenamefont {Holz}, \citenamefont {Melchiorri},\ and\
  \citenamefont {Renzi}}]{DiValentino:2018jbh}%
  \BibitemOpen
  \bibfield  {author} {\bibinfo {author} {\bibfnamefont {E.}~\bibnamefont
  {Di~Valentino}}, \bibinfo {author} {\bibfnamefont {D.~E.}\ \bibnamefont
  {Holz}}, \bibinfo {author} {\bibfnamefont {A.}~\bibnamefont {Melchiorri}}, \
  and\ \bibinfo {author} {\bibfnamefont {F.}~\bibnamefont {Renzi}},\ }\href
  {\doibase 10.1103/PhysRevD.98.083523} {\bibfield  {journal} {\bibinfo
  {journal} {Phys. Rev. D}\ }\textbf {\bibinfo {volume} {98}},\ \bibinfo
  {pages} {083523} (\bibinfo {year} {2018})},\ \Eprint
  {http://arxiv.org/abs/1806.07463} {arXiv:1806.07463 [astro-ph.CO]}
  \BibitemShut {NoStop}%
\bibitem [{\citenamefont {Gray}\ \emph {et~al.}(2020)\citenamefont {Gray} \emph
  {et~al.}}]{Gray:2019ksv}%
  \BibitemOpen
  \bibfield  {author} {\bibinfo {author} {\bibfnamefont {R.}~\bibnamefont
  {Gray}} \emph {et~al.},\ }\href {\doibase 10.1103/PhysRevD.101.122001}
  {\bibfield  {journal} {\bibinfo  {journal} {Phys. Rev. D}\ }\textbf {\bibinfo
  {volume} {101}},\ \bibinfo {pages} {122001} (\bibinfo {year} {2020})},\
  \Eprint {http://arxiv.org/abs/1908.06050} {arXiv:1908.06050 [gr-qc]}
  \BibitemShut {NoStop}%
\bibitem [{\citenamefont {Chen}(2020)}]{Chen:2020dyt}%
  \BibitemOpen
  \bibfield  {author} {\bibinfo {author} {\bibfnamefont {H.-Y.}\ \bibnamefont
  {Chen}},\ }\href {\doibase 10.1103/PhysRevLett.125.201301} {\bibfield
  {journal} {\bibinfo  {journal} {Phys. Rev. Lett.}\ }\textbf {\bibinfo
  {volume} {125}},\ \bibinfo {pages} {201301} (\bibinfo {year} {2020})},\
  \Eprint {http://arxiv.org/abs/2006.02779} {arXiv:2006.02779 [astro-ph.HE]}
  \BibitemShut {NoStop}%
\bibitem [{\citenamefont {Chen}\ \emph {et~al.}(2021)\citenamefont {Chen},
  \citenamefont {Cowperthwaite}, \citenamefont {Metzger},\ and\ \citenamefont
  {Berger}}]{Chen:2020zoq}%
  \BibitemOpen
  \bibfield  {author} {\bibinfo {author} {\bibfnamefont {H.-Y.}\ \bibnamefont
  {Chen}}, \bibinfo {author} {\bibfnamefont {P.~S.}\ \bibnamefont
  {Cowperthwaite}}, \bibinfo {author} {\bibfnamefont {B.~D.}\ \bibnamefont
  {Metzger}}, \ and\ \bibinfo {author} {\bibfnamefont {E.}~\bibnamefont
  {Berger}},\ }\href {\doibase 10.3847/2041-8213/abdab0} {\bibfield  {journal}
  {\bibinfo  {journal} {Astrophys. J. Lett.}\ }\textbf {\bibinfo {volume}
  {908}},\ \bibinfo {pages} {L4} (\bibinfo {year} {2021})},\ \Eprint
  {http://arxiv.org/abs/2011.01211} {arXiv:2011.01211 [astro-ph.CO]}
  \BibitemShut {NoStop}%
\bibitem [{\citenamefont {Chen}\ \emph {et~al.}(2020)\citenamefont {Chen},
  \citenamefont {Haster}, \citenamefont {Vitale}, \citenamefont {Farr},\ and\
  \citenamefont {Isi}}]{Chen:2020gek}%
  \BibitemOpen
  \bibfield  {author} {\bibinfo {author} {\bibfnamefont {H.-Y.}\ \bibnamefont
  {Chen}}, \bibinfo {author} {\bibfnamefont {C.-J.}\ \bibnamefont {Haster}},
  \bibinfo {author} {\bibfnamefont {S.}~\bibnamefont {Vitale}}, \bibinfo
  {author} {\bibfnamefont {W.~M.}\ \bibnamefont {Farr}}, \ and\ \bibinfo
  {author} {\bibfnamefont {M.}~\bibnamefont {Isi}},\ }\href@noop {} {\
  (\bibinfo {year} {2020})},\ \Eprint {http://arxiv.org/abs/2009.14057}
  {arXiv:2009.14057 [astro-ph.CO]} \BibitemShut {NoStop}%
\bibitem [{\citenamefont {Zhang}(2019)}]{Zhang:2019ylr}%
  \BibitemOpen
  \bibfield  {author} {\bibinfo {author} {\bibfnamefont {X.}~\bibnamefont
  {Zhang}},\ }\href {\doibase 10.1007/s11433-019-9445-7} {\bibfield  {journal}
  {\bibinfo  {journal} {Sci. China Phys. Mech. Astron.}\ }\textbf {\bibinfo
  {volume} {62}},\ \bibinfo {pages} {110431} (\bibinfo {year} {2019})},\
  \Eprint {http://arxiv.org/abs/1905.11122} {arXiv:1905.11122 [astro-ph.CO]}
  \BibitemShut {NoStop}%
\bibitem [{\citenamefont {Wang}\ \emph {et~al.}(2018)\citenamefont {Wang},
  \citenamefont {Zhang}, \citenamefont {Zhang},\ and\ \citenamefont
  {Zhang}}]{Wang:2018lun}%
  \BibitemOpen
  \bibfield  {author} {\bibinfo {author} {\bibfnamefont {L.-F.}\ \bibnamefont
  {Wang}}, \bibinfo {author} {\bibfnamefont {X.-N.}\ \bibnamefont {Zhang}},
  \bibinfo {author} {\bibfnamefont {J.-F.}\ \bibnamefont {Zhang}}, \ and\
  \bibinfo {author} {\bibfnamefont {X.}~\bibnamefont {Zhang}},\ }\href
  {\doibase 10.1016/j.physletb.2018.05.027} {\bibfield  {journal} {\bibinfo
  {journal} {Phys. Lett. B}\ }\textbf {\bibinfo {volume} {782}},\ \bibinfo
  {pages} {87} (\bibinfo {year} {2018})},\ \Eprint
  {http://arxiv.org/abs/1802.04720} {arXiv:1802.04720 [astro-ph.CO]}
  \BibitemShut {NoStop}%
\bibitem [{\citenamefont {Zhang}\ \emph
  {et~al.}(2019{\natexlab{a}})\citenamefont {Zhang}, \citenamefont {Wang},
  \citenamefont {Zhang},\ and\ \citenamefont {Zhang}}]{Zhang:2018byx}%
  \BibitemOpen
  \bibfield  {author} {\bibinfo {author} {\bibfnamefont {X.-N.}\ \bibnamefont
  {Zhang}}, \bibinfo {author} {\bibfnamefont {L.-F.}\ \bibnamefont {Wang}},
  \bibinfo {author} {\bibfnamefont {J.-F.}\ \bibnamefont {Zhang}}, \ and\
  \bibinfo {author} {\bibfnamefont {X.}~\bibnamefont {Zhang}},\ }\href
  {\doibase 10.1103/PhysRevD.99.063510} {\bibfield  {journal} {\bibinfo
  {journal} {Phys. Rev. D}\ }\textbf {\bibinfo {volume} {99}},\ \bibinfo
  {pages} {063510} (\bibinfo {year} {2019}{\natexlab{a}})},\ \Eprint
  {http://arxiv.org/abs/1804.08379} {arXiv:1804.08379 [astro-ph.CO]}
  \BibitemShut {NoStop}%
\bibitem [{\citenamefont {Li}\ \emph {et~al.}(2020)\citenamefont {Li},
  \citenamefont {He}, \citenamefont {Zhang},\ and\ \citenamefont
  {Zhang}}]{Li:2019ajo}%
  \BibitemOpen
  \bibfield  {author} {\bibinfo {author} {\bibfnamefont {H.-L.}\ \bibnamefont
  {Li}}, \bibinfo {author} {\bibfnamefont {D.-Z.}\ \bibnamefont {He}}, \bibinfo
  {author} {\bibfnamefont {J.-F.}\ \bibnamefont {Zhang}}, \ and\ \bibinfo
  {author} {\bibfnamefont {X.}~\bibnamefont {Zhang}},\ }\href {\doibase
  10.1088/1475-7516/2020/06/038} {\bibfield  {journal} {\bibinfo  {journal}
  {JCAP}\ }\textbf {\bibinfo {volume} {06}},\ \bibinfo {pages} {038} (\bibinfo
  {year} {2020})},\ \Eprint {http://arxiv.org/abs/1908.03098} {arXiv:1908.03098
  [astro-ph.CO]} \BibitemShut {NoStop}%
\bibitem [{\citenamefont {Zhang}\ \emph
  {et~al.}(2019{\natexlab{b}})\citenamefont {Zhang}, \citenamefont {Zhang},
  \citenamefont {Jin}, \citenamefont {Qi},\ and\ \citenamefont
  {Zhang}}]{Zhang:2019loq}%
  \BibitemOpen
  \bibfield  {author} {\bibinfo {author} {\bibfnamefont {J.-F.}\ \bibnamefont
  {Zhang}}, \bibinfo {author} {\bibfnamefont {M.}~\bibnamefont {Zhang}},
  \bibinfo {author} {\bibfnamefont {S.-J.}\ \bibnamefont {Jin}}, \bibinfo
  {author} {\bibfnamefont {J.-Z.}\ \bibnamefont {Qi}}, \ and\ \bibinfo {author}
  {\bibfnamefont {X.}~\bibnamefont {Zhang}},\ }\href {\doibase
  10.1088/1475-7516/2019/09/068} {\bibfield  {journal} {\bibinfo  {journal}
  {JCAP}\ }\textbf {\bibinfo {volume} {09}},\ \bibinfo {pages} {068} (\bibinfo
  {year} {2019}{\natexlab{b}})},\ \Eprint {http://arxiv.org/abs/1907.03238}
  {arXiv:1907.03238 [astro-ph.CO]} \BibitemShut {NoStop}%
\bibitem [{\citenamefont {Zhang}\ \emph
  {et~al.}(2020{\natexlab{a}})\citenamefont {Zhang}, \citenamefont {Dong},
  \citenamefont {Qi},\ and\ \citenamefont {Zhang}}]{Zhang:2019ple}%
  \BibitemOpen
  \bibfield  {author} {\bibinfo {author} {\bibfnamefont {J.-F.}\ \bibnamefont
  {Zhang}}, \bibinfo {author} {\bibfnamefont {H.-Y.}\ \bibnamefont {Dong}},
  \bibinfo {author} {\bibfnamefont {J.-Z.}\ \bibnamefont {Qi}}, \ and\ \bibinfo
  {author} {\bibfnamefont {X.}~\bibnamefont {Zhang}},\ }\href {\doibase
  10.1140/epjc/s10052-020-7767-3} {\bibfield  {journal} {\bibinfo  {journal}
  {Eur. Phys. J. C}\ }\textbf {\bibinfo {volume} {80}},\ \bibinfo {pages} {217}
  (\bibinfo {year} {2020}{\natexlab{a}})},\ \Eprint
  {http://arxiv.org/abs/1906.07504} {arXiv:1906.07504 [astro-ph.CO]}
  \BibitemShut {NoStop}%
\bibitem [{\citenamefont {Wang}\ \emph {et~al.}(2020)\citenamefont {Wang},
  \citenamefont {Zhao}, \citenamefont {Zhang},\ and\ \citenamefont
  {Zhang}}]{Wang:2019tto}%
  \BibitemOpen
  \bibfield  {author} {\bibinfo {author} {\bibfnamefont {L.-F.}\ \bibnamefont
  {Wang}}, \bibinfo {author} {\bibfnamefont {Z.-W.}\ \bibnamefont {Zhao}},
  \bibinfo {author} {\bibfnamefont {J.-F.}\ \bibnamefont {Zhang}}, \ and\
  \bibinfo {author} {\bibfnamefont {X.}~\bibnamefont {Zhang}},\ }\href
  {\doibase 10.1088/1475-7516/2020/11/012} {\bibfield  {journal} {\bibinfo
  {journal} {JCAP}\ }\textbf {\bibinfo {volume} {11}},\ \bibinfo {pages} {012}
  (\bibinfo {year} {2020})},\ \Eprint {http://arxiv.org/abs/1907.01838}
  {arXiv:1907.01838 [astro-ph.CO]} \BibitemShut {NoStop}%
\bibitem [{\citenamefont {Zhao}\ \emph
  {et~al.}(2020{\natexlab{a}})\citenamefont {Zhao}, \citenamefont {Wang},
  \citenamefont {Zhang},\ and\ \citenamefont {Zhang}}]{Zhao:2019gyk}%
  \BibitemOpen
  \bibfield  {author} {\bibinfo {author} {\bibfnamefont {Z.-W.}\ \bibnamefont
  {Zhao}}, \bibinfo {author} {\bibfnamefont {L.-F.}\ \bibnamefont {Wang}},
  \bibinfo {author} {\bibfnamefont {J.-F.}\ \bibnamefont {Zhang}}, \ and\
  \bibinfo {author} {\bibfnamefont {X.}~\bibnamefont {Zhang}},\ }\href
  {\doibase 10.1016/j.scib.2020.04.032} {\bibfield  {journal} {\bibinfo
  {journal} {Sci. Bull.}\ }\textbf {\bibinfo {volume} {65}},\ \bibinfo {pages}
  {1340} (\bibinfo {year} {2020}{\natexlab{a}})},\ \Eprint
  {http://arxiv.org/abs/1912.11629} {arXiv:1912.11629 [astro-ph.CO]}
  \BibitemShut {NoStop}%
\bibitem [{\citenamefont {Jin}\ \emph {et~al.}(2020)\citenamefont {Jin},
  \citenamefont {He}, \citenamefont {Xu}, \citenamefont {Zhang},\ and\
  \citenamefont {Zhang}}]{Jin:2020hmc}%
  \BibitemOpen
  \bibfield  {author} {\bibinfo {author} {\bibfnamefont {S.-J.}\ \bibnamefont
  {Jin}}, \bibinfo {author} {\bibfnamefont {D.-Z.}\ \bibnamefont {He}},
  \bibinfo {author} {\bibfnamefont {Y.}~\bibnamefont {Xu}}, \bibinfo {author}
  {\bibfnamefont {J.-F.}\ \bibnamefont {Zhang}}, \ and\ \bibinfo {author}
  {\bibfnamefont {X.}~\bibnamefont {Zhang}},\ }\href {\doibase
  10.1088/1475-7516/2020/03/051} {\bibfield  {journal} {\bibinfo  {journal}
  {JCAP}\ }\textbf {\bibinfo {volume} {03}},\ \bibinfo {pages} {051} (\bibinfo
  {year} {2020})},\ \Eprint {http://arxiv.org/abs/2001.05393} {arXiv:2001.05393
  [astro-ph.CO]} \BibitemShut {NoStop}%
\bibitem [{\citenamefont {Wang}\ \emph
  {et~al.}(2021{\natexlab{b}})\citenamefont {Wang}, \citenamefont {Jin},
  \citenamefont {Zhang},\ and\ \citenamefont {Zhang}}]{Wang:2021srv}%
  \BibitemOpen
  \bibfield  {author} {\bibinfo {author} {\bibfnamefont {L.-F.}\ \bibnamefont
  {Wang}}, \bibinfo {author} {\bibfnamefont {S.-J.}\ \bibnamefont {Jin}},
  \bibinfo {author} {\bibfnamefont {J.-F.}\ \bibnamefont {Zhang}}, \ and\
  \bibinfo {author} {\bibfnamefont {X.}~\bibnamefont {Zhang}},\ }\href@noop {}
  {\  (\bibinfo {year} {2021}{\natexlab{b}})},\ \Eprint
  {http://arxiv.org/abs/2101.11882} {arXiv:2101.11882 [gr-qc]} \BibitemShut
  {NoStop}%
\bibitem [{\citenamefont {Qi}\ \emph {et~al.}(2021)\citenamefont {Qi},
  \citenamefont {Jin}, \citenamefont {Fan}, \citenamefont {Zhang},\ and\
  \citenamefont {Zhang}}]{Qi:2021iic}%
  \BibitemOpen
  \bibfield  {author} {\bibinfo {author} {\bibfnamefont {J.-Z.}\ \bibnamefont
  {Qi}}, \bibinfo {author} {\bibfnamefont {S.-J.}\ \bibnamefont {Jin}},
  \bibinfo {author} {\bibfnamefont {X.-L.}\ \bibnamefont {Fan}}, \bibinfo
  {author} {\bibfnamefont {J.-F.}\ \bibnamefont {Zhang}}, \ and\ \bibinfo
  {author} {\bibfnamefont {X.}~\bibnamefont {Zhang}},\ }\href@noop {} {\
  (\bibinfo {year} {2021})},\ \Eprint {http://arxiv.org/abs/2102.01292}
  {arXiv:2102.01292 [astro-ph.CO]} \BibitemShut {NoStop}%
\bibitem [{\citenamefont {Yang}(2021)}]{Yang:2021qge}%
  \BibitemOpen
  \bibfield  {author} {\bibinfo {author} {\bibfnamefont {T.}~\bibnamefont
  {Yang}},\ }\href {\doibase 10.1088/1475-7516/2021/05/044} {\bibfield
  {journal} {\bibinfo  {journal} {JCAP}\ }\textbf {\bibinfo {volume} {05}},\
  \bibinfo {pages} {044} (\bibinfo {year} {2021})},\ \Eprint
  {http://arxiv.org/abs/2103.01923} {arXiv:2103.01923 [astro-ph.CO]}
  \BibitemShut {NoStop}%
\bibitem [{\citenamefont {Yu}\ \emph {et~al.}(2021)\citenamefont {Yu},
  \citenamefont {Song}, \citenamefont {Ai}, \citenamefont {Gao}, \citenamefont
  {Wang}, \citenamefont {Wang}, \citenamefont {Lu}, \citenamefont {Fang},\ and\
  \citenamefont {Zhao}}]{Yu:2021nvx}%
  \BibitemOpen
  \bibfield  {author} {\bibinfo {author} {\bibfnamefont {J.}~\bibnamefont
  {Yu}}, \bibinfo {author} {\bibfnamefont {H.}~\bibnamefont {Song}}, \bibinfo
  {author} {\bibfnamefont {S.}~\bibnamefont {Ai}}, \bibinfo {author}
  {\bibfnamefont {H.}~\bibnamefont {Gao}}, \bibinfo {author} {\bibfnamefont
  {F.}~\bibnamefont {Wang}}, \bibinfo {author} {\bibfnamefont {Y.}~\bibnamefont
  {Wang}}, \bibinfo {author} {\bibfnamefont {Y.}~\bibnamefont {Lu}}, \bibinfo
  {author} {\bibfnamefont {W.}~\bibnamefont {Fang}}, \ and\ \bibinfo {author}
  {\bibfnamefont {W.}~\bibnamefont {Zhao}},\ }\href@noop {} {\  (\bibinfo
  {year} {2021})},\ \Eprint {http://arxiv.org/abs/2104.12374} {arXiv:2104.12374
  [astro-ph.HE]} \BibitemShut {NoStop}%
\bibitem [{\citenamefont {Abbott}\ \emph
  {et~al.}(2017{\natexlab{e}})\citenamefont {Abbott} \emph
  {et~al.}}]{Evans:2016mbw}%
  \BibitemOpen
  \bibfield  {author} {\bibinfo {author} {\bibfnamefont {B.~P.}\ \bibnamefont
  {Abbott}} \emph {et~al.} (\bibinfo {collaboration} {LIGO Scientific}),\
  }\href {\doibase 10.1088/1361-6382/aa51f4} {\bibfield  {journal} {\bibinfo
  {journal} {Class. Quant. Grav.}\ }\textbf {\bibinfo {volume} {34}},\ \bibinfo
  {pages} {044001} (\bibinfo {year} {2017}{\natexlab{e}})},\ \Eprint
  {http://arxiv.org/abs/1607.08697} {arXiv:1607.08697 [astro-ph.IM]}
  \BibitemShut {NoStop}%
\bibitem [{\citenamefont {Punturo}\ \emph {et~al.}(2010)\citenamefont {Punturo}
  \emph {et~al.}}]{Punturo:2010zz}%
  \BibitemOpen
  \bibfield  {author} {\bibinfo {author} {\bibfnamefont {M.}~\bibnamefont
  {Punturo}} \emph {et~al.},\ }\href {\doibase 10.1088/0264-9381/27/19/194002}
  {\bibfield  {journal} {\bibinfo  {journal} {Class. Quant. Grav.}\ }\textbf
  {\bibinfo {volume} {27}},\ \bibinfo {pages} {194002} (\bibinfo {year}
  {2010})}\BibitemShut {NoStop}%
\bibitem [{\citenamefont {Amaro-Seoane}\ \emph {et~al.}(2017)\citenamefont
  {Amaro-Seoane} \emph {et~al.}}]{Audley:2017drz}%
  \BibitemOpen
  \bibfield  {author} {\bibinfo {author} {\bibfnamefont {P.}~\bibnamefont
  {Amaro-Seoane}} \emph {et~al.} (\bibinfo {collaboration} {LISA}),\
  }\href@noop {} {\  (\bibinfo {year} {2017})},\ \Eprint
  {http://arxiv.org/abs/1702.00786} {arXiv:1702.00786 [astro-ph.IM]}
  \BibitemShut {NoStop}%
\bibitem [{\citenamefont {Luo}\ \emph {et~al.}(2020)\citenamefont {Luo} \emph
  {et~al.}}]{Luo:2020bls}%
  \BibitemOpen
  \bibfield  {author} {\bibinfo {author} {\bibfnamefont {J.}~\bibnamefont
  {Luo}} \emph {et~al.},\ }\href {\doibase 10.1088/1361-6382/aba66a} {\bibfield
   {journal} {\bibinfo  {journal} {Class. Quant. Grav.}\ }\textbf {\bibinfo
  {volume} {37}},\ \bibinfo {pages} {185013} (\bibinfo {year} {2020})},\
  \Eprint {http://arxiv.org/abs/2008.09534} {arXiv:2008.09534
  [physics.ins-det]} \BibitemShut {NoStop}%
\bibitem [{\citenamefont {Mei}\ \emph {et~al.}(2021)\citenamefont {Mei} \emph
  {et~al.}}]{mei2021tianqin}%
  \BibitemOpen
  \bibfield  {author} {\bibinfo {author} {\bibfnamefont {J.}~\bibnamefont
  {Mei}} \emph {et~al.},\ }\href {\doibase 10.1093/ptep/ptaa114} {\bibfield
  {journal} {\bibinfo  {journal} {Prog. Theor. Exp. Phys.}\ }\textbf {\bibinfo
  {volume} {2021}},\ \bibinfo {pages} {05A107} (\bibinfo {year}
  {2021})}\BibitemShut {NoStop}%
\bibitem [{\citenamefont {Hu}\ and\ \citenamefont {Wu}(2017)}]{Hu:2017mde}%
  \BibitemOpen
  \bibfield  {author} {\bibinfo {author} {\bibfnamefont {W.-R.}\ \bibnamefont
  {Hu}}\ and\ \bibinfo {author} {\bibfnamefont {Y.-L.}\ \bibnamefont {Wu}},\
  }\href {\doibase 10.1093/nsr/nwx116} {\bibfield  {journal} {\bibinfo
  {journal} {Natl. Sci. Rev.}\ }\textbf {\bibinfo {volume} {4}},\ \bibinfo
  {pages} {685} (\bibinfo {year} {2017})}\BibitemShut {NoStop}%
\bibitem [{\citenamefont {Ruan}\ \emph
  {et~al.}(2020{\natexlab{a}})\citenamefont {Ruan}, \citenamefont {Guo},
  \citenamefont {Cai},\ and\ \citenamefont {Zhang}}]{Guo:2018npi}%
  \BibitemOpen
  \bibfield  {author} {\bibinfo {author} {\bibfnamefont {W.-H.}\ \bibnamefont
  {Ruan}}, \bibinfo {author} {\bibfnamefont {Z.-K.}\ \bibnamefont {Guo}},
  \bibinfo {author} {\bibfnamefont {R.-G.}\ \bibnamefont {Cai}}, \ and\
  \bibinfo {author} {\bibfnamefont {Y.-Z.}\ \bibnamefont {Zhang}},\ }\href
  {\doibase 10.1142/S0217751X2050075X} {\bibfield  {journal} {\bibinfo
  {journal} {Int. J. Mod. Phys. A}\ }\textbf {\bibinfo {volume} {35}},\
  \bibinfo {pages} {2050075} (\bibinfo {year} {2020}{\natexlab{a}})},\ \Eprint
  {http://arxiv.org/abs/1807.09495} {arXiv:1807.09495 [gr-qc]} \BibitemShut
  {NoStop}%
\bibitem [{\citenamefont {Wu}\ \emph {et~al.}(2021)\citenamefont {Wu},
  \citenamefont {Luo}, \citenamefont {Wang} \emph {et~al.}}]{Taiji-1}%
  \BibitemOpen
  \bibfield  {author} {\bibinfo {author} {\bibfnamefont {Y.-L.}\ \bibnamefont
  {Wu}}, \bibinfo {author} {\bibfnamefont {Z.-R.}\ \bibnamefont {Luo}},
  \bibinfo {author} {\bibfnamefont {J.-Y.}\ \bibnamefont {Wang}},  \emph
  {et~al.} (\bibinfo {collaboration} {Taiji Sci Collaboration}),\ }\href
  {\doibase 10.1038/s42005-021-00529-z} {\bibfield  {journal} {\bibinfo
  {journal} {COMMUNICATIONS PHYSICS}\ }\textbf {\bibinfo {volume} {4}},\
  \bibinfo {pages} {34} (\bibinfo {year} {2021})}\BibitemShut {NoStop}%
\bibitem [{\citenamefont {Battye}\ \emph {et~al.}(2013)\citenamefont {Battye},
  \citenamefont {Browne}, \citenamefont {Dickinson}, \citenamefont {Heron},
  \citenamefont {Maffei},\ and\ \citenamefont {Pourtsidou}}]{Battye:2012tg}%
  \BibitemOpen
  \bibfield  {author} {\bibinfo {author} {\bibfnamefont {R.~A.}\ \bibnamefont
  {Battye}}, \bibinfo {author} {\bibfnamefont {I.~W.~A.}\ \bibnamefont
  {Browne}}, \bibinfo {author} {\bibfnamefont {C.}~\bibnamefont {Dickinson}},
  \bibinfo {author} {\bibfnamefont {G.}~\bibnamefont {Heron}}, \bibinfo
  {author} {\bibfnamefont {B.}~\bibnamefont {Maffei}}, \ and\ \bibinfo {author}
  {\bibfnamefont {A.}~\bibnamefont {Pourtsidou}},\ }\href {\doibase
  10.1093/mnras/stt1082} {\bibfield  {journal} {\bibinfo  {journal} {Mon. Not.
  Roy. Astron. Soc.}\ }\textbf {\bibinfo {volume} {434}},\ \bibinfo {pages}
  {1239} (\bibinfo {year} {2013})},\ \Eprint {http://arxiv.org/abs/1209.0343}
  {arXiv:1209.0343 [astro-ph.CO]} \BibitemShut {NoStop}%
\bibitem [{\citenamefont {Dickinson}(2014)}]{Dickinson:2014wda}%
  \BibitemOpen
  \bibfield  {author} {\bibinfo {author} {\bibfnamefont {C.}~\bibnamefont
  {Dickinson}}\ }(\bibinfo {year} {2014})\ \Eprint
  {http://arxiv.org/abs/1405.7936} {arXiv:1405.7936 [astro-ph.IM]} \BibitemShut
  {NoStop}%
\bibitem [{\citenamefont {Nan}\ \emph {et~al.}(2011)\citenamefont {Nan},
  \citenamefont {Li}, \citenamefont {Jin}, \citenamefont {Wang}, \citenamefont
  {Zhu}, \citenamefont {Zhu}, \citenamefont {Zhang}, \citenamefont {Yue},\ and\
  \citenamefont {Qian}}]{Nan:2011um}%
  \BibitemOpen
  \bibfield  {author} {\bibinfo {author} {\bibfnamefont {R.}~\bibnamefont
  {Nan}}, \bibinfo {author} {\bibfnamefont {D.}~\bibnamefont {Li}}, \bibinfo
  {author} {\bibfnamefont {C.}~\bibnamefont {Jin}}, \bibinfo {author}
  {\bibfnamefont {Q.}~\bibnamefont {Wang}}, \bibinfo {author} {\bibfnamefont
  {L.}~\bibnamefont {Zhu}}, \bibinfo {author} {\bibfnamefont {W.}~\bibnamefont
  {Zhu}}, \bibinfo {author} {\bibfnamefont {H.}~\bibnamefont {Zhang}}, \bibinfo
  {author} {\bibfnamefont {Y.}~\bibnamefont {Yue}}, \ and\ \bibinfo {author}
  {\bibfnamefont {L.}~\bibnamefont {Qian}},\ }\href {\doibase
  10.1142/S0218271811019335} {\bibfield  {journal} {\bibinfo  {journal} {Int.
  J. Mod. Phys. D}\ }\textbf {\bibinfo {volume} {20}},\ \bibinfo {pages} {989}
  (\bibinfo {year} {2011})},\ \Eprint {http://arxiv.org/abs/1105.3794}
  {arXiv:1105.3794 [astro-ph.IM]} \BibitemShut {NoStop}%
\bibitem [{\citenamefont {Li}\ \emph {et~al.}(2013)\citenamefont {Li},
  \citenamefont {Nan},\ and\ \citenamefont {Pan}}]{Li:2012ub}%
  \BibitemOpen
  \bibfield  {author} {\bibinfo {author} {\bibfnamefont {D.}~\bibnamefont
  {Li}}, \bibinfo {author} {\bibfnamefont {R.}~\bibnamefont {Nan}}, \ and\
  \bibinfo {author} {\bibfnamefont {Z.}~\bibnamefont {Pan}},\ }\href {\doibase
  10.1017/S1743921312024015} {\bibfield  {journal} {\bibinfo  {journal} {IAU
  Symp.}\ }\textbf {\bibinfo {volume} {291}},\ \bibinfo {pages} {325} (\bibinfo
  {year} {2013})},\ \Eprint {http://arxiv.org/abs/1210.5785} {arXiv:1210.5785
  [astro-ph.IM]} \BibitemShut {NoStop}%
\bibitem [{\citenamefont {Yu}\ \emph {et~al.}(2017)\citenamefont {Yu},
  \citenamefont {Pen}, \citenamefont {Zhang}, \citenamefont {Li},\ and\
  \citenamefont {Chen}}]{Yu:2017hqz}%
  \BibitemOpen
  \bibfield  {author} {\bibinfo {author} {\bibfnamefont {H.-R.}\ \bibnamefont
  {Yu}}, \bibinfo {author} {\bibfnamefont {U.-L.}\ \bibnamefont {Pen}},
  \bibinfo {author} {\bibfnamefont {T.-J.}\ \bibnamefont {Zhang}}, \bibinfo
  {author} {\bibfnamefont {D.}~\bibnamefont {Li}}, \ and\ \bibinfo {author}
  {\bibfnamefont {X.}~\bibnamefont {Chen}},\ }\href {\doibase
  10.1088/1674-4527/17/6/49} {\bibfield  {journal} {\bibinfo  {journal} {Res.
  Astron. Astrophys.}\ }\textbf {\bibinfo {volume} {17}},\ \bibinfo {pages}
  {049} (\bibinfo {year} {2017})},\ \Eprint {http://arxiv.org/abs/1704.04338}
  {arXiv:1704.04338 [astro-ph.GA]} \BibitemShut {NoStop}%
\bibitem [{\citenamefont {Braun}\ \emph {et~al.}(2015)\citenamefont {Braun},
  \citenamefont {Bourke}, \citenamefont {Green}, \citenamefont {Keane},\ and\
  \citenamefont {Wagg}}]{Braun:2015zta}%
  \BibitemOpen
  \bibfield  {author} {\bibinfo {author} {\bibfnamefont {R.}~\bibnamefont
  {Braun}}, \bibinfo {author} {\bibfnamefont {T.}~\bibnamefont {Bourke}},
  \bibinfo {author} {\bibfnamefont {J.~A.}\ \bibnamefont {Green}}, \bibinfo
  {author} {\bibfnamefont {E.}~\bibnamefont {Keane}}, \ and\ \bibinfo {author}
  {\bibfnamefont {J.}~\bibnamefont {Wagg}},\ }\href {\doibase
  10.22323/1.215.0174} {\bibfield  {journal} {\bibinfo  {journal} {PoS}\
  }\textbf {\bibinfo {volume} {AASKA14}},\ \bibinfo {pages} {174} (\bibinfo
  {year} {2015})}\BibitemShut {NoStop}%
\bibitem [{\citenamefont {Bull}\ \emph
  {et~al.}(2015{\natexlab{a}})\citenamefont {Bull}, \citenamefont {Camera},
  \citenamefont {Raccanelli}, \citenamefont {Blake}, \citenamefont {Ferreira},
  \citenamefont {Santos},\ and\ \citenamefont {Schwarz}}]{Bull:2015esa}%
  \BibitemOpen
  \bibfield  {author} {\bibinfo {author} {\bibfnamefont {P.}~\bibnamefont
  {Bull}}, \bibinfo {author} {\bibfnamefont {S.}~\bibnamefont {Camera}},
  \bibinfo {author} {\bibfnamefont {A.}~\bibnamefont {Raccanelli}}, \bibinfo
  {author} {\bibfnamefont {C.}~\bibnamefont {Blake}}, \bibinfo {author}
  {\bibfnamefont {P.}~\bibnamefont {Ferreira}}, \bibinfo {author}
  {\bibfnamefont {M.}~\bibnamefont {Santos}}, \ and\ \bibinfo {author}
  {\bibfnamefont {D.~J.}\ \bibnamefont {Schwarz}},\ }\href {\doibase
  10.22323/1.215.0024} {\bibfield  {journal} {\bibinfo  {journal} {PoS}\
  }\textbf {\bibinfo {volume} {AASKA14}},\ \bibinfo {pages} {024} (\bibinfo
  {year} {2015}{\natexlab{a}})}\BibitemShut {NoStop}%
\bibitem [{\citenamefont {Bacon}\ \emph {et~al.}(2020)\citenamefont {Bacon}
  \emph {et~al.}}]{Bacon:2018dui}%
  \BibitemOpen
  \bibfield  {author} {\bibinfo {author} {\bibfnamefont {D.~J.}\ \bibnamefont
  {Bacon}} \emph {et~al.} (\bibinfo {collaboration} {SKA}),\ }\href {\doibase
  10.1017/pasa.2019.51} {\bibfield  {journal} {\bibinfo  {journal} {Publ.
  Astron. Soc. Austral.}\ }\textbf {\bibinfo {volume} {37}},\ \bibinfo {pages}
  {e007} (\bibinfo {year} {2020})},\ \Eprint {http://arxiv.org/abs/1811.02743}
  {arXiv:1811.02743 [astro-ph.CO]} \BibitemShut {NoStop}%
\bibitem [{\citenamefont {Braun}\ \emph {et~al.}(2019)\citenamefont {Braun},
  \citenamefont {Bonaldi}, \citenamefont {Bourke}, \citenamefont {Keane},\ and\
  \citenamefont {Wagg}}]{Braun:2019gdo}%
  \BibitemOpen
  \bibfield  {author} {\bibinfo {author} {\bibfnamefont {R.}~\bibnamefont
  {Braun}}, \bibinfo {author} {\bibfnamefont {A.}~\bibnamefont {Bonaldi}},
  \bibinfo {author} {\bibfnamefont {T.}~\bibnamefont {Bourke}}, \bibinfo
  {author} {\bibfnamefont {E.}~\bibnamefont {Keane}}, \ and\ \bibinfo {author}
  {\bibfnamefont {J.}~\bibnamefont {Wagg}},\ }\href@noop {} {\  (\bibinfo
  {year} {2019})},\ \Eprint {http://arxiv.org/abs/1912.12699} {arXiv:1912.12699
  [astro-ph.IM]} \BibitemShut {NoStop}%
\bibitem [{\citenamefont {{Chen}}(2011)}]{2011SSPMA..41.1358C}%
  \BibitemOpen
  \bibfield  {author} {\bibinfo {author} {\bibfnamefont {X.}~\bibnamefont
  {{Chen}}},\ }\href {\doibase 10.1360/132011-972} {\bibfield  {journal}
  {\bibinfo  {journal} {Scientia Sinica Physica, Mechanica \& Astronomica}\
  }\textbf {\bibinfo {volume} {41}},\ \bibinfo {pages} {1358} (\bibinfo {year}
  {2011})}\BibitemShut {NoStop}%
\bibitem [{\citenamefont {Chen}(2012)}]{Chen:2012xu}%
  \BibitemOpen
  \bibfield  {author} {\bibinfo {author} {\bibfnamefont {X.}~\bibnamefont
  {Chen}},\ }\href {\doibase 10.1142/S2010194512006459} {\bibfield  {journal}
  {\bibinfo  {journal} {Int. J. Mod. Phys. Conf. Ser.}\ }\textbf {\bibinfo
  {volume} {12}},\ \bibinfo {pages} {256} (\bibinfo {year} {2012})},\ \Eprint
  {http://arxiv.org/abs/1212.6278} {arXiv:1212.6278 [astro-ph.IM]} \BibitemShut
  {NoStop}%
\bibitem [{\citenamefont {Xu}\ \emph {et~al.}(2015)\citenamefont {Xu},
  \citenamefont {Wang},\ and\ \citenamefont {Chen}}]{Xu:2014bya}%
  \BibitemOpen
  \bibfield  {author} {\bibinfo {author} {\bibfnamefont {Y.}~\bibnamefont
  {Xu}}, \bibinfo {author} {\bibfnamefont {X.}~\bibnamefont {Wang}}, \ and\
  \bibinfo {author} {\bibfnamefont {X.}~\bibnamefont {Chen}},\ }\href {\doibase
  10.1088/0004-637X/798/1/40} {\bibfield  {journal} {\bibinfo  {journal}
  {Astrophys. J.}\ }\textbf {\bibinfo {volume} {798}},\ \bibinfo {pages} {40}
  (\bibinfo {year} {2015})},\ \Eprint {http://arxiv.org/abs/1410.7794}
  {arXiv:1410.7794 [astro-ph.CO]} \BibitemShut {NoStop}%
\bibitem [{\citenamefont {Zhang}\ \emph
  {et~al.}(2019{\natexlab{c}})\citenamefont {Zhang}, \citenamefont {Gao},
  \citenamefont {He},\ and\ \citenamefont {Zhang}}]{Zhang:2019dyq}%
  \BibitemOpen
  \bibfield  {author} {\bibinfo {author} {\bibfnamefont {J.-F.}\ \bibnamefont
  {Zhang}}, \bibinfo {author} {\bibfnamefont {L.-Y.}\ \bibnamefont {Gao}},
  \bibinfo {author} {\bibfnamefont {D.-Z.}\ \bibnamefont {He}}, \ and\ \bibinfo
  {author} {\bibfnamefont {X.}~\bibnamefont {Zhang}},\ }\href {\doibase
  10.1016/j.physletb.2019.135064} {\bibfield  {journal} {\bibinfo  {journal}
  {Phys. Lett. B}\ }\textbf {\bibinfo {volume} {799}},\ \bibinfo {pages}
  {135064} (\bibinfo {year} {2019}{\natexlab{c}})},\ \Eprint
  {http://arxiv.org/abs/1908.03732} {arXiv:1908.03732 [astro-ph.CO]}
  \BibitemShut {NoStop}%
\bibitem [{\citenamefont {Zhang}\ \emph
  {et~al.}(2020{\natexlab{b}})\citenamefont {Zhang}, \citenamefont {Wang},\
  and\ \citenamefont {Zhang}}]{Zhang:2019ipd}%
  \BibitemOpen
  \bibfield  {author} {\bibinfo {author} {\bibfnamefont {J.-F.}\ \bibnamefont
  {Zhang}}, \bibinfo {author} {\bibfnamefont {B.}~\bibnamefont {Wang}}, \ and\
  \bibinfo {author} {\bibfnamefont {X.}~\bibnamefont {Zhang}},\ }\href
  {\doibase 10.1007/s11433-019-1516-y} {\bibfield  {journal} {\bibinfo
  {journal} {Sci. China Phys. Mech. Astron.}\ }\textbf {\bibinfo {volume}
  {63}},\ \bibinfo {pages} {280411} (\bibinfo {year} {2020}{\natexlab{b}})},\
  \Eprint {http://arxiv.org/abs/1907.00179} {arXiv:1907.00179 [astro-ph.CO]}
  \BibitemShut {NoStop}%
\bibitem [{\citenamefont {Zhang}\ \emph {et~al.}(2021)\citenamefont {Zhang},
  \citenamefont {Wang}, \citenamefont {Wu}, \citenamefont {Qi}, \citenamefont
  {Xu}, \citenamefont {Zhang},\ and\ \citenamefont {Zhang}}]{Zhang:2021yof}%
  \BibitemOpen
  \bibfield  {author} {\bibinfo {author} {\bibfnamefont {M.}~\bibnamefont
  {Zhang}}, \bibinfo {author} {\bibfnamefont {B.}~\bibnamefont {Wang}},
  \bibinfo {author} {\bibfnamefont {P.-J.}\ \bibnamefont {Wu}}, \bibinfo
  {author} {\bibfnamefont {J.-Z.}\ \bibnamefont {Qi}}, \bibinfo {author}
  {\bibfnamefont {Y.}~\bibnamefont {Xu}}, \bibinfo {author} {\bibfnamefont
  {J.-F.}\ \bibnamefont {Zhang}}, \ and\ \bibinfo {author} {\bibfnamefont
  {X.}~\bibnamefont {Zhang}},\ }\href {\doibase 10.3847/1538-4357/ac0ef5}
  {\bibfield  {journal} {\bibinfo  {journal} {Astrophys. J.}\ }\textbf
  {\bibinfo {volume} {918}},\ \bibinfo {pages} {56} (\bibinfo {year} {2021})},\
  \Eprint {http://arxiv.org/abs/2102.03979} {arXiv:2102.03979 [astro-ph.CO]}
  \BibitemShut {NoStop}%
\bibitem [{\citenamefont {Xu}\ and\ \citenamefont {Zhang}(2020)}]{Xu:2020uws}%
  \BibitemOpen
  \bibfield  {author} {\bibinfo {author} {\bibfnamefont {Y.}~\bibnamefont
  {Xu}}\ and\ \bibinfo {author} {\bibfnamefont {X.}~\bibnamefont {Zhang}},\
  }\href {\doibase 10.1007/s11433-020-1544-3} {\bibfield  {journal} {\bibinfo
  {journal} {Sci. China Phys. Mech. Astron.}\ }\textbf {\bibinfo {volume}
  {63}},\ \bibinfo {pages} {270431} (\bibinfo {year} {2020})},\ \Eprint
  {http://arxiv.org/abs/2002.00572} {arXiv:2002.00572 [astro-ph.CO]}
  \BibitemShut {NoStop}%
\bibitem [{\citenamefont {Zhao}\ \emph {et~al.}(2011)\citenamefont {Zhao},
  \citenamefont {Van Den~Broeck}, \citenamefont {Baskaran},\ and\ \citenamefont
  {Li}}]{Zhao:2010sz}%
  \BibitemOpen
  \bibfield  {author} {\bibinfo {author} {\bibfnamefont {W.}~\bibnamefont
  {Zhao}}, \bibinfo {author} {\bibfnamefont {C.}~\bibnamefont {Van
  Den~Broeck}}, \bibinfo {author} {\bibfnamefont {D.}~\bibnamefont {Baskaran}},
  \ and\ \bibinfo {author} {\bibfnamefont {T.}~\bibnamefont {Li}},\ }\href
  {\doibase 10.1103/PhysRevD.83.023005} {\bibfield  {journal} {\bibinfo
  {journal} {Phys. Rev. D}\ }\textbf {\bibinfo {volume} {83}},\ \bibinfo
  {pages} {023005} (\bibinfo {year} {2011})},\ \Eprint
  {http://arxiv.org/abs/1009.0206} {arXiv:1009.0206 [astro-ph.CO]} \BibitemShut
  {NoStop}%
\bibitem [{\citenamefont {Sathyaprakash}\ and\ \citenamefont
  {Schutz}(2009)}]{Sathyaprakash:2009xs}%
  \BibitemOpen
  \bibfield  {author} {\bibinfo {author} {\bibfnamefont {B.}~\bibnamefont
  {Sathyaprakash}}\ and\ \bibinfo {author} {\bibfnamefont {B.}~\bibnamefont
  {Schutz}},\ }\href {\doibase 10.12942/lrr-2009-2} {\bibfield  {journal}
  {\bibinfo  {journal} {Living Rev. Rel.}\ }\textbf {\bibinfo {volume} {12}},\
  \bibinfo {pages} {2} (\bibinfo {year} {2009})},\ \Eprint
  {http://arxiv.org/abs/0903.0338} {arXiv:0903.0338 [gr-qc]} \BibitemShut
  {NoStop}%
\bibitem [{\citenamefont {Blanchet}\ and\ \citenamefont
  {Iyer}(2005)}]{Blanchet:2004bb}%
  \BibitemOpen
  \bibfield  {author} {\bibinfo {author} {\bibfnamefont {L.}~\bibnamefont
  {Blanchet}}\ and\ \bibinfo {author} {\bibfnamefont {B.~R.}\ \bibnamefont
  {Iyer}},\ }\href {\doibase 10.1103/PhysRevD.71.024004} {\bibfield  {journal}
  {\bibinfo  {journal} {Phys. Rev. D}\ }\textbf {\bibinfo {volume} {71}},\
  \bibinfo {pages} {024004} (\bibinfo {year} {2005})},\ \Eprint
  {http://arxiv.org/abs/gr-qc/0409094} {arXiv:gr-qc/0409094} \BibitemShut
  {NoStop}%
\bibitem [{ETc()}]{ETcurve-web}%
  \BibitemOpen
  \href@noop {} {}\bibinfo {howpublished}
  {\url{https://www.et-gw.eu/index.php/etsensitivities/}}\BibitemShut {NoStop}%
\bibitem [{\citenamefont {Hirata}\ \emph {et~al.}(2010)\citenamefont {Hirata},
  \citenamefont {Holz},\ and\ \citenamefont {Cutler}}]{Hirata:2010ba}%
  \BibitemOpen
  \bibfield  {author} {\bibinfo {author} {\bibfnamefont {C.~M.}\ \bibnamefont
  {Hirata}}, \bibinfo {author} {\bibfnamefont {D.~E.}\ \bibnamefont {Holz}}, \
  and\ \bibinfo {author} {\bibfnamefont {C.}~\bibnamefont {Cutler}},\ }\href
  {\doibase 10.1103/PhysRevD.81.124046} {\bibfield  {journal} {\bibinfo
  {journal} {Phys. Rev. D}\ }\textbf {\bibinfo {volume} {81}},\ \bibinfo
  {pages} {124046} (\bibinfo {year} {2010})},\ \Eprint
  {http://arxiv.org/abs/1004.3988} {arXiv:1004.3988 [astro-ph.CO]} \BibitemShut
  {NoStop}%
\bibitem [{\citenamefont {Speri}\ \emph {et~al.}(2021)\citenamefont {Speri},
  \citenamefont {Tamanini}, \citenamefont {Caldwell}, \citenamefont {Gair},\
  and\ \citenamefont {Wang}}]{Speri:2020hwc}%
  \BibitemOpen
  \bibfield  {author} {\bibinfo {author} {\bibfnamefont {L.}~\bibnamefont
  {Speri}}, \bibinfo {author} {\bibfnamefont {N.}~\bibnamefont {Tamanini}},
  \bibinfo {author} {\bibfnamefont {R.~R.}\ \bibnamefont {Caldwell}}, \bibinfo
  {author} {\bibfnamefont {J.~R.}\ \bibnamefont {Gair}}, \ and\ \bibinfo
  {author} {\bibfnamefont {B.}~\bibnamefont {Wang}},\ }\href {\doibase
  10.1103/PhysRevD.103.083526} {\bibfield  {journal} {\bibinfo  {journal}
  {Phys. Rev. D}\ }\textbf {\bibinfo {volume} {103}},\ \bibinfo {pages}
  {083526} (\bibinfo {year} {2021})},\ \Eprint
  {http://arxiv.org/abs/2010.09049} {arXiv:2010.09049 [astro-ph.CO]}
  \BibitemShut {NoStop}%
\bibitem [{\citenamefont {Kocsis}\ \emph {et~al.}(2006)\citenamefont {Kocsis},
  \citenamefont {Frei}, \citenamefont {Haiman},\ and\ \citenamefont
  {Menou}}]{Kocsis:2005vv}%
  \BibitemOpen
  \bibfield  {author} {\bibinfo {author} {\bibfnamefont {B.}~\bibnamefont
  {Kocsis}}, \bibinfo {author} {\bibfnamefont {Z.}~\bibnamefont {Frei}},
  \bibinfo {author} {\bibfnamefont {Z.}~\bibnamefont {Haiman}}, \ and\ \bibinfo
  {author} {\bibfnamefont {K.}~\bibnamefont {Menou}},\ }\href {\doibase
  10.1086/498236} {\bibfield  {journal} {\bibinfo  {journal} {Astrophys. J.}\
  }\textbf {\bibinfo {volume} {637}},\ \bibinfo {pages} {27} (\bibinfo {year}
  {2006})},\ \Eprint {http://arxiv.org/abs/astro-ph/0505394}
  {arXiv:astro-ph/0505394} \BibitemShut {NoStop}%
\bibitem [{\citenamefont {Li}(2015)}]{li2015extracting}%
  \BibitemOpen
  \bibfield  {author} {\bibinfo {author} {\bibfnamefont {T.~G.}\ \bibnamefont
  {Li}},\ }\href@noop {} {\emph {\bibinfo {title} {Extracting physics from
  gravitational waves: testing the strong-field dynamics of general relativity
  and inferring the large-scale structure of the Universe}}}\ (\bibinfo
  {publisher} {Springer},\ \bibinfo {year} {2015})\BibitemShut {NoStop}%
\bibitem [{\citenamefont {Klein}\ \emph {et~al.}(2016)\citenamefont {Klein}
  \emph {et~al.}}]{Klein:2015hvg}%
  \BibitemOpen
  \bibfield  {author} {\bibinfo {author} {\bibfnamefont {A.}~\bibnamefont
  {Klein}} \emph {et~al.},\ }\href {\doibase 10.1103/PhysRevD.93.024003}
  {\bibfield  {journal} {\bibinfo  {journal} {Phys. Rev. D}\ }\textbf {\bibinfo
  {volume} {93}},\ \bibinfo {pages} {024003} (\bibinfo {year} {2016})},\
  \Eprint {http://arxiv.org/abs/1511.05581} {arXiv:1511.05581 [gr-qc]}
  \BibitemShut {NoStop}%
\bibitem [{\citenamefont {Ruan}\ \emph
  {et~al.}(2020{\natexlab{b}})\citenamefont {Ruan}, \citenamefont {Liu},
  \citenamefont {Guo}, \citenamefont {Wu},\ and\ \citenamefont
  {Cai}}]{Ruan:2020smc}%
  \BibitemOpen
  \bibfield  {author} {\bibinfo {author} {\bibfnamefont {W.-H.}\ \bibnamefont
  {Ruan}}, \bibinfo {author} {\bibfnamefont {C.}~\bibnamefont {Liu}}, \bibinfo
  {author} {\bibfnamefont {Z.-K.}\ \bibnamefont {Guo}}, \bibinfo {author}
  {\bibfnamefont {Y.-L.}\ \bibnamefont {Wu}}, \ and\ \bibinfo {author}
  {\bibfnamefont {R.-G.}\ \bibnamefont {Cai}},\ }\href {\doibase
  10.1038/s41550-019-1008-4} {\bibfield  {journal} {\bibinfo  {journal} {Nature
  Astron.}\ }\textbf {\bibinfo {volume} {4}},\ \bibinfo {pages} {108} (\bibinfo
  {year} {2020}{\natexlab{b}})},\ \Eprint {http://arxiv.org/abs/2002.03603}
  {arXiv:2002.03603 [gr-qc]} \BibitemShut {NoStop}%
\bibitem [{\citenamefont {Cutler}(1998)}]{Cutler:1997ta}%
  \BibitemOpen
  \bibfield  {author} {\bibinfo {author} {\bibfnamefont {C.}~\bibnamefont
  {Cutler}},\ }\href {\doibase 10.1103/PhysRevD.57.7089} {\bibfield  {journal}
  {\bibinfo  {journal} {Phys. Rev. D}\ }\textbf {\bibinfo {volume} {57}},\
  \bibinfo {pages} {7089} (\bibinfo {year} {1998})},\ \Eprint
  {http://arxiv.org/abs/gr-qc/9703068} {arXiv:gr-qc/9703068} \BibitemShut
  {NoStop}%
\bibitem [{\citenamefont {Krolak}\ \emph {et~al.}(1995)\citenamefont {Krolak},
  \citenamefont {Kokkotas},\ and\ \citenamefont {Schaefer}}]{Krolak:1995md}%
  \BibitemOpen
  \bibfield  {author} {\bibinfo {author} {\bibfnamefont {A.}~\bibnamefont
  {Krolak}}, \bibinfo {author} {\bibfnamefont {K.~D.}\ \bibnamefont
  {Kokkotas}}, \ and\ \bibinfo {author} {\bibfnamefont {G.}~\bibnamefont
  {Schaefer}},\ }\href {\doibase 10.1103/PhysRevD.52.2089} {\bibfield
  {journal} {\bibinfo  {journal} {Phys. Rev. D}\ }\textbf {\bibinfo {volume}
  {52}},\ \bibinfo {pages} {2089} (\bibinfo {year} {1995})},\ \Eprint
  {http://arxiv.org/abs/gr-qc/9503013} {arXiv:gr-qc/9503013} \BibitemShut
  {NoStop}%
\bibitem [{\citenamefont {Buonanno}\ \emph {et~al.}(2009)\citenamefont
  {Buonanno}, \citenamefont {Iyer}, \citenamefont {Ochsner}, \citenamefont
  {Pan},\ and\ \citenamefont {Sathyaprakash}}]{Buonanno:2009zt}%
  \BibitemOpen
  \bibfield  {author} {\bibinfo {author} {\bibfnamefont {A.}~\bibnamefont
  {Buonanno}}, \bibinfo {author} {\bibfnamefont {B.}~\bibnamefont {Iyer}},
  \bibinfo {author} {\bibfnamefont {E.}~\bibnamefont {Ochsner}}, \bibinfo
  {author} {\bibfnamefont {Y.}~\bibnamefont {Pan}}, \ and\ \bibinfo {author}
  {\bibfnamefont {B.}~\bibnamefont {Sathyaprakash}},\ }\href {\doibase
  10.1103/PhysRevD.80.084043} {\bibfield  {journal} {\bibinfo  {journal} {Phys.
  Rev. D}\ }\textbf {\bibinfo {volume} {80}},\ \bibinfo {pages} {084043}
  (\bibinfo {year} {2009})},\ \Eprint {http://arxiv.org/abs/0907.0700}
  {arXiv:0907.0700 [gr-qc]} \BibitemShut {NoStop}%
\bibitem [{\citenamefont {Feng}\ \emph {et~al.}(2019)\citenamefont {Feng},
  \citenamefont {Wang}, \citenamefont {Hu}, \citenamefont {Hu},\ and\
  \citenamefont {Wang}}]{Feng:2019wgq}%
  \BibitemOpen
  \bibfield  {author} {\bibinfo {author} {\bibfnamefont {W.-F.}\ \bibnamefont
  {Feng}}, \bibinfo {author} {\bibfnamefont {H.-T.}\ \bibnamefont {Wang}},
  \bibinfo {author} {\bibfnamefont {X.-C.}\ \bibnamefont {Hu}}, \bibinfo
  {author} {\bibfnamefont {Y.-M.}\ \bibnamefont {Hu}}, \ and\ \bibinfo {author}
  {\bibfnamefont {Y.}~\bibnamefont {Wang}},\ }\href {\doibase
  10.1103/PhysRevD.99.123002} {\bibfield  {journal} {\bibinfo  {journal} {Phys.
  Rev. D}\ }\textbf {\bibinfo {volume} {99}},\ \bibinfo {pages} {123002}
  (\bibinfo {year} {2019})},\ \Eprint {http://arxiv.org/abs/1901.02159}
  {arXiv:1901.02159 [astro-ph.IM]} \BibitemShut {NoStop}%
\bibitem [{\citenamefont {Tamanini}(2017)}]{Tamanini:2016uin}%
  \BibitemOpen
  \bibfield  {author} {\bibinfo {author} {\bibfnamefont {N.}~\bibnamefont
  {Tamanini}},\ }\href {\doibase 10.1088/1742-6596/840/1/012029} {\bibfield
  {journal} {\bibinfo  {journal} {J. Phys. Conf. Ser.}\ }\textbf {\bibinfo
  {volume} {840}},\ \bibinfo {pages} {012029} (\bibinfo {year} {2017})},\
  \Eprint {http://arxiv.org/abs/1612.02634} {arXiv:1612.02634 [astro-ph.CO]}
  \BibitemShut {NoStop}%
\bibitem [{\citenamefont {Dahlen}\ \emph {et~al.}(2013)\citenamefont {Dahlen}
  \emph {et~al.}}]{Dahlen:2013fea}%
  \BibitemOpen
  \bibfield  {author} {\bibinfo {author} {\bibfnamefont {T.}~\bibnamefont
  {Dahlen}} \emph {et~al.},\ }\href {\doibase 10.1088/0004-637X/775/2/93}
  {\bibfield  {journal} {\bibinfo  {journal} {Astrophys. J.}\ }\textbf
  {\bibinfo {volume} {775}},\ \bibinfo {pages} {93} (\bibinfo {year} {2013})},\
  \Eprint {http://arxiv.org/abs/1308.5353} {arXiv:1308.5353 [astro-ph.CO]}
  \BibitemShut {NoStop}%
\bibitem [{\citenamefont {Ilbert}\ \emph {et~al.}(2013)\citenamefont {Ilbert}
  \emph {et~al.}}]{Ilbert:2013bf}%
  \BibitemOpen
  \bibfield  {author} {\bibinfo {author} {\bibfnamefont {O.}~\bibnamefont
  {Ilbert}} \emph {et~al.},\ }\href {\doibase 10.1051/0004-6361/201321100}
  {\bibfield  {journal} {\bibinfo  {journal} {Astron. Astrophys.}\ }\textbf
  {\bibinfo {volume} {556}},\ \bibinfo {pages} {A55} (\bibinfo {year}
  {2013})},\ \Eprint {http://arxiv.org/abs/1301.3157} {arXiv:1301.3157
  [astro-ph.CO]} \BibitemShut {NoStop}%
\bibitem [{\citenamefont {Bull}\ \emph
  {et~al.}(2015{\natexlab{b}})\citenamefont {Bull}, \citenamefont {Ferreira},
  \citenamefont {Patel},\ and\ \citenamefont {Santos}}]{Bull:2014rha}%
  \BibitemOpen
  \bibfield  {author} {\bibinfo {author} {\bibfnamefont {P.}~\bibnamefont
  {Bull}}, \bibinfo {author} {\bibfnamefont {P.~G.}\ \bibnamefont {Ferreira}},
  \bibinfo {author} {\bibfnamefont {P.}~\bibnamefont {Patel}}, \ and\ \bibinfo
  {author} {\bibfnamefont {M.~G.}\ \bibnamefont {Santos}},\ }\href {\doibase
  10.1088/0004-637X/803/1/21} {\bibfield  {journal} {\bibinfo  {journal}
  {Astrophys. J.}\ }\textbf {\bibinfo {volume} {803}},\ \bibinfo {pages} {21}
  (\bibinfo {year} {2015}{\natexlab{b}})},\ \Eprint
  {http://arxiv.org/abs/1405.1452} {arXiv:1405.1452 [astro-ph.CO]} \BibitemShut
  {NoStop}%
\bibitem [{\citenamefont {Kaiser}(1987)}]{Kaiser:1987qv}%
  \BibitemOpen
  \bibfield  {author} {\bibinfo {author} {\bibfnamefont {N.}~\bibnamefont
  {Kaiser}},\ }\href@noop {} {\bibfield  {journal} {\bibinfo  {journal} {Mon.
  Not. Roy. Astron. Soc.}\ }\textbf {\bibinfo {volume} {227}},\ \bibinfo
  {pages} {1} (\bibinfo {year} {1987})}\BibitemShut {NoStop}%
\bibitem [{\citenamefont {Seo}\ and\ \citenamefont
  {Eisenstein}(2003)}]{Seo:2003pu}%
  \BibitemOpen
  \bibfield  {author} {\bibinfo {author} {\bibfnamefont {H.-J.}\ \bibnamefont
  {Seo}}\ and\ \bibinfo {author} {\bibfnamefont {D.~J.}\ \bibnamefont
  {Eisenstein}},\ }\href {\doibase 10.1086/379122} {\bibfield  {journal}
  {\bibinfo  {journal} {Astrophys. J.}\ }\textbf {\bibinfo {volume} {598}},\
  \bibinfo {pages} {720} (\bibinfo {year} {2003})},\ \Eprint
  {http://arxiv.org/abs/astro-ph/0307460} {arXiv:astro-ph/0307460} \BibitemShut
  {NoStop}%
\bibitem [{\citenamefont {Li}\ \emph {et~al.}(2007)\citenamefont {Li},
  \citenamefont {Jing}, \citenamefont {Kauffmann}, \citenamefont {Boerner},
  \citenamefont {Kang},\ and\ \citenamefont {Wang}}]{Li:2007rpa}%
  \BibitemOpen
  \bibfield  {author} {\bibinfo {author} {\bibfnamefont {C.}~\bibnamefont
  {Li}}, \bibinfo {author} {\bibfnamefont {Y.~P.}\ \bibnamefont {Jing}},
  \bibinfo {author} {\bibfnamefont {G.}~\bibnamefont {Kauffmann}}, \bibinfo
  {author} {\bibfnamefont {G.}~\bibnamefont {Boerner}}, \bibinfo {author}
  {\bibfnamefont {X.}~\bibnamefont {Kang}}, \ and\ \bibinfo {author}
  {\bibfnamefont {L.}~\bibnamefont {Wang}},\ }\href {\doibase
  10.1111/j.1365-2966.2007.11518.x} {\bibfield  {journal} {\bibinfo  {journal}
  {Mon. Not. Roy. Astron. Soc.}\ }\textbf {\bibinfo {volume} {376}},\ \bibinfo
  {pages} {984} (\bibinfo {year} {2007})},\ \Eprint
  {http://arxiv.org/abs/astro-ph/0701218} {arXiv:astro-ph/0701218} \BibitemShut
  {NoStop}%
\bibitem [{\citenamefont {Lewis}\ \emph {et~al.}(2000)\citenamefont {Lewis},
  \citenamefont {Challinor},\ and\ \citenamefont {Lasenby}}]{Lewis:1999bs}%
  \BibitemOpen
  \bibfield  {author} {\bibinfo {author} {\bibfnamefont {A.}~\bibnamefont
  {Lewis}}, \bibinfo {author} {\bibfnamefont {A.}~\bibnamefont {Challinor}}, \
  and\ \bibinfo {author} {\bibfnamefont {A.}~\bibnamefont {Lasenby}},\ }\href
  {\doibase 10.1086/309179} {\bibfield  {journal} {\bibinfo  {journal}
  {Astrophys. J.}\ }\textbf {\bibinfo {volume} {538}},\ \bibinfo {pages} {473}
  (\bibinfo {year} {2000})},\ \Eprint {http://arxiv.org/abs/astro-ph/9911177}
  {arXiv:astro-ph/9911177} \BibitemShut {NoStop}%
\bibitem [{\citenamefont {Tegmark}(1997)}]{Tegmark:1997rp}%
  \BibitemOpen
  \bibfield  {author} {\bibinfo {author} {\bibfnamefont {M.}~\bibnamefont
  {Tegmark}},\ }\href {\doibase 10.1103/PhysRevLett.79.3806} {\bibfield
  {journal} {\bibinfo  {journal} {Phys. Rev. Lett.}\ }\textbf {\bibinfo
  {volume} {79}},\ \bibinfo {pages} {3806} (\bibinfo {year} {1997})},\ \Eprint
  {http://arxiv.org/abs/astro-ph/9706198} {arXiv:astro-ph/9706198} \BibitemShut
  {NoStop}%
\bibitem [{\citenamefont {Pourtsidou}\ \emph {et~al.}(2017)\citenamefont
  {Pourtsidou}, \citenamefont {Bacon},\ and\ \citenamefont
  {Crittenden}}]{Pourtsidou:2016dzn}%
  \BibitemOpen
  \bibfield  {author} {\bibinfo {author} {\bibfnamefont {A.}~\bibnamefont
  {Pourtsidou}}, \bibinfo {author} {\bibfnamefont {D.}~\bibnamefont {Bacon}}, \
  and\ \bibinfo {author} {\bibfnamefont {R.}~\bibnamefont {Crittenden}},\
  }\href {\doibase 10.1093/mnras/stx1479} {\bibfield  {journal} {\bibinfo
  {journal} {Mon. Not. Roy. Astron. Soc.}\ }\textbf {\bibinfo {volume} {470}},\
  \bibinfo {pages} {4251} (\bibinfo {year} {2017})},\ \Eprint
  {http://arxiv.org/abs/1610.04189} {arXiv:1610.04189 [astro-ph.CO]}
  \BibitemShut {NoStop}%
\bibitem [{\citenamefont {Smith}\ \emph {et~al.}(2003)\citenamefont {Smith},
  \citenamefont {Peacock}, \citenamefont {Jenkins}, \citenamefont {White},
  \citenamefont {Frenk}, \citenamefont {Pearce}, \citenamefont {Thomas},
  \citenamefont {Efstathiou},\ and\ \citenamefont {Couchmann}}]{Smith:2002dz}%
  \BibitemOpen
  \bibfield  {author} {\bibinfo {author} {\bibfnamefont {R.}~\bibnamefont
  {Smith}}, \bibinfo {author} {\bibfnamefont {J.}~\bibnamefont {Peacock}},
  \bibinfo {author} {\bibfnamefont {A.}~\bibnamefont {Jenkins}}, \bibinfo
  {author} {\bibfnamefont {S.}~\bibnamefont {White}}, \bibinfo {author}
  {\bibfnamefont {C.}~\bibnamefont {Frenk}}, \bibinfo {author} {\bibfnamefont
  {F.}~\bibnamefont {Pearce}}, \bibinfo {author} {\bibfnamefont
  {P.}~\bibnamefont {Thomas}}, \bibinfo {author} {\bibfnamefont
  {G.}~\bibnamefont {Efstathiou}}, \ and\ \bibinfo {author} {\bibfnamefont
  {H.}~\bibnamefont {Couchmann}} (\bibinfo {collaboration} {VIRGO
  Consortium}),\ }\href {\doibase 10.1046/j.1365-8711.2003.06503.x} {\bibfield
  {journal} {\bibinfo  {journal} {Mon. Not. Roy. Astron. Soc.}\ }\textbf
  {\bibinfo {volume} {341}},\ \bibinfo {pages} {1311} (\bibinfo {year}
  {2003})},\ \Eprint {http://arxiv.org/abs/astro-ph/0207664}
  {arXiv:astro-ph/0207664} \BibitemShut {NoStop}%
\bibitem [{\citenamefont {Witzemann}\ \emph {et~al.}(2018)\citenamefont
  {Witzemann}, \citenamefont {Bull}, \citenamefont {Clarkson}, \citenamefont
  {Santos}, \citenamefont {Spinelli},\ and\ \citenamefont
  {Weltman}}]{Witzemann:2017lhi}%
  \BibitemOpen
  \bibfield  {author} {\bibinfo {author} {\bibfnamefont {A.}~\bibnamefont
  {Witzemann}}, \bibinfo {author} {\bibfnamefont {P.}~\bibnamefont {Bull}},
  \bibinfo {author} {\bibfnamefont {C.}~\bibnamefont {Clarkson}}, \bibinfo
  {author} {\bibfnamefont {M.~G.}\ \bibnamefont {Santos}}, \bibinfo {author}
  {\bibfnamefont {M.}~\bibnamefont {Spinelli}}, \ and\ \bibinfo {author}
  {\bibfnamefont {A.}~\bibnamefont {Weltman}},\ }\href {\doibase
  10.1093/mnrasl/sly062} {\bibfield  {journal} {\bibinfo  {journal} {Mon. Not.
  Roy. Astron. Soc.}\ }\textbf {\bibinfo {volume} {477}},\ \bibinfo {pages}
  {L122} (\bibinfo {year} {2018})},\ \Eprint {http://arxiv.org/abs/1711.02179}
  {arXiv:1711.02179 [astro-ph.CO]} \BibitemShut {NoStop}%
\bibitem [{\citenamefont {Lewis}\ and\ \citenamefont
  {Bridle}(2002)}]{Lewis:2002ah}%
  \BibitemOpen
  \bibfield  {author} {\bibinfo {author} {\bibfnamefont {A.}~\bibnamefont
  {Lewis}}\ and\ \bibinfo {author} {\bibfnamefont {S.}~\bibnamefont {Bridle}},\
  }\href {\doibase 10.1103/PhysRevD.66.103511} {\bibfield  {journal} {\bibinfo
  {journal} {Phys. Rev. D}\ }\textbf {\bibinfo {volume} {66}},\ \bibinfo
  {pages} {103511} (\bibinfo {year} {2002})},\ \Eprint
  {http://arxiv.org/abs/astro-ph/0205436} {arXiv:astro-ph/0205436} \BibitemShut
  {NoStop}%
\bibitem [{\citenamefont {Zhao}\ \emph
  {et~al.}(2020{\natexlab{b}})\citenamefont {Zhao}, \citenamefont {Li},
  \citenamefont {Qi}, \citenamefont {Gao}, \citenamefont {Zhang},\ and\
  \citenamefont {Zhang}}]{Zhao:2020ole}%
  \BibitemOpen
  \bibfield  {author} {\bibinfo {author} {\bibfnamefont {Z.-W.}\ \bibnamefont
  {Zhao}}, \bibinfo {author} {\bibfnamefont {Z.-X.}\ \bibnamefont {Li}},
  \bibinfo {author} {\bibfnamefont {J.-Z.}\ \bibnamefont {Qi}}, \bibinfo
  {author} {\bibfnamefont {H.}~\bibnamefont {Gao}}, \bibinfo {author}
  {\bibfnamefont {J.-F.}\ \bibnamefont {Zhang}}, \ and\ \bibinfo {author}
  {\bibfnamefont {X.}~\bibnamefont {Zhang}},\ }\href {\doibase
  10.3847/1538-4357/abb8ce} {\bibfield  {journal} {\bibinfo  {journal}
  {Astrophys. J.}\ }\textbf {\bibinfo {volume} {903}},\ \bibinfo {pages} {83}
  (\bibinfo {year} {2020}{\natexlab{b}})},\ \Eprint
  {http://arxiv.org/abs/2006.01450} {arXiv:2006.01450 [astro-ph.CO]}
  \BibitemShut {NoStop}%
\bibitem [{\citenamefont {Suyu}\ \emph {et~al.}(2020)\citenamefont {Suyu} \emph
  {et~al.}}]{Suyu:2020opl}%
  \BibitemOpen
  \bibfield  {author} {\bibinfo {author} {\bibfnamefont {S.~H.}\ \bibnamefont
  {Suyu}} \emph {et~al.},\ }\href {\doibase 10.1051/0004-6361/202037757}
  {\bibfield  {journal} {\bibinfo  {journal} {Astron. Astrophys.}\ }\textbf
  {\bibinfo {volume} {644}},\ \bibinfo {pages} {A162} (\bibinfo {year}
  {2020})},\ \Eprint {http://arxiv.org/abs/2002.08378} {arXiv:2002.08378
  [astro-ph.CO]} \BibitemShut {NoStop}%
\bibitem [{\citenamefont {Ili\'c}\ \emph {et~al.}(2021)\citenamefont {Ili\'c}
  \emph {et~al.}}]{Euclid:2021qvm}%
  \BibitemOpen
  \bibfield  {author} {\bibinfo {author} {\bibfnamefont {S.}~\bibnamefont
  {Ili\'c}} \emph {et~al.} (\bibinfo {collaboration} {Euclid}),\ }\href
  {\doibase 10.1051/0004-6361/202141556} {\  (\bibinfo {year} {2021}),\
  10.1051/0004-6361/202141556},\ \Eprint {http://arxiv.org/abs/2106.08346}
  {arXiv:2106.08346 [astro-ph.CO]} \BibitemShut {NoStop}%
\bibitem [{\citenamefont {Abell}\ \emph {et~al.}(2009)\citenamefont {Abell}
  \emph {et~al.}}]{Abell:2009aa}%
  \BibitemOpen
  \bibfield  {author} {\bibinfo {author} {\bibfnamefont {P.~A.}\ \bibnamefont
  {Abell}} \emph {et~al.} (\bibinfo {collaboration} {LSST Science, LSST
  Project}),\ }\href@noop {} {\  (\bibinfo {year} {2009})},\ \Eprint
  {http://arxiv.org/abs/0912.0201} {arXiv:0912.0201 [astro-ph.IM]} \BibitemShut
  {NoStop}%
\bibitem [{\citenamefont {Laureijs}\ \emph {et~al.}(2011)\citenamefont
  {Laureijs} \emph {et~al.}}]{Laureijs:2011gra}%
  \BibitemOpen
  \bibfield  {author} {\bibinfo {author} {\bibfnamefont {R.}~\bibnamefont
  {Laureijs}} \emph {et~al.} (\bibinfo {collaboration} {EUCLID}),\ }\href@noop
  {} {\  (\bibinfo {year} {2011})},\ \Eprint {http://arxiv.org/abs/1110.3193}
  {arXiv:1110.3193 [astro-ph.CO]} \BibitemShut {NoStop}%
\bibitem [{\citenamefont {Spergel}\ \emph {et~al.}(2013)\citenamefont {Spergel}
  \emph {et~al.}}]{Spergel:2013tha}%
  \BibitemOpen
  \bibfield  {author} {\bibinfo {author} {\bibfnamefont {D.}~\bibnamefont
  {Spergel}} \emph {et~al.},\ }\href@noop {} {\  (\bibinfo {year} {2013})},\
  \Eprint {http://arxiv.org/abs/1305.5422} {arXiv:1305.5422 [astro-ph.IM]}
  \BibitemShut {NoStop}%
\bibitem [{\citenamefont {Rajagopal}\ and\ \citenamefont
  {Romani}(1995)}]{Rajagopal:1994zj}%
  \BibitemOpen
  \bibfield  {author} {\bibinfo {author} {\bibfnamefont {M.}~\bibnamefont
  {Rajagopal}}\ and\ \bibinfo {author} {\bibfnamefont {R.~W.}\ \bibnamefont
  {Romani}},\ }\href {\doibase 10.1086/175813} {\bibfield  {journal} {\bibinfo
  {journal} {Astrophys. J.}\ }\textbf {\bibinfo {volume} {446}},\ \bibinfo
  {pages} {543} (\bibinfo {year} {1995})},\ \Eprint
  {http://arxiv.org/abs/astro-ph/9412038} {arXiv:astro-ph/9412038} \BibitemShut
  {NoStop}%
\end{thebibliography}%

\end{document}